\documentclass{aa}  

\usepackage{graphicx}
\usepackage{amsmath}
\usepackage{txfonts}
\def\referee{ }

\usepackage{subcaption}

\begin{document}

   \title{Estimating stellar ages and metallicities from parallaxes and broadband photometry - successes and shortcomings}
  \titlerunning{Stellar ages and metallicities from parallaxes and broadband photometry}
  \authorrunning{L. M. Howes et al.}

   \author{Louise M. Howes\inst{1}, Lennart Lindegren\inst{1}, Sofia Feltzing\inst{1}, Ross P. Church\inst{1},
          \and
          Thomas Bensby\inst{1}
          }

   \institute{$^{1}$Lund Observatory, Department of Astronomy and Theoretical Physics, Box 43, SE-221 00 Lund, Sweden\\
              \email{[louise; lennart; sofia; ross; tbensby]@astro.lu.se}
             }

   \date{Received XXXX; accepted XXXX}

 
  \abstract{A deep understanding of the Milky Way galaxy, its formation and evolution requires observations of huge numbers of stars. Stellar photometry, therefore, provides an economical method to obtain intrinsic stellar parameters. With the addition of distance information -- a prospect made real for more than a billion stars with the second {\it Gaia} data release -- deriving reliable ages from photometry is a possibility. 

 We have developed a Bayesian method that generates 2D probability maps of a star’s age and metallicity from photometry and parallax using isochrones. Our synthetic tests show that including a near-UV passband enables us to break the degeneracy between a star's age and metallicity for certain evolutionary stages. It is possible to find well-constrained ages and metallicities for turn-off and sub-giant stars with colours including a $U$ band and a parallax with uncertainty less than $\sim20\%$. Metallicities alone are possible for the main sequence and giant branch. 

{\referee We find good agreement with the literature when we apply our method to the {\it Gaia} benchmark stars, particularly for turn-off and young stars.} Further tests on the old open cluster NGC 188, however, reveal significant limitations in the stellar isochrones. The ages derived for the cluster stars vary with evolutionary stage, such that turn-off ages disagree with those on the sub-giant branch, and metallicities vary significantly throughout. Furthermore, the parameters vary appreciably depending on which colour combinations are used in the derivation. {\referee We identify the causes of these mismatches and show that improvements are needed in the modelling of giant branch stars and in the creation and calibration of synthetic near-UV photometry.}

Our results warn against applying isochrone fitting indiscriminately. {\referee In particular, the uncertainty on the stellar models should be quantitatively taken into account.} Further efforts to improve the models will result in significant advancements in our ability to study the Galaxy.}


   \maketitle
%

\section{Introduction} 
{\referee Galactic archaeology, concerned with understanding the Milky Way as a galaxy \citep{2002ARA&A..40..487F}, allows us to explore a large range of astrophysical processes found in the Universe.} Within Galactic archaeology we are especially interested in understanding how the mass of the Galaxy was assembled. Although the larger part of the mass in a galaxy like the Milky Way is in the form of dark matter \citep{2016ARA&A..54..529B} the stellar mass provides an excellent tracer of past events, such as mergers and secular evolution (see, e.g., discussions in \citealt{2016AN....337..703M} and \citealt{2018arXiv180107720A}). 

To obtain constraints on the stellar populations it is important to study the Milky Way on as large a scale as possible, which means collecting information and characterising as many stars as possible across all regions of the Galaxy. Close to the Sun we are able to obtain extremely detailed information about the stars, but as we move further away the targets get fainter and precise data such as high resolution spectroscopy become prohibitively expensive. It is therefore natural that astronomers have sought alternative ways to determine stellar parameters such as effective temperature and metallicity, which characterise a star and allow us to investigate its origin. 

Photometric measurements are relatively economic compared to stellar spectroscopy, and with the advent of the large CCD camera it became possible to obtain data for all the stars in the field-of-view of the telescope. \citet{2005ARA&A..43..293B} provides an extensive review of the (broadband) photometric systems in use today. Different passband systems have been designed with specific needs in mind; for example surveys such as the Sloan Digital Sky Survey \citep{2000AJ....120.1579Y} and SkyMapper \citep{2018PASA...35...10W}, have customised their passbands in order to optimise their scientific value. 

To study the Milky Way as a galaxy and trace its assembly and evolution over cosmic time, it is desirable to have not only metallicities but also ages for the stars. Deriving stellar ages is a notoriously hard problem, even when good data are available. \citet{2010ARA&A..48..581S} provides an exhaustive review of the various methods available to the Galactic archaeologist. As broadband photometry is a relatively affordable commodity it is interesting to explore how well we can use it to derive metallicities and ages for stars. 

To derive the age of a star from broadband photometry we need to know its distance, so that we can compare its colour and luminosity to stellar evolutionary models such as the PARSEC \citep{2012MNRAS.427..127B} or MIST \citep{2016ApJS..222....8D} models. Until recently this has only been possible for stars that are  within about 100\,pc of the Sun, where the \textsc{Hipparcos} satellite provided stellar parallaxes that can be used to obtain the absolute magnitudes of the stars. With the current and upcoming releases of {\it Gaia} data, however, this situation is completely changing \citep{2001A&A...369..339P, 2016A&A...595A...2G, 2016A&A...595A...4L, 2018A&A...616A...1G}. \textit{Gaia} is providing parallaxes and broadband photometry for more than one billion stars across the whole Milky Way, providing an unprecedented sample of stars. 

\textit{Gaia}'s parallaxes and photometry are complemented with spectra from the onboard Radial Velocity Spectrograph \citep[RVS,][]{2016A&A...585A..93R} and groundbased massively multiplex surveys \citep[e.g., RAVE, SEGUE, Gaia-ESO Survey, APOGEE-I and II, LAMOST, GALAH, DESI, WEAVE and 4MOST: ][respectively]{2006AJ....132.1645S,2009AJ....137.4377Y,Gilmore:2012wv,2017AJ....154...94M,2012RAA....12..723Z,2015MNRAS.449.2604D,2016arXiv161100036D,2014SPIE.9147E..0LD,2016SPIE.9908E..1OD}. However, only stars brighter than magnitude 15 in the \textit{Gaia} $G$ band will have RVS spectra and from the ground we will only be able to obtain data for perhaps up to 50 million stars, leaving the vast majority of the fainter \textit{Gaia} stars without stellar spectra. Hence it is interesting to obtain stellar properties directly from the available broadband photometry combined with stellar parallaxes. 

To do this we have written code based on the Bayesian age estimation code first described by \citet{2005A&A...436..127J}. Our code derives a 2D probability map of metallicity and age for each star. We employ these maps to find combinations of photometric passbands and stellar type where unique solutions in this 2D space are possible. {\referee We test our code in two separate ways. Firstly we use theoretical data to explore the limits of the technique. We explore different types of star, combinations of photometric passbands, and astrometric uncertainties, in order to determine where the technique can provide useful results. Secondly we apply the knowledge learned to real observational data. These observational tests determine how feasible the method is in reality and where efforts are needed in order to make progress.} Our investigation has highlighted a number of successes and shortcomings. Although not all of these issues are unknown, with the newly available {\it Gaia} astrometry it becomes important to re-address them. Of particular interest for us is their impact on Galactic archaeology and provide pointers to where it would be particularly pertinent to put in (substantial) efforts for those developing stellar photometric surveys as well as those calculating stellar isochrones. 

 The structure of the paper is as follows. In Section~2, we describe the theoretical investigations and their results; then in Section~3 we discuss the physical reasons behind the successful combinations that we found. Section~4 goes on to test the method and chosen passbands on real data from the {\it Gaia} benchmark stars \citep{2014A&A...566A..98B} and the open cluster NGC 188. {\referee We highlight areas where the technique works well, and where there are discrepancies between the theory and the observations.} Section~5 discusses the problems raised by these tests, and how progress can be made towards fixing them. Finally in Section~6 we summarise the results found.

\section{Determination of stellar ages and metallicities -- theoretical tests}
\label{sec:two}

\subsection{Combining astrometry and photometry}

In order to see what information about the intrinsic stellar parameters can be gained from astrometry and photometry, we have developed the $\mathcal{G}$ function\footnote{The symbol $\mathcal{G}$ is used here to avoid a possible confusion with the Gaia magnitude, denoted $G$.}, $\mathcal{G}(\tau,\zeta\,|\,\vec{x})$, which is a two-dimensional map describing the marginal likelihood of different stellar ages $\tau$ and metallicities $\zeta$, given a set of observations $\vec{x}$, and a given set of isochrones. The theory behind the Bayesian technique of the code, and in particular the method for incorporating the parallax measurement into the probability, is described in Appendix~\ref{section:maths}. 

{\referee  We have chosen this two-dimensional format in order to demonstrate the correlations between ages and metallicities, and the difficulty in determining each uniquely. Our aim is to demonstrate visually how much information on these two parameters photometry and astrometry can provide. Because of this, we decided not to marginalise over [Fe/H] (or conversely age) to produce a best-fitting age (or [Fe/H]). Furthermore, marginalisation would require the input of a prior on that parameter -- and this is best done when considering the properties of the stellar population being studied (e.g., studying stars in a known cluster would have a different prior than one used for Galactic disc stars in the field). In our case, we wish to demonstrate the method without restricting it to a particular scientific use.}

We note that for the purposes of the tests in this paper, we have not considered reddening as a parameter in our likelihood calculations. This simplifies the problem, allowing us to see what is possible and where the problems lie. In order to use this code on large quantities of stars in the Galaxy, one would first need to make the appropriate reddening corrections. {\referee As reddening would be a further variable to include, we did not want to skew our results or hide problems with age and metallicity determination by also trying to solve for it here.}

With the mathematical framework in place to produce a multi-dimensional likelihood function for the stellar parameters of interest, we describe the isochrones used, and then go on to explore the results using simulated data.

\subsection{Stellar isochrones}
The model relies on using isochrones to produce the intrinsic stellar properties, so we created a large grid of precomputed isochrones such that the spacing between each parameter is considerably smaller than any required uncertainty level. {\referee This allows us to avoid the lengthy computations caused by on-the-fly interpolations.}

As one goal of this paper is to test which photometric passbands provide useful information about intrinsic stellar parameters, we chose to use the PARSEC isochrones (\citealt{2012MNRAS.427..127B}, using also the extensions made available by \citealt{2014MNRAS.445.4287T, 2014MNRAS.444.2525C, 2015MNRAS.452.1068C}) that have been calculated for a wide variety of passbands, including crucially the {\it Gaia} $G$ band. {\referee These tests were performed pre-{\it Gaia} DR2, and so the isochrones are based on the pre-launch $G$ band filter curve. This is not a perfect representation of the now-available photometry \citep{2016A&A...595A...7C}, but for our theoretical tests this is unimportant.} From the PARSEC interpolator available on the website\footnote{http://stev.oapd.inaf.it/cgi-bin/cmd}, we created a grid spanning ages from $0.1$ to $13.5$\,Gyr in intervals of $0.1$\,Gyr and metallicities of [M/H]$=-2.2$ to $+0.5$\,dex in intervals of $0.05$\,dex.

The PARSEC isochrones do not include variation in $\alpha$ and [Fe/H] separately, rather incorporating the total metallicity in [M/H], so we have also treated metallicity as one parameter, where $\zeta=$ [M/H] $=\log (Z/Z_\odot) - \log (X/X_\odot)$.  When comparing to observations, we take the approximation [Fe/H] $\approx$ [M/H]. As the code is adaptable to different grids of isochrones, we retain the capability to reintroduce the $\alpha$ abundance and study the $\mathcal{G}$ function in 3D.

\subsection{Demonstration of the method}
\label{sec:demo}

\begin{figure}
  \centering
  \includegraphics[width=0.99\columnwidth]{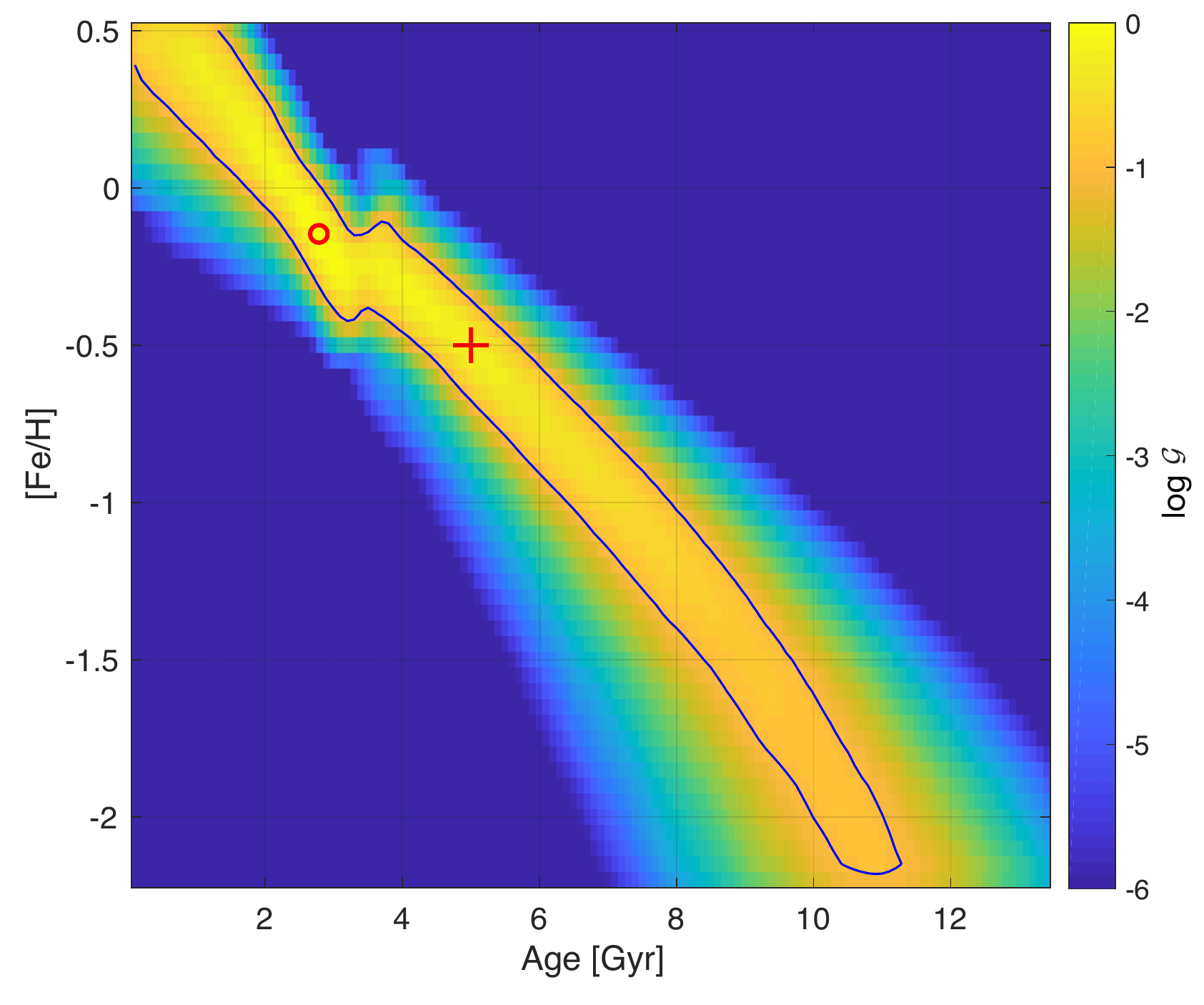}
  \caption{The $\mathcal{G}$ function, a 2D age-metallicity probability map, calculated for the example star in Sect. \ref{sec:demo}, using the {\it Gaia} $G$ band, and the colours $(G-G_\text{BP})$ and $(G-G_\text{RP})$. The colour bar shows the log of the marginalised likelihood, normalised to 1 at its highest point, with the $90\%$ confidence interval highlighted with a solid dark line. The red circle represents the point of highest probability, and the red cross represents the star's true [Fe/H] and age.}
  \label{fig:ExamplegFunc}
\end{figure}
\begin{figure}
  \centering
  \includegraphics[width=0.99\columnwidth]{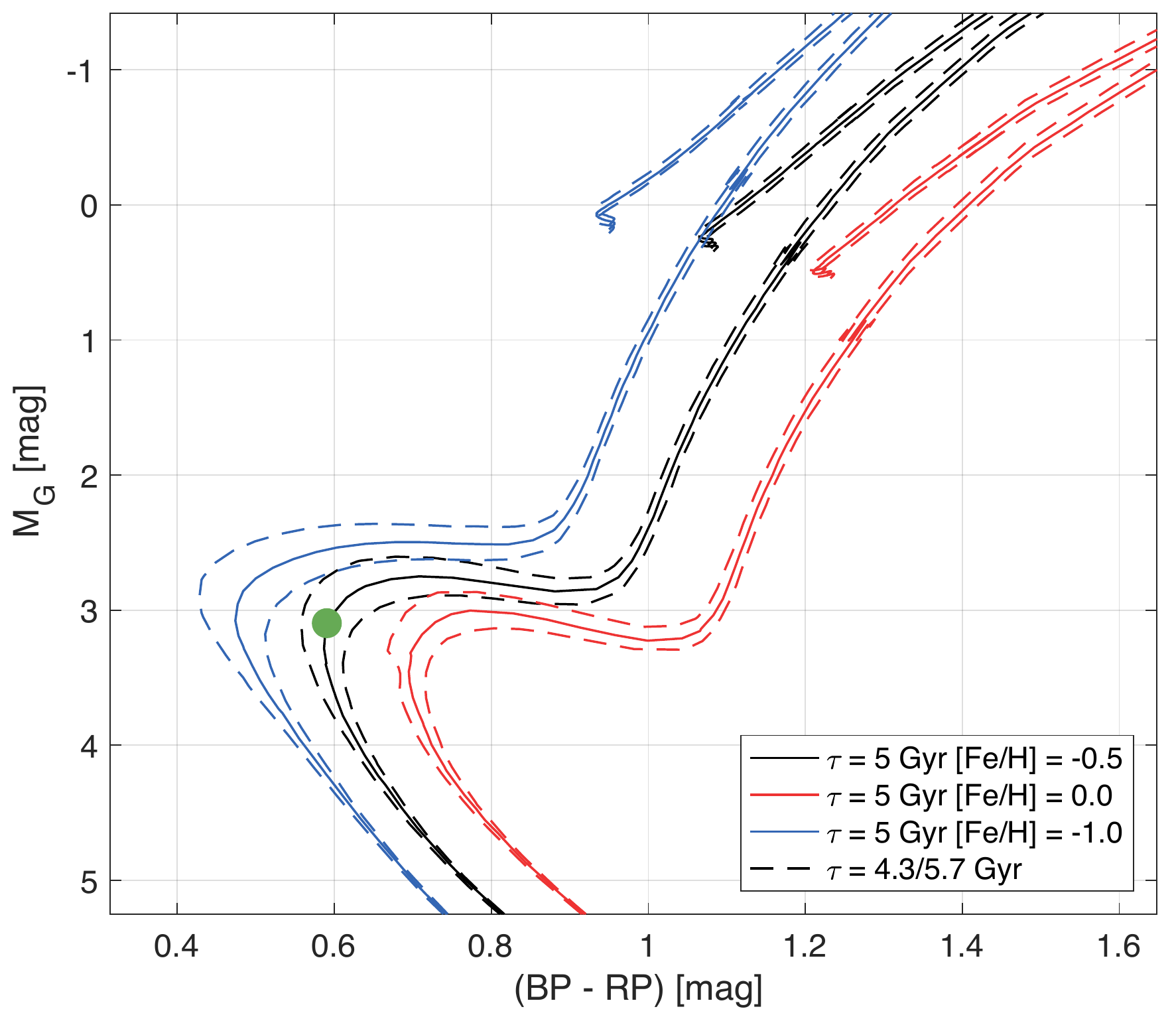}
  \caption{The colour-magnitude diagram of our example star (Fig.~\ref{fig:ExamplegFunc}), where the colour is the {\it Gaia} $(G_\text{BP}-G_\text{RP})$, and the absolute magnitude is calculated for {\it Gaia} $G$. The filled green circle represents our example star, with age $\tau=5$\,Gyr and metallicity [Fe/H]$=-0.5$\,dex. The black solid line shows the isochrone with the true age and metallicity of the star. The blue and red solid lines show isochrones with [Fe/H]$=-1.0$ and $+0.0$\,dex respectively, and the dashed lines on the left and right sides of the solid lines are for ages offset by $-0.7$ and $+0.7$\,Gyr.}
  \label{fig:ExampleCMD}
\end{figure}

We show in Fig.~\ref{fig:ExamplegFunc} the age-metallicity $\mathcal{G}$ function for the best-case scenario, a typical main-sequence turnoff (MSTO) star. Such a star lies in a region of the colour-magnitude diagram (CMD\footnote{The term CMD is often used to mean either plots of colour vs. apparent magnitude, or colour vs. absolute magnitude. The second of these is also often referred to as the observational HR diagram. Throughout the paper we do not show any plots with apparent magnitude, so we will use `CMD' when we are discussing colour vs. absolute magnitude plots.}) where the respective isochrones show the widest separation (Fig.~\ref{fig:ExampleCMD}). The assumed ``true'' parameters in this case are an age of $\tau = 5.0$\,Gyr, metallicity [Fe/H]$ = -0.5$\,dex, and parallax $\varpi = 5.0$\,mas. In this example, to illustrate what is possible with photometry and astrometry both coming exclusively from {\it Gaia}, we calculate the $\mathcal{G}$ function using only the parallax, $G$ magnitude, and the colours $(G-G_\text{RP})$ and $(G-G_\text{BP})$. We assume that the observed values coincide with the expected ones calculated from the true parameters and isochrones, with uncertainties of 0.3\,mas in parallax and 0.01\,mag in each of the three photometric quantities. Only the relative parallax uncertainty matters for the distance information, and the assumed relative {\referee uncertainty} of 6\% may therefore be representative for stars at much larger distances when $\sigma_{\varpi}$ is much smaller than in this example.

Diagrams like Fig.~\ref{fig:ExamplegFunc} are used throughout this paper to examine the ability of different colour combinations to constrain the age and metallicity of a star, given its parallax, and a short explanation should be given about its interpretation. The diagram shows, for age versus metallicity, the value of the $\mathcal{G}$ function on a logarithmic colour scale ranging from bright yellow for $\mathcal{G}=1$ to dark blue for numerically insignificant values (in this case, $\mathcal{G}< 10^{-15}$). The red cross marks the true age and metallicity, here 5\,Gyr and $-0.5$\,dex. The $\mathcal{G}$ function is always normalised to 1 at its maximum, which is marked by the red circle. The yellow area within the black contour is the region with $\mathcal{G}>0.1$. As explained in Appendix~\ref{section:maths}, the $\mathcal{G}$ function is the likelihood of the age and metallicity marginalised (i.e.\ averaged) over the remaining parameters weighted by their respective prior densities. This means that the $\mathcal{G}$ function, multiplied by the prior (joint) density of age and metallicity, equals the posterior density of the two parameters. In other words, for a uniform prior in age and metallicity, the $\mathcal{G}$ function is simply the Bayesian posterior density of these parameters (up to some normalisation factor); in this case the red circle is the maximum a posteriori probability (MAP) estimate, and the black contour can be interpreted equivalent to a 90\% confidence region of this estimate. With other priors in age and/or metallicity, both the MAP estimate and the confidence region may be very different.

As is clear from Fig.~\ref{fig:ExamplegFunc}, the most likely age and [Fe/H] lie in a narrow region of the probability space, however there is a strong degeneracy between the two properties. Without further information, the star could have almost any age or metallicity, and the most likely values are not equal to the true values. With additional information, such as prior knowledge of either of the two parameters, it would be possible to constrain the other to a very narrow confidence interval. Another source of crucial information could be other common photometric passbands, and so here we test a variety of combinations.

\subsection{Testing available photometric passbands}
\label{sec:tests}

Large scale surveys have made photometry publicly available across the optical and infrared spectrum for large numbers of stars. We endeavoured to test which of these provide key information on the stars' parameters, and so here we will consider combinations of a variety of frequently used passbands, in the infrared, optical, and near-UV. All passbands are included in the model as $(G-x)$ colours, where $x$ is the passband in question.

\begin{figure}
  \centering
  \includegraphics[width=0.99\columnwidth]{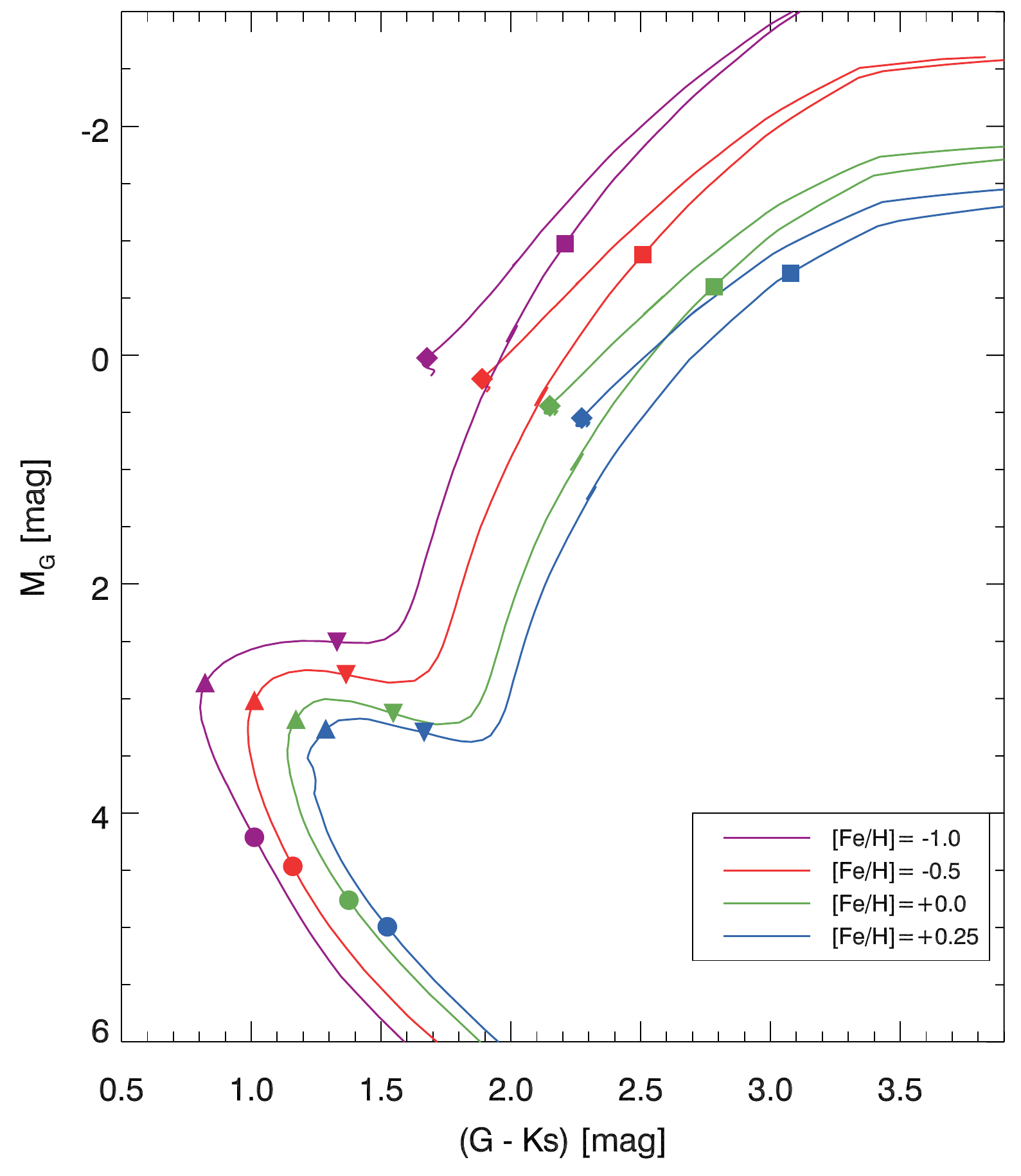}
  \caption{The $(G-K\text{s})$, $M_{G}$ colour-magnitude diagram for the isochrones of 5\,Gyr at the four metallicities ([Fe/H] $=-1.0$ in purple, [Fe/H] $=-0.5$ in red, [Fe/H] $=+0.0$ in green, [Fe/H] $=+0.25$ in blue) of the theoretical stars chosen in Sect.~\ref{sec:tests}. The five evolutionary states with parameters listed in Table~\ref{tab:stellarmasses} are marked with different symbols; circle - dwarf, upward triangle - MSTO, downward triangle - SGB, diamond - red clump, square - high RGB.}
  \label{fig:5allCMD}
\end{figure}

In order to test the suitability of available photometric data for deriving ages and metallicities of stars, we have created a grid of 20 different theoretical stars. These stars come from one of five different evolutionary stages -- dwarfs, MSTO stars, sub-giant branch stars, stars high on the red giant branch (RGB), and red clump stars. The stars have four different metallicities, ranging from [Fe/H] $=-1.0$ to $+0.25$, and all have an age of $\tau=5$\,Gyr. To get the `observed' magnitudes and colours for each star, initial masses were taken from an isochrone of the correct metallicity and age, estimating a suitable position for that evolutionary state by eye. The masses were then used in the isochrones of each photometric passband to calculate the magnitudes. These isochrones are plotted in the CMD shown in Fig.~\ref{fig:5allCMD}, with the synthetic stars' locations marked. The initial mass of each star, along with its $T_{\rm{eff}}$ and $\log{g}$, are given in Table~\ref{tab:stellarmasses}.

\begin{table}
	\caption{The stellar parameters of the 20 synthetic test stars chosen to cover a range of evolutionary stages and metallicities with age $\tau = 5$\,Gyr. The five evolutionary stages are: dwarf, main-sequence turn-off (MSTO), sub-giant branch (SGB), red clump (RC), and higher on the red giant branch (high-RGB). The initial mass, $T_{\rm{eff}}$, and $\log{g}$ are taken from the isochrones, see further details in Sect.~\ref{sec:tests}.}
	\centering
	\label{tab:stellarmasses}
	\begin{tabular}{lllll}
		\hline
	 	\noalign{\smallskip}
		[Fe/H] & Evolutionary & Initial & $T_{\rm{eff}}$ & $\log{g}$\\
		(dex) & Stage & Mass ($M_{\odot}$) & (K) & (dex)\\
		\noalign{\smallskip}
		\hline
		\noalign{\smallskip}
        $-1.00$ & Dwarf & 0.900 & 6524 & 4.40\\
              & MSTO &1.020 & 6957 & 4.11\\
              & SGB & 1.055 & 5994 & 3.64\\
              & RC & 1.080 & 5265 & 2.39\\
              & High-RGB & 1.074 & 4643 & 1.75\\
 		\noalign{\smallskip}
		\hline
		\noalign{\smallskip}
       $-0.50$ & Dwarf & 0.930 & 6177 & 4.43\\
              & MSTO & 1.080 & 6491 & 4.04\\
              & SGB & 1.115 & 5784 & 3.73\\
              & RC & 1.147 & 5000 & 2.41\\
              & High-RGB & 1.141 & 4391 & 1.72\\
		\noalign{\smallskip}
		\hline
		\noalign{\smallskip}
        $+0.00$ & Dwarf & 0.960 & 5752 & 4.46\\
              & MSTO & 1.170 & 6139 & 4.06\\
              & SGB & 1.217 & 5474 & 3.81\\
              & RC & 1.262 & 4711 & 2.42\\
              & High-RGB & 1.254 & 4147 & 1.69\\
 		\noalign{\smallskip}
		\hline
		\noalign{\smallskip}
       $+0.25$ & Dwarf & 1.000 & 5600 & 4.44\\
              & MSTO & 1.210 & 5980 & 4.10\\
              & SGB & 1.260 & 5291 & 3.83\\
              & RC & 1.316 & 4574 & 2.40\\
              & High-RGB & 1.307 & 3890 & 1.40\\
        \noalign{\smallskip}
        \hline
	\end{tabular}
\end{table}

5\,Gyr was chosen as the test age as it is an intermediate age and thus we are less likely to run up against the edges of the isochrone grid. Furthermore it is close to the age of our Sun and many of the stars in the disc of the Galaxy. Later we discuss the impact of changing the age of the test stars. It is necessary to test all important evolutionary stages due to the complex nature of the isochrones; as mentioned before, the MSTO is the point on the CMD where the most precise ages can be determined due to the separation of isochrones (e.g., \citealt{1993A&A...275..101E, 2001A&A...377..911F, 2003MNRAS.340..304R, 2015A&A...579A..52N}), but what happens when we try to determine age and metallicity information from giants or dwarfs? {\referee Main-sequence stars make up the bulk of the stellar matter, so testing dwarf stars is essential. Furthermore, giant stars are often employed as tracer populations throughout the Galaxy due to their bright nature, so are also important to test here. In particular, red clump stars are often used for studies of the Galaxy due to their uniformity; we have decided to include both red clump stars and stars higher up the giant branch here.}

We test multiple values of [Fe/H], as the shapes of the isochrones change significantly with varying metallicity. This can be seen most clearly between the different giant branches in Fig.~\ref{fig:5allCMD}. {\referee The chosen range $-1.0<\textrm{[Fe/H]}<+0.25$ covers the metallicity distribution of the thin and thick discs in the Milky Way. We have tested various examples beyond this metallicity range (for example in Sect.~\ref{sec:GBS}), and found that the conclusions are broadly similar to those found for [Fe/H]$=-1.0$ (for more metal-poor) or [Fe/H]$=+0.25$ (for more metal-rich stars), so have not discussed them further.}

Throughout we have used a parallax of $\varpi=5$\,mas, with an uncertainty of $\sigma_{\varpi}=\pm0.3$\,mas, {\referee which is approximately the median standard uncertainty in parallax for sources at $G=19$ in {\it Gaia} DR2 with a full astrometric solution \citep{2018arXiv180409366L}.} This relative parallax uncertainty of $\sim6\%$ will be achieved for all solar-type stars down to approximately $G=16$ in the final {\it Gaia} data release\footnote{Taken from the {\it Gaia} website https://www.cosmos.esa.int/web/gaia/sp-table1}. We also used an uncertainty on the $G$ band photometry of $\pm0.01$\,mag.

\subsubsection{Near-infrared photometric passbands}
We start the investigation with the near-infrared $J$, $H$, and $K\text{s}$ photometric passbands of The Two Micron All Sky Survey (2MASS; \citealt{2006AJ....131.1163S}). We calculated $\mathcal{G}$ functions for all 20 test stars, using the `observables' of parallax, $G$ magnitude, and the colours $(G-J)$, $(G-H)$, and $(G-K\text{s})$. The 2MASS uncertainties are reported to be $<0.03$\,mag in all three passbands for $K\text{s}<13$\,mag \citep{2006AJ....131.1163S}, so we have assumed an uncertainty of 0.03\,mag. 

\begin{figure*}
\begin{minipage}{180mm}
  \centering
  \includegraphics[width=0.85\columnwidth]{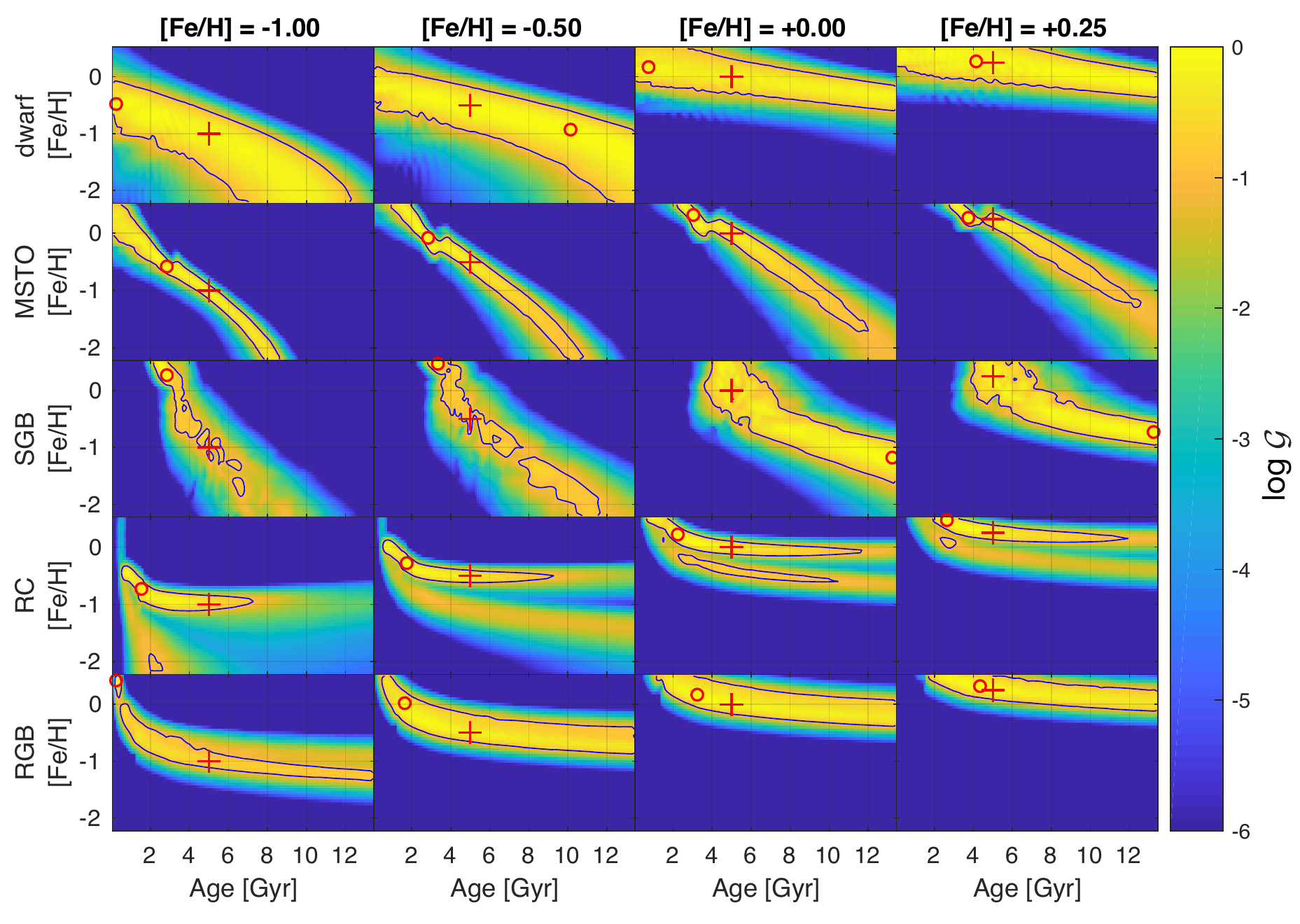}
  \caption{$\mathcal{G}$ functions for the 20 synthetic stars with age 5\,Gyr, calculated using all 2MASS passbands together as input. Each row contains the $\mathcal{G}$ functions for different evolutionary states (labelled on the left hand side), with stars of four different metallicities. The red cross marks the true age and metallicity, and the open red circle marks the most probable point on the map.}
  \label{fig:2massPhotTest}
\end{minipage}
\end{figure*}

\begin{figure*}
\begin{minipage}{180mm}
  \centering
  \includegraphics[width=0.85\columnwidth]{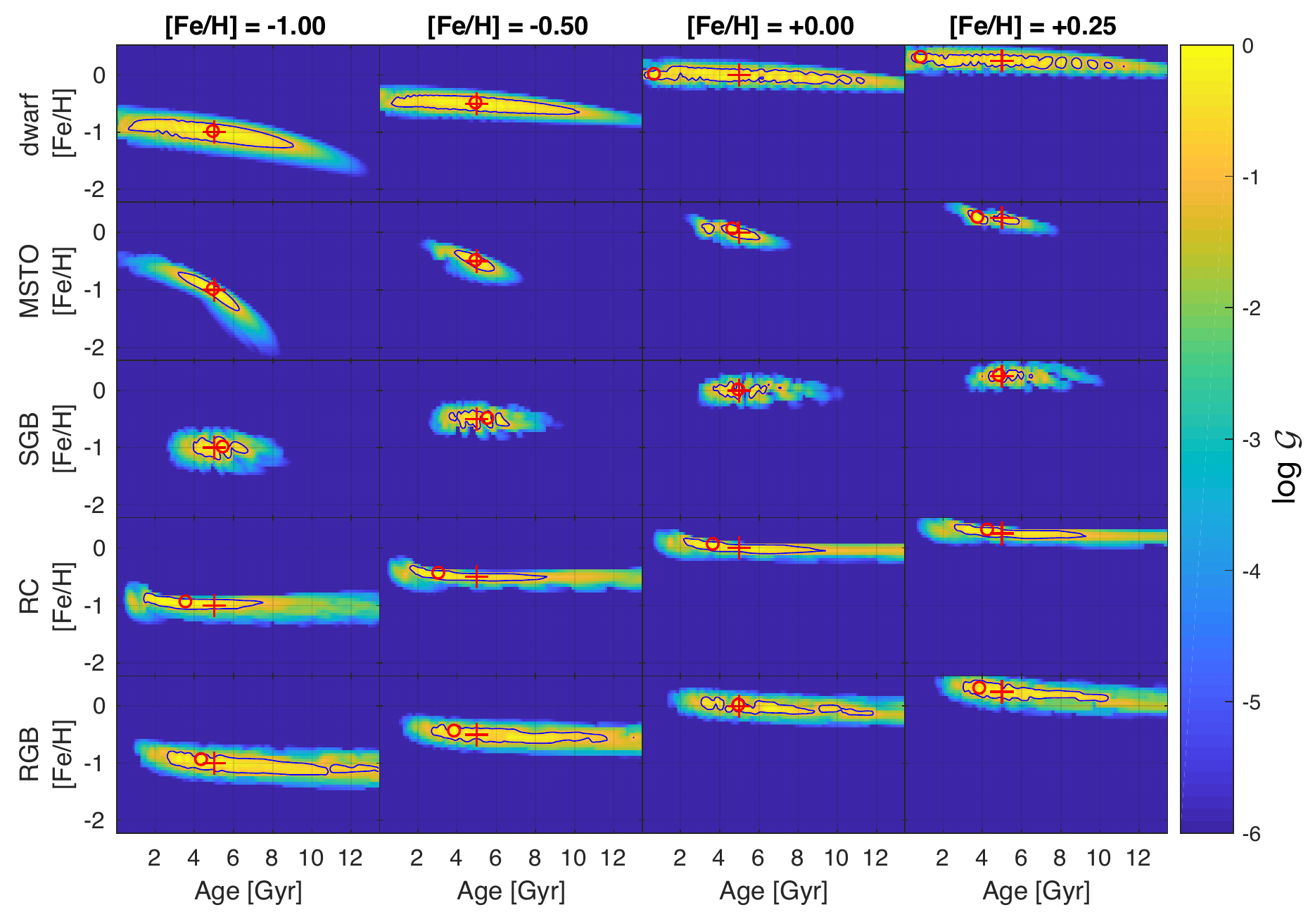}
  \caption{As in Fig. \ref{fig:2massPhotTest}, where the $\mathcal{G}$ functions have been calculated with all 2MASS and $UBVRI$ passbands used together, for a star of age 5\,Gyr.}
  \label{fig:UBVPhotTest}
\end{minipage}
\end{figure*}

\begin{figure*}
\begin{minipage}{180mm}
  \centering
  \includegraphics[width=0.85\columnwidth]{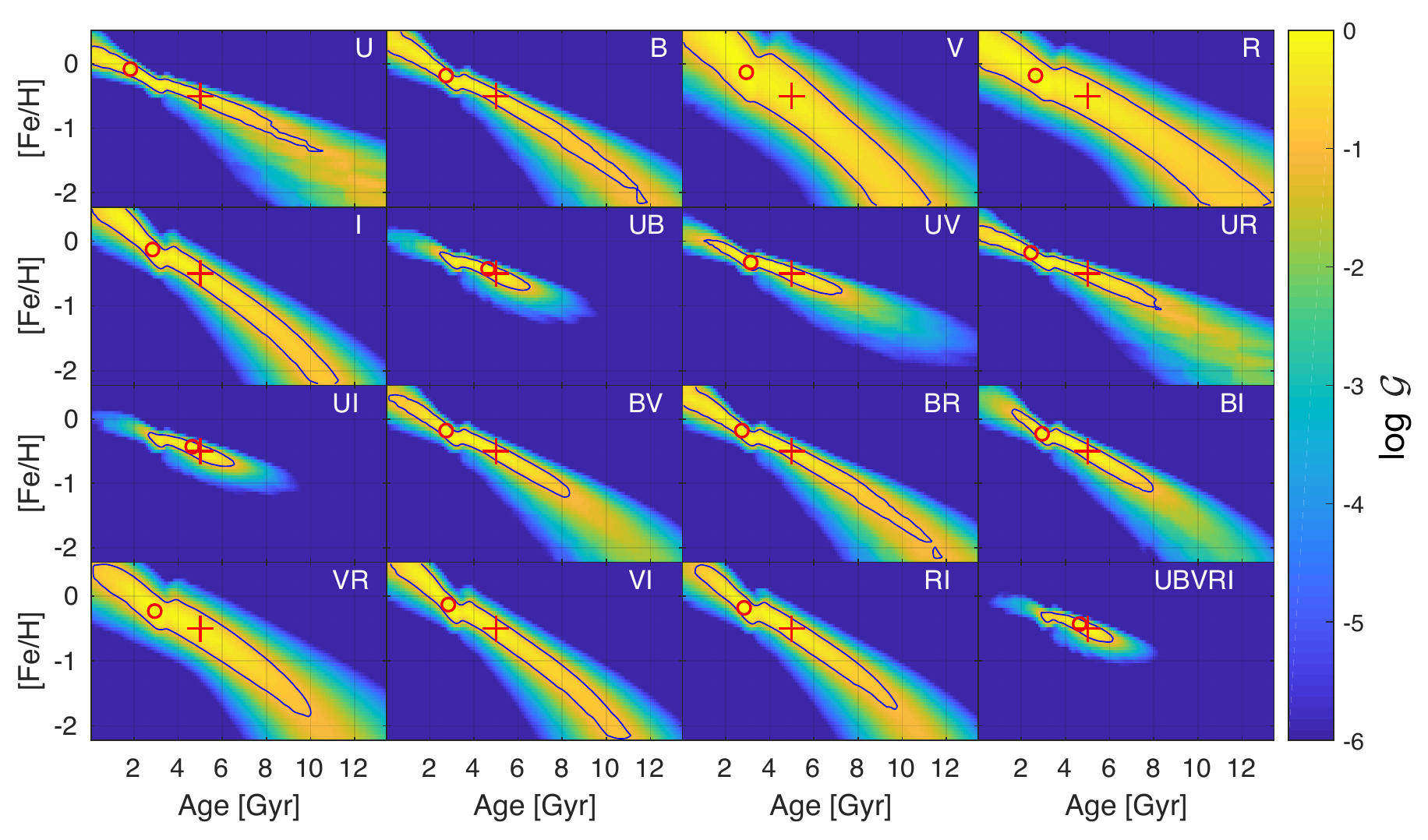}
  \caption{The $\mathcal{G}$ functions calculated for a theoretical MSTO star with age $\tau=5$\,Gyr, metallicity [Fe/H] $=-0.5$, using the individual passbands, and two-band combinations, listed at the top of each plot. The final figure (bottom right) shows the $\mathcal{G}$ function calculated when all 5 passbands are used simultaneously.}
  \label{fig:UBVdemo}
\end{minipage}
\end{figure*}

The results are shown in Fig.~\ref{fig:2massPhotTest}. By adding just the 2MASS colours to {\it Gaia} data, there is not enough information to uniquely determine both the age and metallicity of a star at any evolutionary stage. As was the case in the example in Sect.~\ref{sec:demo}, for MSTO stars there is a strong degeneracy between the two variables, and although there is only a very narrow region of parameter space with a probability greater than 0.1, this space spans almost the whole range of metallicities and ages. Further, the most likely age-metallicity value is not coincident with the true values. The same is true for the sub-giants; albeit with slightly larger uncertainty, and the most probable age-metallicity is in fact further from the true value. 

The ages of both dwarf and giant stars remain unconstrained with these data. The reasons for this are obvious when viewed on the CMD, as the isochrones for different ages are inseparable on the main sequence and almost as close on the giant branch. Interestingly, however, the $\mathcal{G}$ functions are able to derive [Fe/H] values with reasonable precision for most of the dwarf and RGB stars. In particular for the red clump stars, [Fe/H] could be constrained to a region of $\sim 0.5$\,dex. All four red clump stars have a lower true metallicity than the most probable value - although this is accompanied by a much younger age (all predicted to be around 2\,Gyrs).

We tested the effect of using only one of the passbands instead of all three together, and found that the uncertainty contours were smaller and thus slightly better constrained by having all three passbands, but that the shape of the contours was identical in all cases. In the case of missing passbands for a star, no crucial information is lost.

\subsubsection{{\referee Optical and near-UV} photometric passbands}
\label{sec:opt-nUV}
The Johnson $UBV$ photometric system \citep{1953ApJ...117..313J} was the first system to be standardised and was subsequently extended into the Johnson-Cousins $UBVRI$ system \citep{1976MmRAS..81...25C}. For these tests we have assumed a magnitude {\referee uncertainty} of $0.01$\,mag; the large catalogue of $UBVRI$ photometry by \citet{2000PASP..112..925S} quotes {\referee uncertainties} for local stars of significantly less than this value.

Figure~\ref{fig:UBVPhotTest} displays the $\mathcal{G}$ functions created when both 2MASS and $UBVRI$ colours are used as the optional inputs. The improvement gained by adding the $UBVRI$ photometric passbands is immediately obvious. In many cases, the most probable age and metallicity coincides almost exactly with the true values. In particular, in the case of MSTO and SGB stars, the ages are constrained to an uncertainty of approximately 1\,Gyr either side - better in a couple of specific cases. [Fe/H] is also well constrained in these stars, with uncertainties less than $0.2$\,dex. For the other evolutionary states, the probability distributions are still very useful; in all plots the metallicity is well constrained. For dwarf stars, the ages remain uncertain: within the 90\% confidence interval, we can only claim the stars' ages to be $<8-10$\,Gyr. 

The two giant cases, red clump and high-RGB, have similar $\mathcal{G}$ functions to the dwarf stars. The ages have large uncertainties, although the oldest ages are excluded at 90\% confidence, allowing some inference to be made about whether the star is young or old. Unlike in the case of the dwarfs, these $\mathcal{G}$ functions also show some power to discriminate that the star is not younger than 2\,Gyr. Despite this, the most probable ages in both cases are often younger than the true age.

How does the addition of $UBVRI$ passbands resolve the age-metallicity degeneracy? In Fig.~\ref{fig:UBVdemo}, we separate the effect of each passband for an example star with the parameters of a MSTO star ([Fe/H] $=-0.5$ and age $\tau=5$\,Gyr). Each $\mathcal{G}$ function was calculated using the fixed inputs of the {\it Gaia} $G$ band magnitude and parallax, and also one colour; $(G-U)$ in the top left case, and so on with each of the five passbands, then two colours (e.g., $(G-U)$ and $(G-B)$). Finally the last $\mathcal{G}$ function uses all five colours simultaneously. Noticeably, no one passband provides a unique age and metallicity for the star. In particular, the $V$ and $R$ passbands perform least well due to their similarity to the {\it Gaia} $G$ band, creating a colour that holds little information. The crucial passband is the $U$ band, for which the slope of the high-probability region is shallower. As seen in the bottom right panel of Fig.~\ref{fig:UBVdemo}, the results are almost as good as those in Fig.~\ref{fig:UBVPhotTest} -- in this case, the $I$ band produces almost identical results to $JHK\text{s}$. To conclude, combining {\it Gaia} $G$ and parallax, with $U$ band photometry and one additional optical passband (preferably $B$ or $I$) allows us to break the age-metallicity degeneracy and derive both simultaneously for many stellar types.

\subsubsection{Parallax uncertainties}
 Whilst the {\referee uncertainties} considered in the previous examples are a reasonable estimate for many stars in the current and future {\it Gaia} data releases, some will have much better parallax precision, but many more will have much higher relative uncertainties. Therefore we calculated $\mathcal{G}$ functions for two contrasting cases -- relative parallax uncertainties of 1\% and 50\% (examples can be found in Appendix Figs.~ \ref{fig:highSNPhotTest} and \ref{fig:lowSNPhotTest}). As expected, when the parallax uncertainty is as low as 1\%, the ages and metallicities derived are much better constrained than in our previous examples, allowing subgiant stellar ages to be determined with uncertainties at the 90\% confidence level of less than 0.5\,Gyr. Conversely, with 50\% relative uncertainties almost no age information remains at any evolutionary stage. Despite this, metallicities remain constrained, with uncertainties less than 0.3\,dex when using $UBVRIJHK\text{s}$. At {\it Gaia's} fainter reaches, an accurate metallicity map of the Galaxy would still be possible. Testing a variety of parallax uncertainties showed that determining ages for MSTO stars is limited to parallax uncertainties lower than approximately 20\%; greater than that and the age uncertainties become larger than 3\,Gyr.

\subsubsection{Different stellar ages}
The results discussed throughout this section are mostly indifferent to the age of the test star. We ran the same tests for a range of different stellar ages, and the $\mathcal{G}$ functions are qualitatively similar. The predominant difference is how uncertain the age estimate is; for older stars, the region defined by the 90\% confidence interval increases in size. For example, in Fig.~\ref{fig:UBVPhotTest} the MSTO star at [Fe/H] $=-0.5$ has a confidence interval that stretches approximately $\pm1$\,Gyr -- for a star with the same metallicity but an age of $\tau=10$\,Gyr, the confidence interval grows to $\pm2$\,Gyr.

\begin{figure}
  \centering
  \includegraphics[width=0.99\columnwidth]{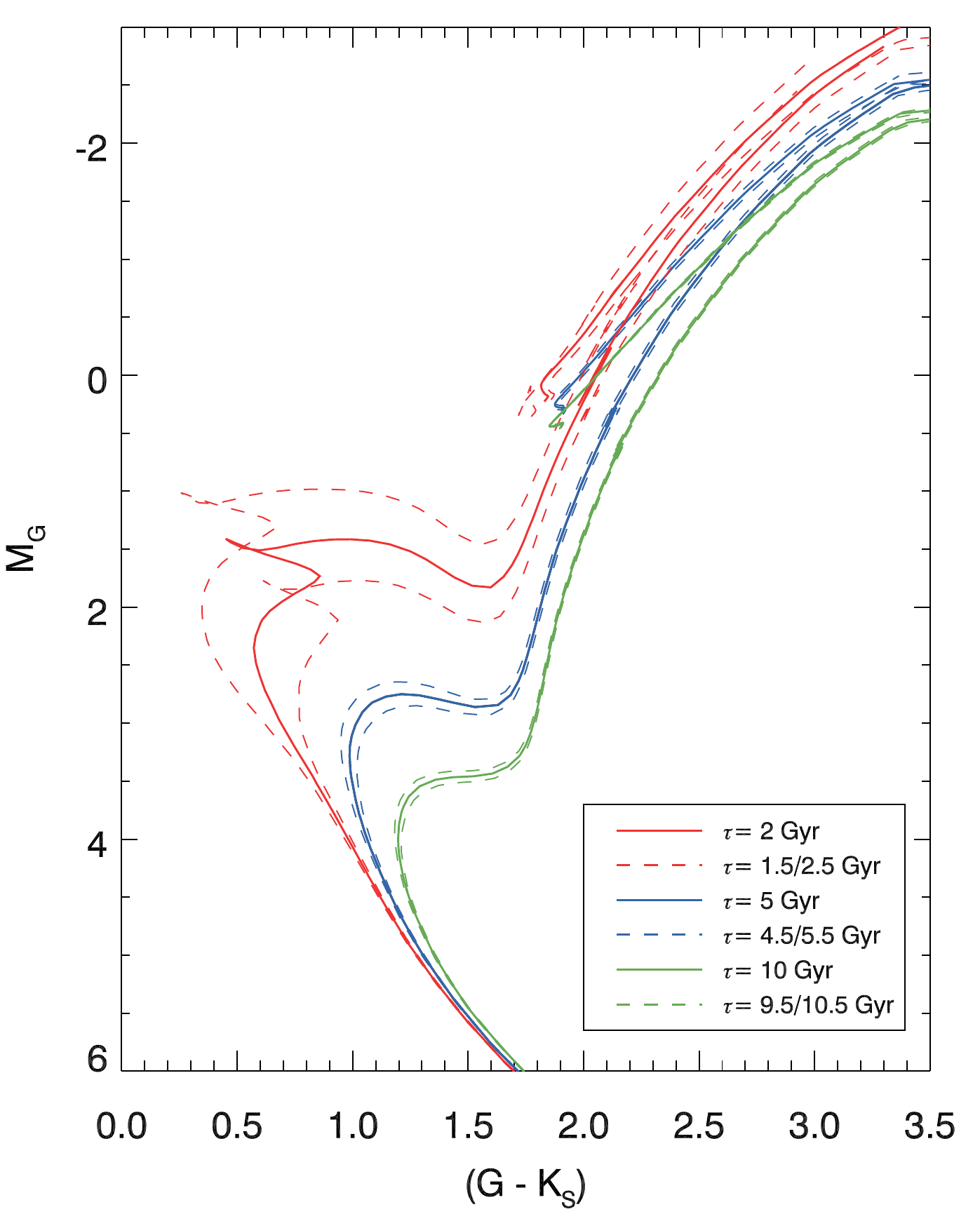}
  \caption{The $(G-K\text{s})$, $G$ colour-magnitude diagrams for isochrones of one metallicity ([Fe/H] $=-0.5$) at three different ages; 2\,Gyr (red), 5\,Gyr (blue), and 10\,Gyr (green). The dashed lines show isochrones that are $\pm0.5$\,Gyr off the given age.}
  \label{fig:agesCMD}
\end{figure}

The reason for this is clear from Fig.~\ref{fig:agesCMD}. The older isochrones are much closer together and so for a given {\referee uncertainty} in the observations, a wider range of ages will be possible. Conversely at 2\,Gyr, the isochrones are spaced further apart, allowing for a more precise age determination. 

Figure~\ref{fig:agesCMD} also explains the other noticeable difference in the $\mathcal{G}$ functions at younger ages; at 2\,Gyr the turn-off has a more complicated shape than that of the isochrones at 5 and 10\,Gyr. This `wiggle' is caused by the different nature of the stellar core in higher mass stars. Stars with mass $\gtrsim1.1$\,M$_{\odot}$, like those on the 2\,Gyr isochrone at the turnoff point, have convective, well-mixed cores. Once the core exhausts the supply of hydrogen, the star stops producing enough energy to support itself, and so the star contracts -- heating up, and moving blueward on the CMD. Eventually the base of the envelope heats up enough to start burning hydrogen in a radiative shell, and the increased radiation pressure causes the entire envelope to expand, cool, and shift red-ward onto the subgiant branch. At masses $\lesssim1.1$\,M$_{\odot}$,
however, the core is radiative, with a smooth clear boundary to the convective envelope. The transition to shell burning is smooth, without the contraction, and so there is no `wiggle' at the turnoff \citep{1942ApJ....96..161S, 2012sse..book.....K}.

Since this effect causes a significant change to the shape of the isochrone for higher mass stars, we have investigated the $\mathcal{G}$ functions for stars with $\tau= 2$\,Gyr in all the photometric passbands. Examples can be found in the Appendix, Figs.~\ref{fig:2mass2gyrPhotTest} and \ref{fig:UBV2gyrPhotTest}. These tests show that much younger stars result in smaller uncertainties in age; in fact, when using $UBVRIJHK\text{s}$ the age of the stars can be restricted to below 4 to 6\,Gyr in all 20 cases. For young enough MSTO stars, the `wiggle' shown in the CMD provides a sharp lower limit on a star's mass, and so the age can be determined reasonably with only 2MASS data (Fig.~\ref{fig:2mass2gyrPhotTest}); however, unlike in older stars, [Fe/H] is not constrained. Again, addition of the $UBVRI$ bands results in a unique age and metallicity solution for these stars.

\subsection{Availability of broadband photometry}
{\referee The results of these tests show that it could be possible to obtain ages and metallicities of many millions of stars, provided the right photometry exists, so we shall briefly mention the availability of suitable photometric data.

No all-sky modern-day survey using the Johnson-Cousins system has been made, however the majority of the brightest stars have archival data for some, if not all, of these passbands. Additionally, many stellar clusters and special fields have comprehensive data in this passband system (e.g., \citealt{2005PASP..117.1325S}). There are large photometric surveys which include some kind of $U$ passband, such as the SkyMapper Southern Sky Survey \cite{2007PASA...24....1K}, and the most prominent of these is the SDSS survey \citep{2000AJ....120.1579Y}. There are over 260 million stars with photometry in the SDSS $ugriz$ bands, covering a brightness range of $13\lesssim g \lesssim 22$\,mag, making SDSS the single largest source of ultraviolet photometry for stars in {\it Gaia} DR2 and later catalogues. In Fig.~\ref{fig:ugrizPhotTest}, the $\mathcal{G}$ functions for the test stars using $ugriz+JHK\text{s}$ are shown, assuming an uncertainty of $0.02$ on the $ugriz$ magnitudes \citep{2003MmSAI..74..978I}. As expected, the addition of the $u$-band leads to very similar results as found in Section \ref{sec:opt-nUV}.

2MASS observed effectively the entire sky in the $J$, $H$, and $K\text{s}$ wavelength passbands, down to a magnitude of 15.8 in $J$. It is perhaps the survey with the largest overlap with {\it Gaia} -- 39.2\% of stars observed in {\it Gaia} DR1 have been matched to a 2MASS observation \citep{Marrese:2017ey}. We also performed tests using the WISE passbands \citep{2010AJ....140.1868W}, which span the near and mid infrared. The four passbands at 3.4, 4.6, 12.0, and 22.0\,$\mu$m have been used as a metallicity indicator for very metal-poor stars \citep{2014ApJ...797...13S}. Qualitatively, there is little difference between $\mathcal{G}$ functions produced using 2MASS and WISE; the uncertainty contours are almost identical, and the WISE passbands do not break the age-metallicity degeneracy. It would appear that using either 2MASS or WISE data would be equally helpful, but there is no real benefit in using both. As 2MASS is more complete and easier to match to the optical {\it Gaia} data, it is the obvious choice. There may also be further benefit in using only one or two 2MASS passbands, and using the remaining data, along with WISE, to determine the reddening. This could be done by employing, for example, the Rayleigh-Jeans Colour Excess method \citep{2011ApJ...739...25M}, which requires at least one colour composed of a near- and mid-infrared passband.

Since all-sky broadband photometry is now available from {\it Gaia}, we further investigated using the {\it Gaia} $G_\text{BP}$ and $G_\text{RP}$ bands, like the example in Fig. \ref{fig:ExamplegFunc}. We have tested using these passbands alone, and combining them with 2MASS passbands, shown in Appendix~Figs.~\ref{fig:GaiaNo2MASSPhotTest}
and \ref{fig:GaiaPhotTest} respectively. The resulting $\mathcal{G}$ functions are very similar to those found when using 2MASS alone and the confidence region is much larger than when using $UBVRI$. 

The final {\it Gaia} data release will contain photometry very similar to the optical $griz$ passbands, which are also used in the PAN-STARRs photometric survey \citep{2016arXiv161205560C}, which covers 75\% of the sky down to magnitudes of $g=23.3$. We tested these along with the 2MASS passbands, shown in Fig.~\ref{fig:grizPhotTest}. The lack of a near-UV passband, however, means again that these surveys can not overcome the degeneracy between age and metallicity.
  }

\section{On the usefulness of the $U$ band to obtain stellar parameters}
\label{sec:four}

\begin{figure*}
\begin{minipage}{180mm}
  \centering
  \includegraphics[width=0.5\columnwidth]{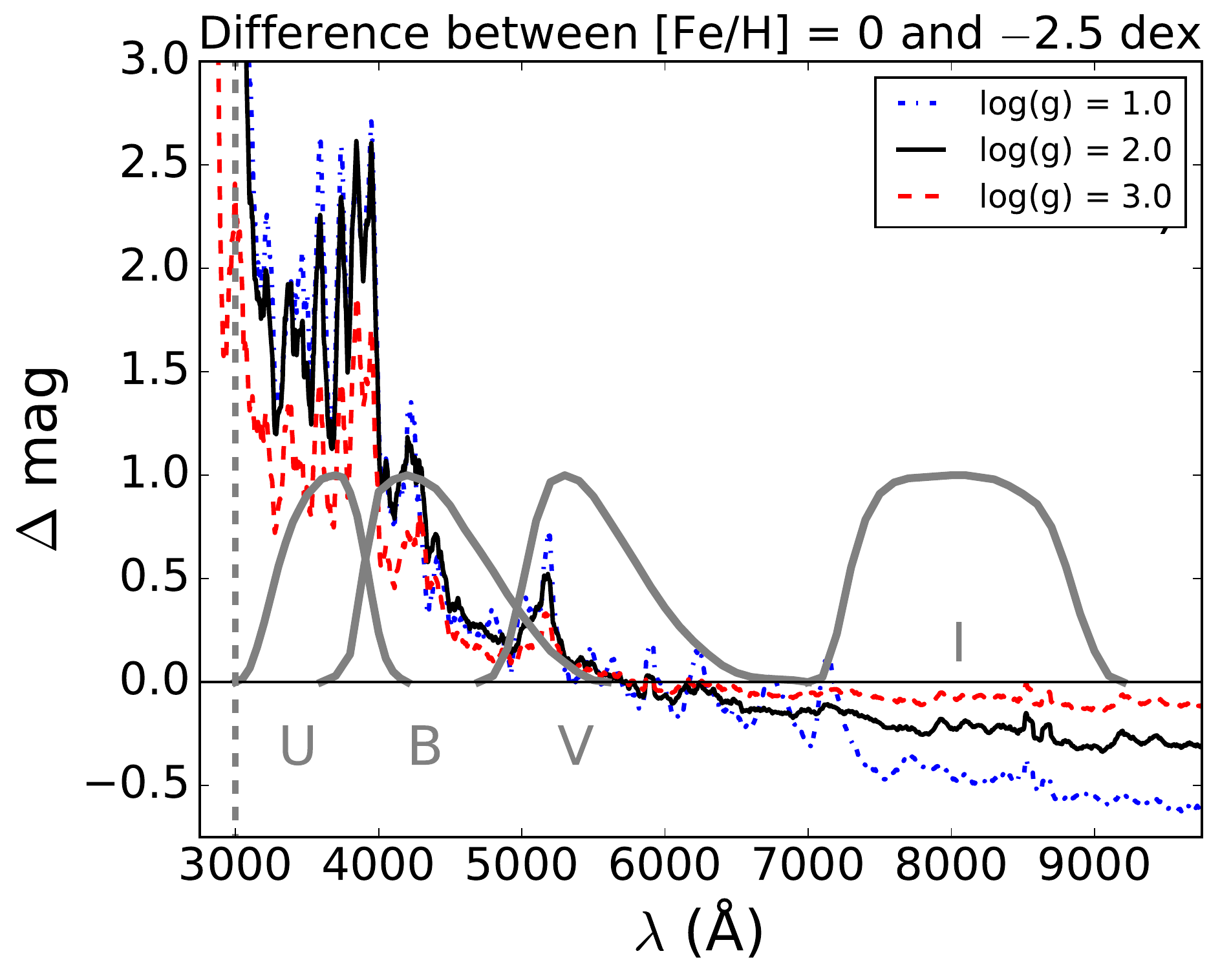}\includegraphics[width=0.5\columnwidth]{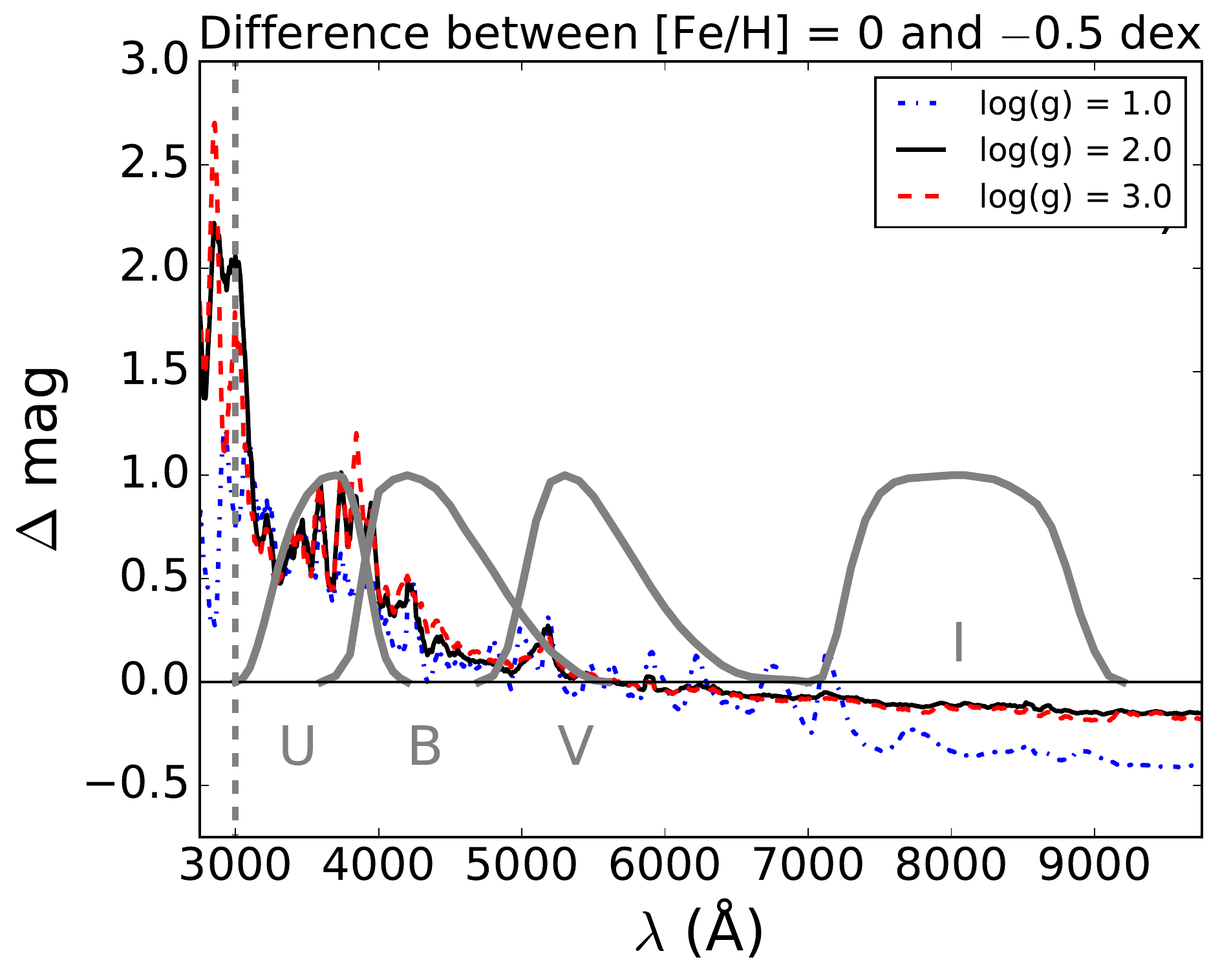}
  \caption{Illustration of the sensitivity of different passbands to stellar metallicity. Plotted is the difference in magnitude between a star with solar $T_{\rm eff}$, $\log{g}$, and metallicity, and one with [Fe/H]$=-2.5$\,dex (left) or one with [Fe/H]$=-0.5$\,dex (right) for three different $\log{g}$ values as indicated in the legend. The $T_{\rm eff}$ values for the models are chosen to agree with the metallicity and $\log{g}$ of the isochrones. See Sect.~\ref{sec:four} for further details on how $\Delta  mag$ was calculated, which has been normalised at 5550\,{\AA} (as in Bond 1999). The four broad Johnson-Cousins passbands ($UBVI$) have been indicated (thick grey lines), as well as the Earth's atmospheric cutoff (vertical dashed grey line).}
\label{fig:deltaspectra}
\end{minipage}
\end{figure*}

\begin{figure*}
\begin{minipage}{180mm}
  \centering
  \includegraphics[width=0.99\columnwidth]{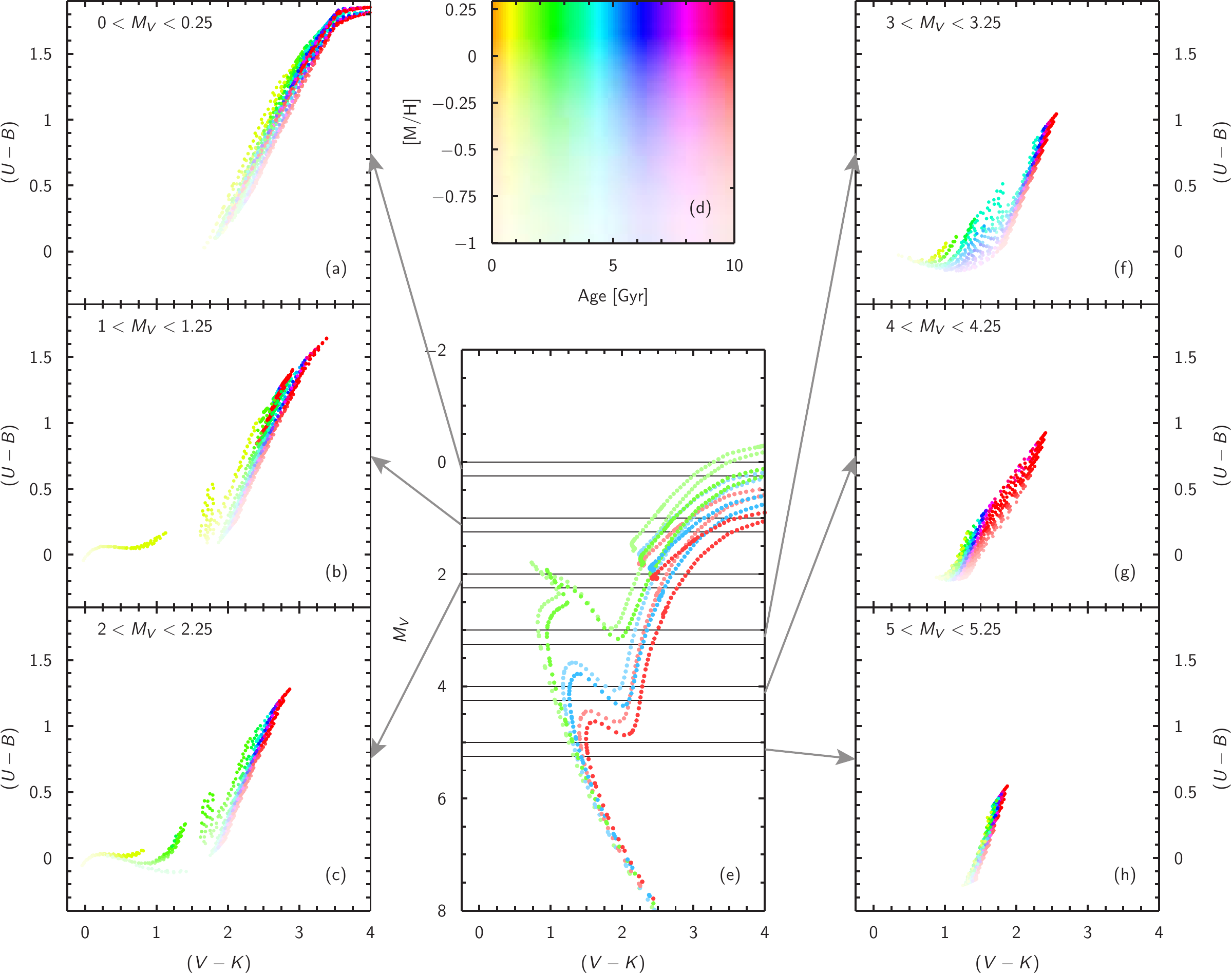}
  \caption{The $(U-B)$ vs. $(V-K)$ colour-colour plots for a variety of absolute magnitude ranges. Panels (a), (b), (c), (f), (g), and (h) show the six colour-colour plots, with data from the isochrone grid. Panel (d) shows the colour scheme used in all panels, with different colours representing different ages, and the brightness of the colour representing the metallicity. Panel (e) shows the various slices in absolute magnitude used in the outer panels (arrows indicate which slice is shown in which panel), along with six isochrones plotted for demonstration. These isochrones have ages of 2 (green), 5 (blue) and 10 (red) Gyr, and both [Fe/H]$=0.0$ (bright) and [Fe/H]$=-0.25$ (faint).}
  \label{fig:ross}
\end{minipage}
\end{figure*}

Different parts of the stellar spectrum contain information about the star, encoded in the overall spectral energy distribution as well as in the strength and shapes of individual spectral lines. The type of information varies for different types of stars. Here we discuss the spectral information available in the near-UV part of the spectrum, which we have shown to be especially helpful in breaking the age-metallicity degeneracy (see Sect. \ref{sec:two}).

Figure~\ref{fig:deltaspectra} shows the sensitivity to metallicity for the spectral region around the Balmer jump. For three different evolutionary stages, the magnitude difference between a star with [Fe/H]$=0.0$ and a star with [Fe/H]$=-2.5$ have been calculated, making use of the synthetic stellar spectral library by \citet{2005A&A...442.1127M}. We have converted the fluxes in the stellar library to magnitudes and taken the difference between them. We follow Bond (1999)\footnote{H.E. Bond, 1999, Where is the information located in stellar spectra?

Unpublished study prepared for the Wide Field Camera 3 Scientific Oversight Commitee, 1999. Available at 
http://www.stsci.edu/$\sim$bond/whereistheinfo.pdf} and normalize the magnitude difference at 5550\,{\AA}. From this figure it is easy to understand why near-UV colours carry more metallicity information than redder colours. For example, if we derive the standard Johnson magnitudes from these spectra we find that  $(U-B)_{+0.0} - (U-B)_{-2.5}$ is large (where the subscript denotes the [Fe/H] of the spectrum), while $(V-I)_{+0.0} - (V-I)_{-2.5}$ is small, almost zero for higher $\log{g}$ values. Hence, if filter and atmospheric throughput were equal it would be much easier to get photometry that is capable to measure metallicity using the near-UV part of the spectrum. 

However, the $U$-band is not only sensitive to metallicity, it is also sensitive to gravity. For hotter stars on the main-sequence it is sensitive to effective temperature, but for stars of A-type and later on the main sequence the region below the Balmer jump changes due to the gravity of the star. 

\citet{2005AJ....129.2914B} discusses how to best define a photometric passband in the region below the Balmer jump such that it is most sensitive for determining the surface gravity of the star (which is equivalent to measuring its luminosity).  Figure~2 in \citet{2005AJ....129.2914B} illustrates some of the available choices: the broad standard Johnson $U$ with a high throughput, the narrower $u'$ used in the SDSS, via the Thuan-Gunn $u$, to the bluer and most narrow Str\"omgren $u$. The latter two are located essentially entirely below the Balmer jump and hence offer very good prospects for determining the luminosity of the star from combining the passband with a redder band. The two broader passbands (Johnson $U$ and SDSS $u'$) both have a significant part of their bandwidth spanning the Balmer jump, hence they are less sensitive to luminosity than the other two. As the Str\"omgren $u$ has a very poor throughput compared to the Thuan-Gunn $u$ (Table~1 in \citealt{2005AJ....129.2914B}), the author argues for a standardized photometric system using $uBVI$ \citep{2005AJ....129.2914B, 2005AJ....129.2924S}.

But if we have the luminosity of the star, which may be the case when we are working with {\it Gaia} parallaxes, then, to quote directly from Bond (1999): \textit{ "Johnson $B$ or Str\"omgren $v$ is useful for isolating the metallicity color changes at about 4000 -- 4500 {\AA} from gravity changes in the $u$ band. If the luminosity is known separately, the $u$ band is extremely sensitive to metallicity."}.

Figure~\ref{fig:ross} further illustrates the sensitivity of the $U$-band combined with visual and infrared passbands to the metallicity and age if the parallax and hence the absolute magnitude of the star are known. We observe that if we are able to obtain the absolute magnitude of the star, i.e. by having access to its parallax and assuming that we can handle the reddening adequately, a combination of a colour including the $U$ band and a colour of visual and infrared passbands should essentially allow us to obtain the metallicity and age simultaneously for all but the least evolved stars.

The quality of the photometry and parallax matters, and we note that on the upper RGB, although the age information is present, small {\referee uncertainties} and offsets will weaken the age determination, while the metallicity determination is very robust because in this evolutionary phase the $(U-B)$ colour is well ``stretched out''. For other ranges of absolute magnitude the young ages are almost trivial to obtain whilst the higher ages for the most metal-poor stars are almost impossible to obtain in detail (e.g., Fig.~\ref{fig:ross}, panel (c)).

Although the $U$ band is important for our ability to use stellar photometry and parallaxes to break the age-metallicity degeneracy for as many stellar evolutionary stages as possible, we note that observations in this and similar passbands are both difficult and time-consuming. Throughput is often poor and the Earth's atmosphere at these wavelengths absorbs a great deal of the stellar light. In addition, the original $U$ band in the $UBV$-system suffers from an incorrect transformation to outside the Earth's atmosphere (see, e.g., \citealt{1992msp..book.....S, 1999BaltA...8..505S}), which in turn has resulted in an ill-defined system of standards. These things taken together have meant that observations in the $U$ band are relatively rare and do not allow precise comparison with theoretical models. For accurate work in the ultraviolet it is necessary to use better-defined and possibly narrower passbands to cope successfully with atmospheric extinction and other transformation issues. However, we note the development of the SDSS $ugriz$ system has renewed the interest in the ultraviolet passband, e.g., the Luau-project at CFHT \citep{2017ApJ...848..129I}, or the SkyMapper telescope \citep{2007PASA...24....1K}.

To conclude, passbands in the near-UV, below the Balmer jump, are sensitive to several stellar parameters. When photometry is combined with knowledge of the star's luminosity, this bluer region provides a powerful measure of the stellar metallicity. This should be considered when designing  photometric stellar surveys. Particular care should be taken when defining the exact photometric passband as well as when establishing its calibration (for new passbands).

\section{Application to real stars}
Having demonstrated the ability to use photometry and parallaxes to find ages and metallicities in synthetic examples, testing on real observational data is a necessary next step. {\referee We use two data sets suitable for testing our results against literature values. First we use the {\it Gaia} benchmark stars. These bright stars have high-quality $UBV$ photometry and accurate distances from \textsc{Hipparcos}. Secondly, we test the open cluster NGC 188, which has available $ugriz$ photometry, as well as a well-known age, metallicity, and distance.}

Up until now we have used the {\it Gaia} $G$ band as the first input for the tests. Whilst {\it Gaia} DR1 released $G$-band data for the whole catalogue, it has been noted that the throughput of the $G$ band observations differs significantly from that predicted pre-launch \citep{2016A&A...595A...7C}. Because the calculated isochrones in the $G$ band are based on the pre-launch predictions, they are not a good match to the observations. An updated passband is available for DR2\footnote{https://www.cosmos.esa.int/web/gaia/iow\_20180316}, solving this problem for future isochrones, but for these tests we omit the $G$ band observations in favour of ground-based $V$ and $g$ band photometry.

For the present tests we assume uniform priors in both age and [Fe/H], and adopt as the most probable values the location of the global maximum in the $\mathcal{G}$ function. 90\% confidence intervals in age and [Fe/H] are obtained from the extreme values of the 90\% confidence region along each axis in the 2D map. {\referee We have chosen to produce these quantities so that straightforward numerical comparisons can be made with the literature, not to provide rigorously quantified estimates of either of the parameters.}

\subsection{The {\it Gaia} benchmark stars}
\label{sec:GBS}

\begin{figure*}
\begin{minipage}{180mm}
  \centering
  \includegraphics[width=0.75\columnwidth]{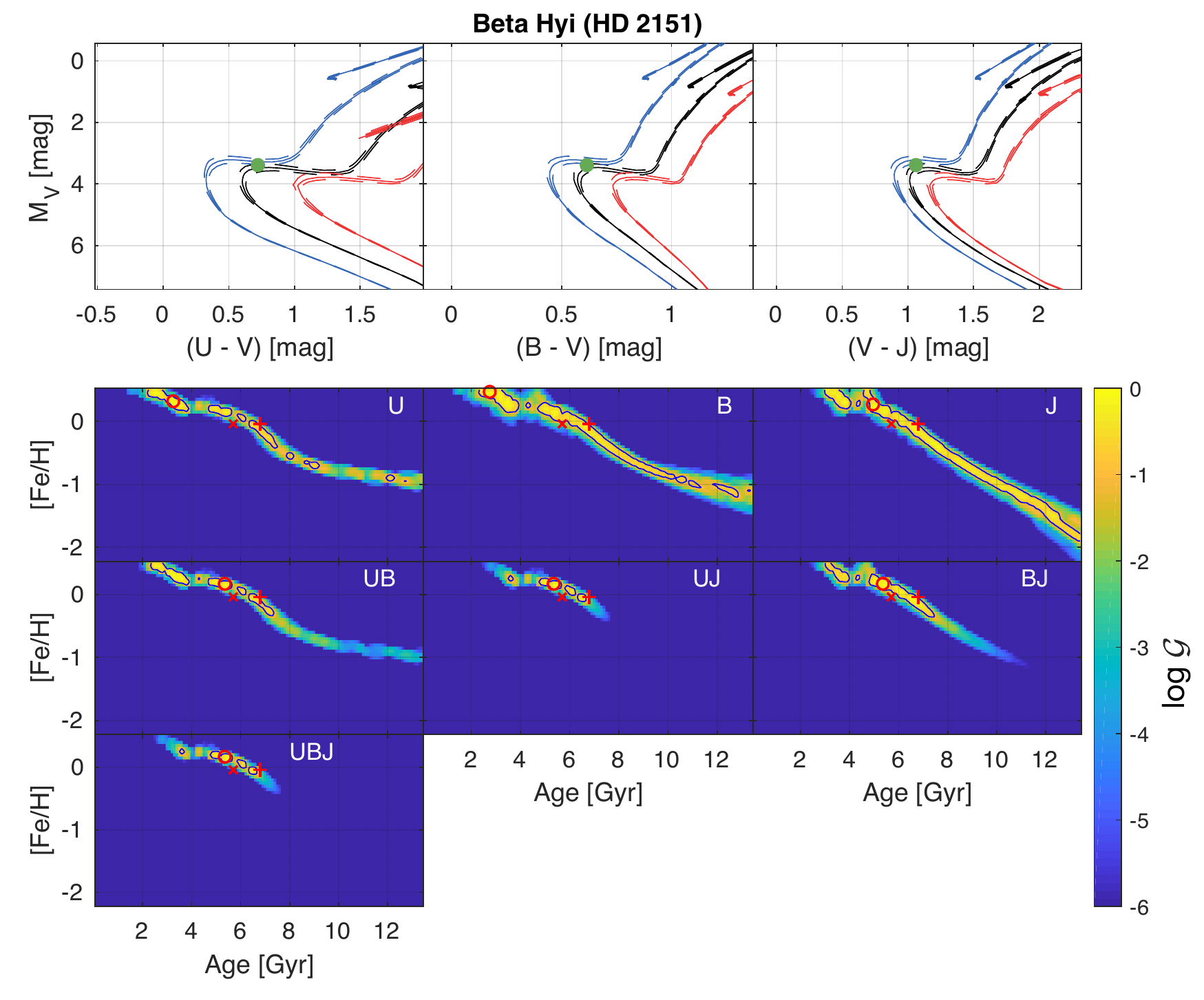}
  \caption{Tests run on the {\it Gaia} benchmark star $\beta$ Hyi. \textit{Top:} The CMDs of the three different colours used; $(U-V)$, $(B-V)$, and $(V-J)$. The black isochrone is at the calculated age for the star (given in Table~\ref{tab:gaiabenchmark}), with the other isochrones defined as in Fig. \ref{fig:ExampleCMD}. The green point shows the observed photometry. \textit{Bottom:} The $\mathcal{G}$ functions calculated using the parallax, $V$ band magnitude, and the colours created from $V$ and each passband denoted in the top right corner of each subplot. Both the literature age (as a red cross) and the calculated spectroscopic age (as a red plus sign) are shown, taken from \citet{Sahlholdt:2018ty}.}
  \label{fig:BetaHyi}
\end{minipage}
\end{figure*}

\begin{figure*}
\begin{minipage}{180mm}
  \centering
  \includegraphics[width=0.75\columnwidth]{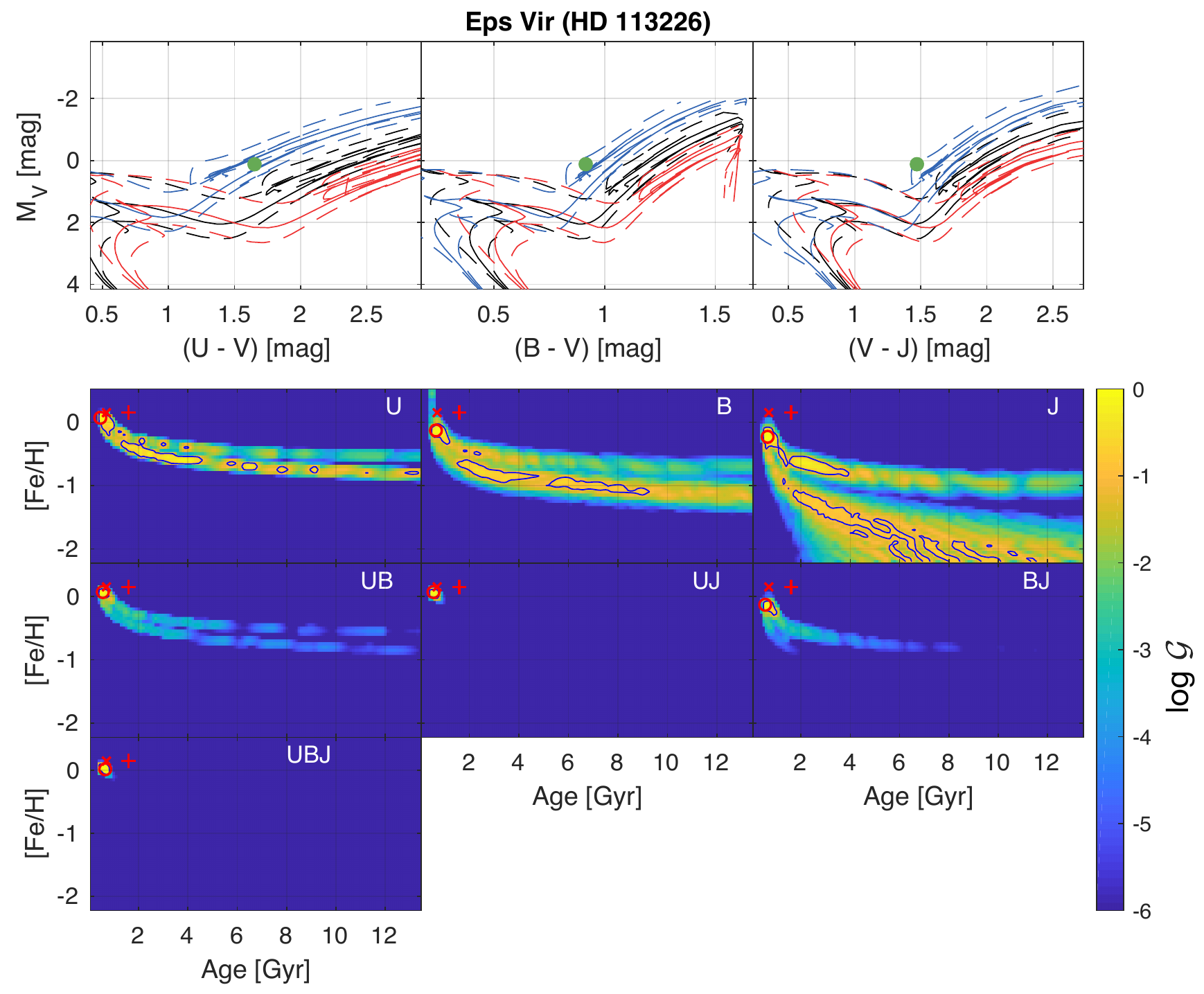}
  \caption{Tests run on the {\it Gaia} benchmark star $\epsilon$ Vir. As in Fig. \ref{fig:BetaHyi}. Due to the metal-rich, young nature of the star, the isochrones plotted in the CMDs are different from in previous plots; here they are $\pm0.3$\,dex in [Fe/H] and $\pm0.5$\,Gyr in age.}
  \label{fig:EpsVir}
\end{minipage}
\end{figure*}

\begin{figure*}
\begin{minipage}{180mm}
  \centering
  \includegraphics[width=0.75\columnwidth]{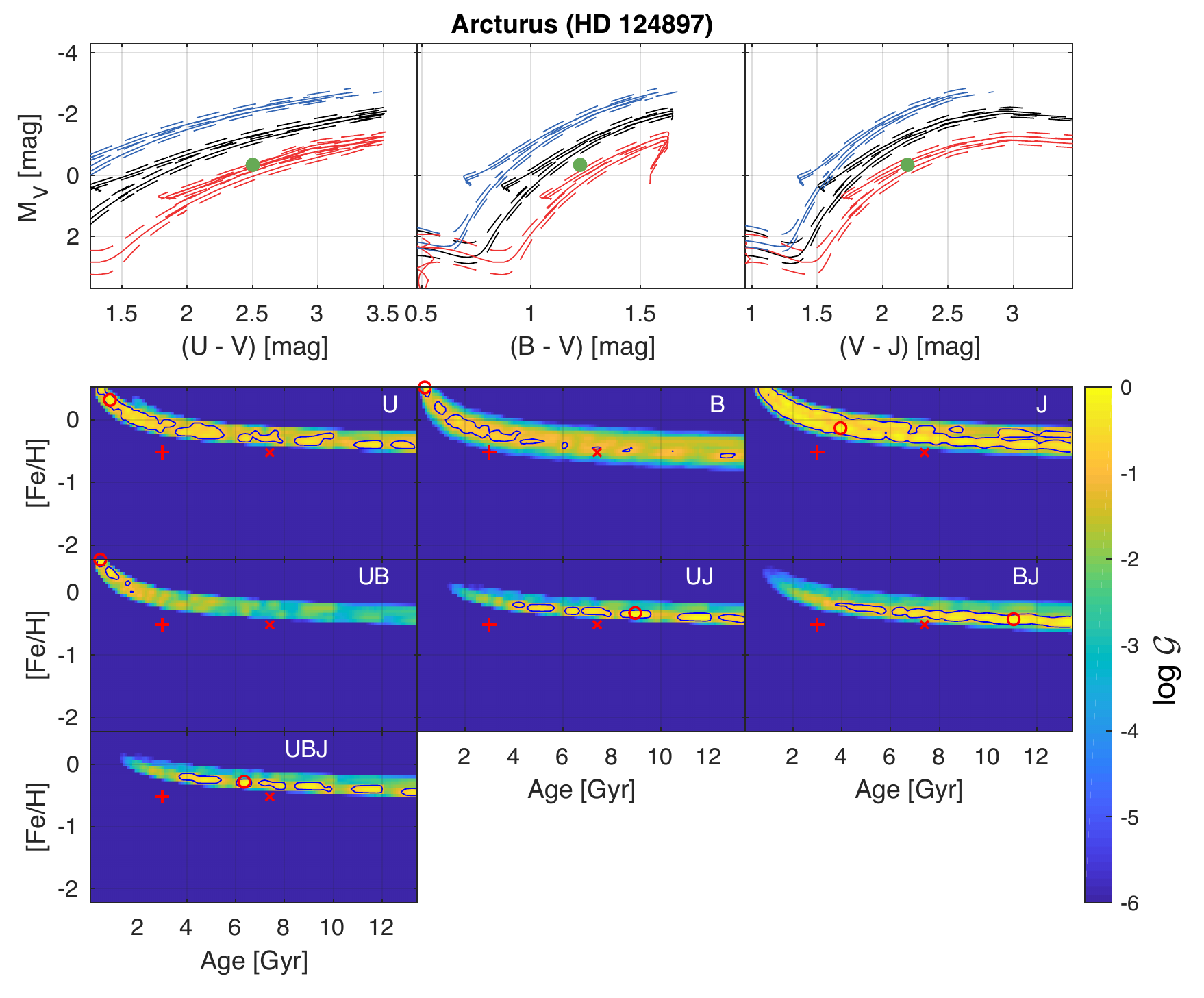}
  \caption{Tests run on the {\it Gaia} benchmark star Arcturus. As in Fig. \ref{fig:BetaHyi}.}
  \label{fig:Arcturus}
\end{minipage}
\end{figure*}

\begin{figure*}
\begin{minipage}{180mm}
  \centering
  \includegraphics[width=0.75\columnwidth]{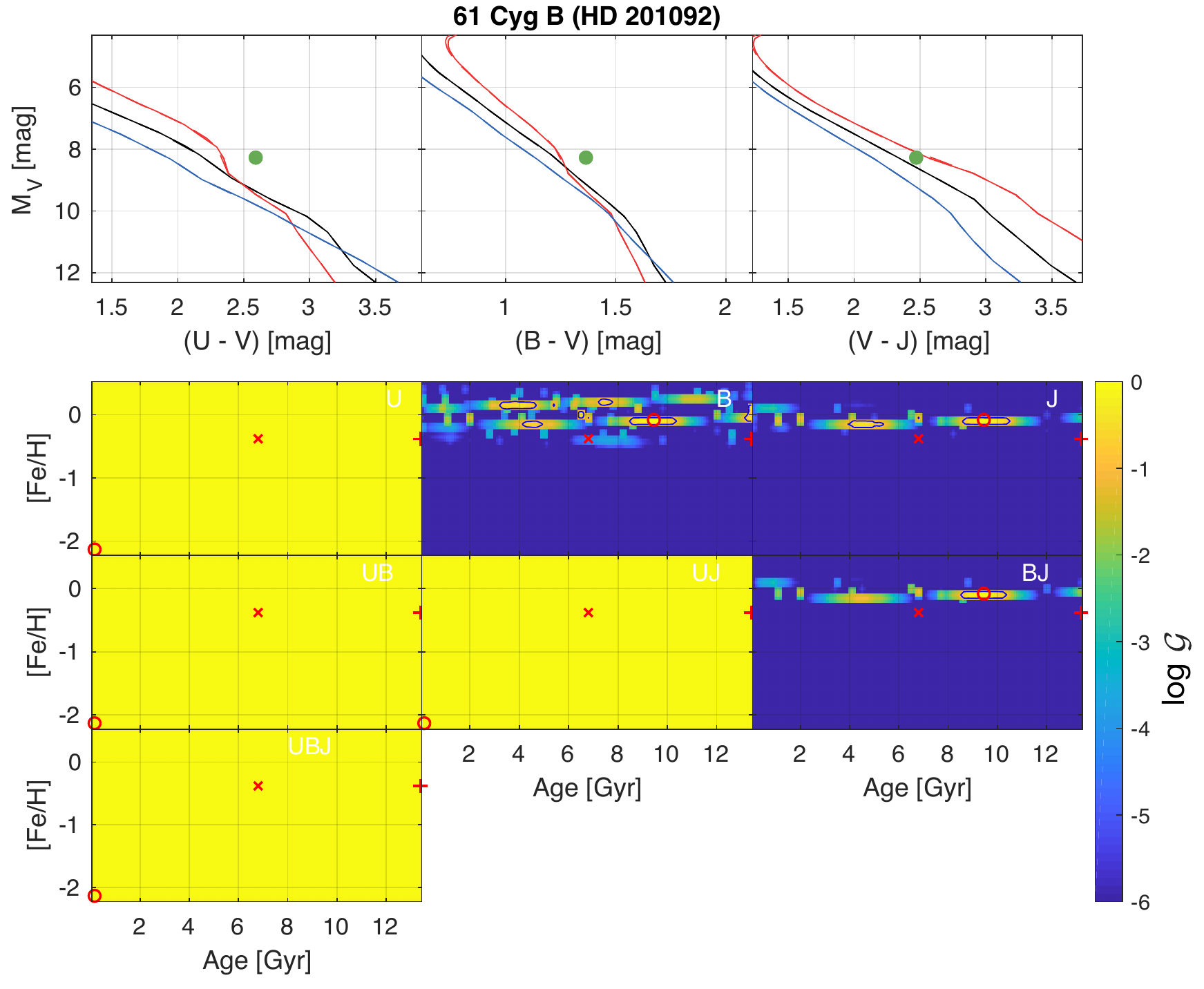}
  \caption{Tests run on the {\it Gaia} benchmark star 61 Cyg B. As in Fig. \ref{fig:BetaHyi}. {\referee The plots that are uniformly yellow have no solution in isochrone space, and so all ages and metallicities are equally likely.}}
  \label{fig:61CygB}
\end{minipage}
\end{figure*}

The {\it Gaia} benchmark stars \citep{2014A&A...566A..98B, 2015A&A...582A..49H} are a sample of more than 30 FGKM type stars of different metallicity and evolutionary type that have been studied in detail in order to provide a set of calibration stars with precisely determined stellar parameters and abundances \citep{2014A&A...564A.133J, 2015A&A...582A..49H, 2015A&A...582A..81J}. As they are a well characterised set of stars, they make a good choice for the initial tests. 

Not all of the stars are suitable for our tests, for example some of them are variable. Further, we would like to test these bright stars with the Johnson-Cousins passbands, especially $U$, but not all of them have suitable photometry available. Eleven of the stars have been chosen as having suitable available data and these are listed, along with their reported parameters, in Table~\ref{tab:gaiabenchmark}.


To assess the resulting estimates from the $\mathcal{G}$ functions, we took ages from \citet{Sahlholdt:2018ty}, which contains a compilation of ages found in the literature for the benchmark stars. {\referee The `literature ages' used here are the mean values of all the isochrone-fitting ages found in the literature over the last 20 years. We also calculate the standard deviation as a measure of the reliability of this value, given as the `error' term in Table~\ref{tab:gaiabenchmark}. To complement this, we include in Table~\ref{tab:gaiabenchmark} a `spectroscopic age', also taken from \citet{Sahlholdt:2018ty}. In that case, the age is calculated from the $\mathcal{G}$ function similar to that used here, but also marginalised over [Fe/H] in order to get a 1D age probability distribution function. These were calculated using the same PARSEC isochrones as in this study, and the stellar parameters ($T_{\rm{eff}}$, spectroscopic $\log{g}$ and [Fe/H]) from \citet{2015A&A...582A..49H}. We have chosen these specific ages from \citet{Sahlholdt:2018ty} rather than the general recommended ages from that paper, because the methods and isochrones are identical to those used here so the ages should differ only because of the use of photometry rather than spectroscopy. We note that these ages with PARSEC isochrones, $T_{\rm{eff}}$, $\log{g}$, and [Fe/H] as input are in general slightly higher than the recommended values. Three of the stars tested here have spectroscopic ages outside of the recommended ages and these have been highlighted in Table~\ref{tab:gaiabenchmark}. Finally we include a rank for each star, also taken from \citet{Sahlholdt:2018ty}, which separates those stars into three categories. `A' stars have ages that are contained to a well-defined range of only a few Gyr. `B' stars have a large range of possible ages, or lack an upper limit to the age. Lastly `C' stars have little to no reliable age information available via current methods.}

The parallaxes were taken from the \textsc{Hipparcos} catalogue \citep{2007A&A...474..653V}. The majority of the photometry used came from the online catalogue of \citet{2002yCat.2237....0D} containing photometry in the Johnson passbands. After some testing, it became clear that the $R$ and $I$ passbands do not have the same transmission curves as those used for the photometric calculations of the isochrones, which were taken from \citet{1990PASP..102.1181B}. Consequently they have been removed from the tests and only $UBVJ$ have been used. In some cases, the $J$ band used in the observations came from the 2MASS passband set, and in these cases the appropriate isochrones were used.

\begin{table*}
\begin{minipage}{180mm}
	\caption{Details of the {\it Gaia} benchmark stars used in the test in Sect.~\ref{sec:GBS}, along with a likely evolutionary stage classification. All parameters are taken from \citet{2014A&A...564A.133J} and \citet{2015A&A...582A..49H}, apart from the literature and {\referee spectroscopic} ages, and the rank for each star, which have all been taken from \citet{Sahlholdt:2018ty}, and are described in more detail in Sect.~\ref{sec:GBS}. The spectroscopic ages are derived using the same PARSEC isochrones as here. The photometry used for each star is given in Appendix Table~\ref{tab:GBSphoto}.}
	\centering
	\label{tab:gaiabenchmark}
	\begin{tabular}{lrlrrrrrr}
		\hline
	 	\noalign{\smallskip}
		Common & HD & Evol. & Parallax & [Fe/H] & Mass & Age (lit.) & Age ({\referee spec.}) & {\referee Rank} \\
		name & number & stage & (mas) & (dex) & ($M_{\odot}$) & (Gyr) & (Gyr) & \\
		\noalign{\smallskip}
		\hline
		\noalign{\smallskip}
		$\beta$ Hyi & 2151 & MSTO & $134.07\pm0.11$ & $-0.04$ & 1.15 & $5.7\pm0.9$ & $6.8$ & A \\
        HD 22879 & 22879 & Dwarf & $39.13\pm0.57$ & $-0.86$ & 0.75 & $10.4\pm3.2$ & $13.5$ & B \\
        $\tau$ Cet & 10700 & Dwarf & $273.96\pm0.17$ & $-0.49$ & 0.78 & $6.2\pm3.5$ & $9.3$ & B \\
        $\beta$ Vir & 102870 & MSTO & $91.50\pm0.22$ & $+0.24$ & 1.34 & $2.9\pm0.4$ & $3.3$ & A \\
        Arcturus & 124897 & High-RGB & $88.83\pm0.53$ & $-0.52$ & 1.03 & $7.4\pm1.9$ & $3.0^{a}$ & B \\
        $\mu$ Leo & 85503 & High-RGB & $26.27\pm0.16$ & $+0.25$ & 1.69 & $3.4\pm0.4$ & $5.3$ & B \\
        $\beta$ Gem & 62509 & RC & $96.52\pm0.24$ & $+0.13$ & 2.30 & $1.1\pm0.2$ & $1.4$ & A \\
        $\epsilon$ Vir & 113226 & RC & $29.75\pm0.14$ & $+0.15$ & 3.02 & $0.7\pm0.3$ & $1.6^{a}$ & A \\
        HD 107328 & 107328 & RC & $10.60\pm0.25$ & $-0.33$ & 1.41 & $6.8\pm1.4$ & $1.7$ & B \\
        Gmb 1830 & 103095 & Dwarf & $109.98\pm0.41$ & $-1.46$ & 0.64 & $9.5\pm4.0$ & $13.5^{b}$ & C \\
        61 Cyg B & 201092 & Dwarf & $285.89\pm0.55$ & $-0.38$ & 0.61 & $6.8\pm0.7$ & $13.4^{a}$ & B \\
        \noalign{\smallskip}
        \hline
        \noalign{\smallskip}
	\end{tabular}
	\caption*{$^{a}$ This value does not fall inside the recommended range from \citet{Sahlholdt:2018ty}, see text for details. \\ $^{b}$ \citet{Sahlholdt:2018ty} does not provide a recommended range for this star. }
\end{minipage}
\end{table*}

\begin{table*}
\begin{minipage}{180mm}
	\caption{A summary of the $\mathcal{G}$ functions for the {\it Gaia} benchmark stars, calculated using the colours $(U-V)$, $(B-V)$, and $(V-J)$. For each star, we list the most probable age and [Fe/H] (highest point in the $\mathcal{G}$ function), the offset of these values from the literature values, and also the maximum and minimum values for each parameter within the 90\% confidence interval (`90\% Age Range', see Sect.~\ref{sec:GBS}). The $\mathcal{G}$ function for 61 Cyg B produced no solution, see text for more details.}
	\centering
	\label{tab:BenchmarkResults}
	\begin{tabular}{lrrrrrrr}
		\hline
	 	\noalign{\smallskip}
		Common & Probable & 90\% Age & Offset from & Offset from & Probable & 90\% [Fe/H] & Offset from \\
		name & age (Gyr) & range (Gyr) & lit. (Gyr) & {\referee spec.} (Gyr) & [Fe/H] (dex) & range (dex) & lit. (dex) \\
		\noalign{\smallskip}
		\hline
		\noalign{\smallskip}
        $\beta$ Hyi & $5.4$ & [$3.5$, $6.7$] & $-0.3$ & $-1.4$ & $+0.15$ & [$-0.11$, $+0.30$] & $+0.19$ \\
        HD 22879 & $8.2$ & [$3.2$, $13.4$] & $-2.2$ & $-5.3$ & $-0.40$ & [$-0.64$, $-0.22$] & $+0.46$ \\
        $\tau$ Cet & $12.4$ & [$6.3$, $12.6$] & $+6.2$ & $+3.1$ & $-0.35$ & [$-0.42$, $-0.23$] & $+0.14$ \\
        $\beta$ Vir & $3.3$ & [$2.3$, $5.0$] & $+0.4$ & $0.0$ & $+0.20$ & [$+0.05$, $+0.44$] & $-0.04$ \\
        Arcturus & $2.9$ & [$1.5$, $10.1$] & $-4.5$ & $-0.1$ & $-0.15$ & [$-0.32$, $+0.07$] & $+0.37$ \\
        $\mu$ Leo & $5.0$ & [$3.8$, $5.9$] & $+1.6$ & $-0.3$ & $+0.50^{a}$ & [$+0.44$, $+0.50^{a}$] & $+0.25$ \\
        $\beta$ Gem & $1.5$ & [$0.9$, $1.7$] & $+0.4$ & $+0.1$ & $+0.25$ & [$+0.15$, $+0.31$] & $+0.12$ \\
        $\epsilon$ Vir & $0.7$ & [$0.4$, $0.8$] & $0.0$ & $-0.9$ & $+0.00$ & [$-0.06$, $+0.11$] & $-0.15$ \\
        HD 107328 & $1.7$ & [$1.3$, $3.5$] & $-4.9$ & $0.0$ & $+0.05$ & [$-0.19$, $+0.21$] & $+0.38$ \\
        Gmb 1830 & $12.7$ & [$0.1$, $13.4$] & $+3.2$ & $-0.8$ & $-1.20$ & [$-1.33$, $-1.01$] & $+0.26$ \\
        61 Cyg B & - & - & - & - & - & - \\
        \noalign{\smallskip}
        \hline
        \noalign{\smallskip}
	\end{tabular}
	\caption*{$^{a}$ This value is at the edge of the isochrone grid and so is likely unreliable.} 
\end{minipage}
\end{table*}

{\referee A summary of our results can be found in Table~\ref{tab:BenchmarkResults}. The $\mathcal{G}$ functions perform best for those stars with the highest ranks, and are broadly consistent with the theoretical results presented earlier} in their ability to accurately determine ages or metallicities. As expected, the turn-off stars resulted not only in very precisely determined ages and metallicities, but also very close matches to the literature values for the two parameters. An example of this is $\beta$ Hyi, shown in Fig.~\ref{fig:BetaHyi}. The $\mathcal{G}$ function created by combining three colours $(U-V)$, $(B-V)$, and $(V-J)$ predicts a most probable age that is only $0.3$\,Gyr smaller than the literature age.

The giant branch and red clump stars fell into two cases; the very young stars (with an age less than $\sim2$\,Gyr) where our analysis was also very successful - see Fig. \ref{fig:EpsVir} for an example of this, and older stars (e.g., Arcturus shown in Fig. \ref{fig:Arcturus}), where the ages were not well constrained. All five giant stars have most probable ages that match very well to the ages calculated in \citet{Sahlholdt:2018ty}, suggesting that the differences between the other literature sources and our ages come from differences in the stellar models used rather than the use of photometry instead of spectroscopy. In both sets of ages, the older giant stars have large uncertainties.

The dwarf stars performed similarly, albeit with larger differences between our ages and the spectroscopic ages, and with much larger uncertainties as predicted theoretically in Sect.~\ref{sec:two}.

We find that the predicted metallicities are constrained to a 90\% confidence interval of less than 0.5\,dex in almost all cases. However, in eight out of the ten cases with $\mathcal{G}$ function solutions, the literature metallicity falls outside this interval. In the majority of these, the metallicity predicted from the $\mathcal{G}$ function is too metal-rich. A clear example of this is Arcturus (Fig. \ref{fig:Arcturus}), and the CMDs demonstrate the reason for the overestimate. In all three colours, the observed photometry is redder than the cluster isochrone, therefore falling on more metal-rich isochrones. This is true for three of the four stars shown here: only in the CMDs of $\beta$ Hyi do the observed colours match the correct isochrone. The size of the offset between the observed and theoretical colours varies between each passband -- e.g. the offset in the $(U-V)$ colour for Arcturus is considerably larger than that for either the $(B-V)$ or $(V-J)$ colours.

This discrepancy is at its most extreme in the case of 61 Cyg B (Fig.~\ref{fig:61CygB}), where the $(U-V)$ offset is so large (and the astrometry so precise, leading to small uncertainties in the absolute magnitude) that the photometry no longer lies on or even close to the isochrone grid, leading to a $\mathcal{G}$ function with no information (shown as a uniform yellow in the $\mathcal{G}$ function plot). The other passbands however are not as offset and do produce solutions -- but as soon as the $(U-V)$ colour is included in the $\mathcal{G}$ function, no solution is produced.

The case of 61 Cyg B highlights a concern -- if for a particular star any of the photometric passbands are badly offset, it would hamper the ability to derive an age or metallicity, even if the other passbands were accurate. For small samples like this, we can examine each colour and exclude those which are discrepant, but if we were to apply this method to large samples of {\it Gaia} data, such detailed examination would become prohibitively time-consuming.


\subsection{NGC 188 -- an old open cluster}
\label{sec:ngc188}
The tests on the {\it Gaia} benchmark stars show the importance of uniform, well understood photometry. In order to test the applicability of $ugriz$ photometry as an age-metallicity indicator, we move away from using measured parallaxes as input, and turn to open clusters, which have well established ages, metallicities, and distances. 

We have chosen to use for our test NGC 188, one of the oldest open clusters in the Galaxy. It is located at a Galactocentric radius of $\sim10$\,kpc and is positioned out of the plane of the Galaxy \citep{2005A&A...433..917B}. It has low dust extinction \citep{1999AJ....118.2894S}, and there are several hundred confirmed members, meaning that the cluster has been extensively studied: from the early studies of \citet{1962ApJ...135..333S}, to many more recent works (e.g., \citealt{1985JRASC..79..230V, 1990AJ....100..710H, 1999AJ....118.2894S, 2002AJ....124.2693F}).

For NGC 188 we adopted the photometry by \citet{2007AJ....133.1409F}, which covers both the main sequence and giant branch. As their data are given in the $u'g'r'i'z'$ passband system, which is slightly different from the SDSS $ugriz$ passband system to which our isochrones refer, we used the equations in \citet{2007AJ....133.1409F} to transform the photometry to the SDSS system. \citet{2007AJ....133.1409F} estimate the age of the cluster to be $7.5\pm 0.7$\,Gyr, distance $1700\pm 100$\,pc, reddening $E(B-V) = 0.025$\,mag, and assume a metallicity of [Fe/H]$= 0.0$ based on an evaluation of available literature values. For the present test we take these values to be our `true' parameters. We correct the photometry for extinction using the \citet{2007AJ....133.1409F} value, and invert the distance into a parallax, with a conservative 10\% relative uncertainty. We assume a uniform prior on age and metallicity. Whilst it is true that a more accurate prior tailored for a specific cluster would provide somewhat different answers, by continuing with uniform priors, we can use the test to directly compare this with what would happen if we applied the code to a large sample of field stars, which is the ultimate goal. 


\begin{figure}
  \centering
  \includegraphics[width=0.85\columnwidth]{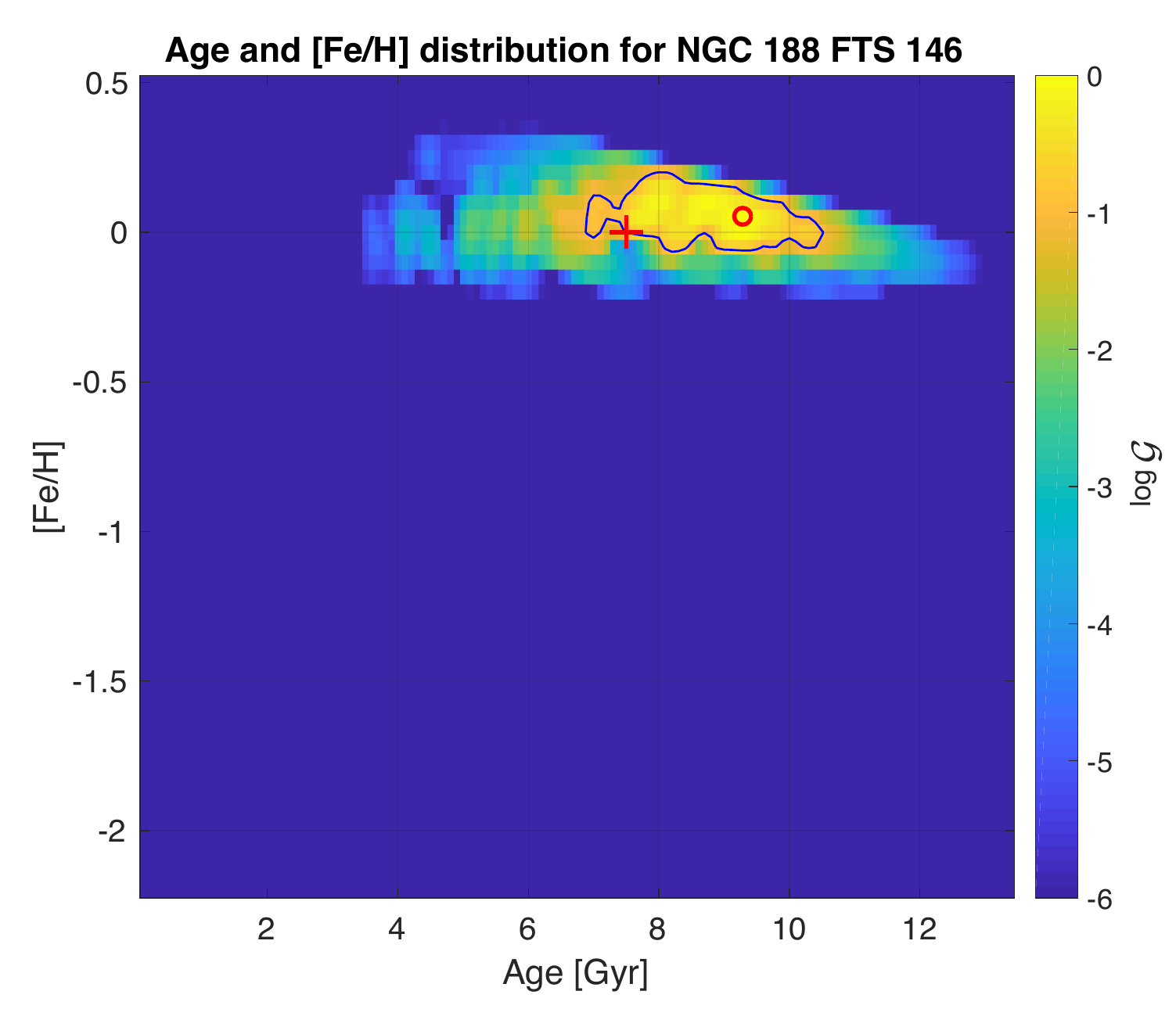}
  \caption{The plot of the $\mathcal{G}$ function calculated for an example MSTO star (NGC 188 FTS 146) in NGC 188, using the $ugriz$ passbands.}
  \label{fig:ExampleNGC188}
\end{figure}

\begin{figure*}
\begin{minipage}{180mm}
  \centering
  \includegraphics[width=0.99\columnwidth]{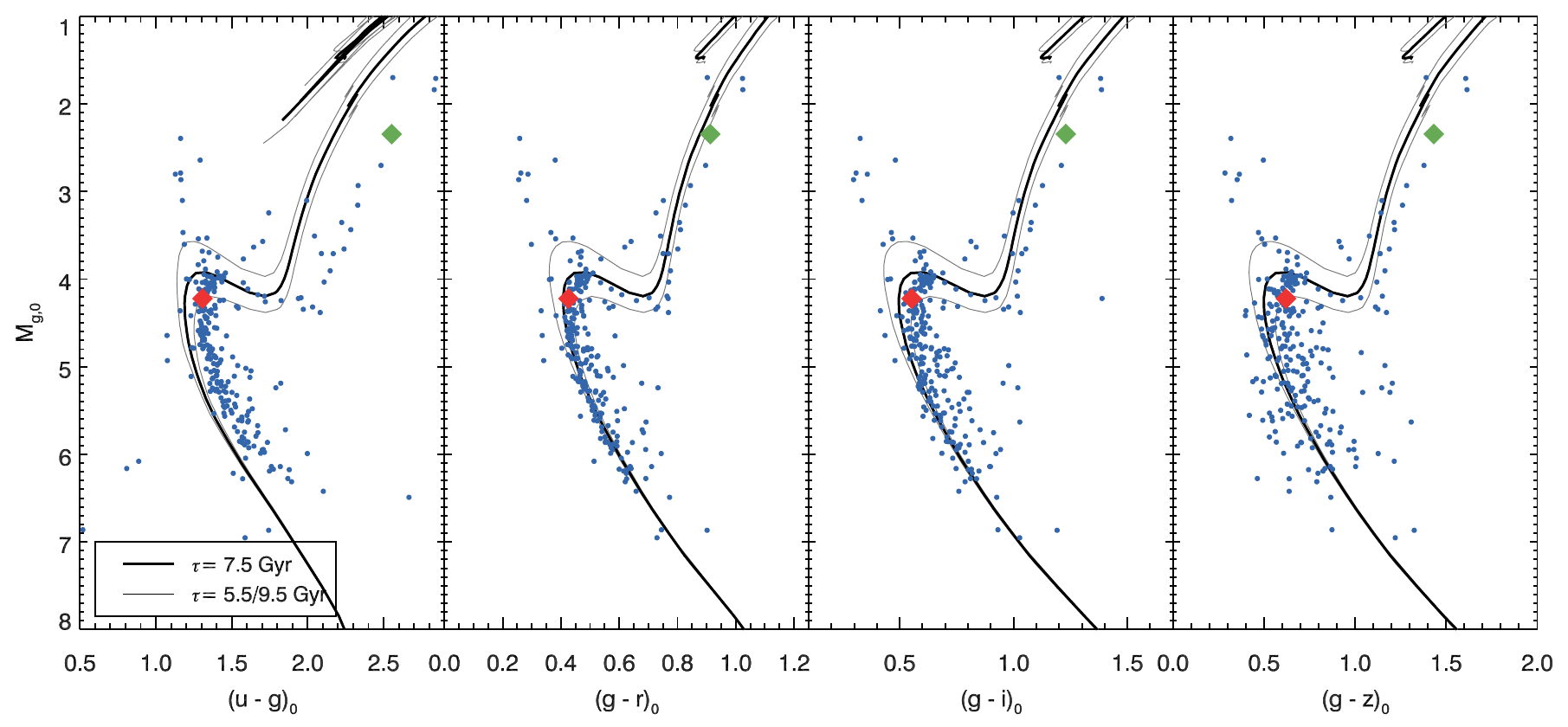}
  \caption{Colour-magnitude diagrams of NGC 188 using the transformed $ugriz$ observations from \citet{2007AJ....133.1409F} (blue points). The black isochrone has [Fe/H]$=0.0$ and an age of 7.5\,Gyr, and the grey isochrones have the same metallicity but ages $\pm2$\,Gyr either side of the black. The red diamond is the MSTO star discussed in the text (NGC 188 FTS 146), and the green diamond is the RGB star discussed (NGC 188 FTS 11).}
  \label{fig:NGC188_CMD}
\end{minipage}
\end{figure*}

In theory, because the age was derived from isochrone fitting to the cluster fiducial sequence in \citet{2007AJ....133.1409F}, by using the same isochrone set, distance, and extinction value we would expect to arrive at the same answer. However, the cluster parameters are not derived from a perfect fit to the cluster data but instead to set points on the fiducial sequence, in particular the turn-off. Furthermore, the fit was made using one colour and magnitude combination ( \citealt{2007AJ....133.1409F} used the $g'$, $(g'-r')$ CMD). We first test the code on NGC 188 FTS 146, identified on the CMD as a MSTO star. Figure~\ref{fig:ExampleNGC188} shows the $\mathcal{G}$ function calculated by using all five passbands - in this case, the $g$ band magnitude, and the four colours $(u-g)$, $(g-r)$, $(g-i)$, $(g-z)$. It is immediately apparent that the result is close to the actual age and metallicity (the red cross), although slightly offset. The estimated most probable age is 9.2\,Gyr instead of the 7.5\,Gyr derived by \citet{2007AJ....133.1409F}, and the [Fe/H] is less than 0.1\,dex higher than the literature value.

Whilst this is promising, it is worth investigating why the solution is offset from the answer that was derived from the same isochrones. Figure~\ref{fig:NGC188_CMD} shows the CMDs for all four colours used in the calculation of the $\mathcal{G}$ function in Fig. \ref{fig:ExampleNGC188}, with the star in question shown as a red diamond in all four plots. The isochrone drawn in black has the same age -- 7.5\,Gyr -- as that estimated by \citet{2007AJ....133.1409F}. Apart from the $g$, $(g-r)$ CMD where the observed colours match the isochrone well, the other observed colours are all redder than the isochrones. The star lies closer to the grey line underneath, which has an age of 9.5\,Gyr. This mismatch between the observation and theory explains why the most likely age of the star in the $\mathcal{G}$ function plot is $\sim2$\,Gyr older than the reported age of the cluster.

\begin{figure}
  \centering
  \includegraphics[width=0.85\columnwidth]{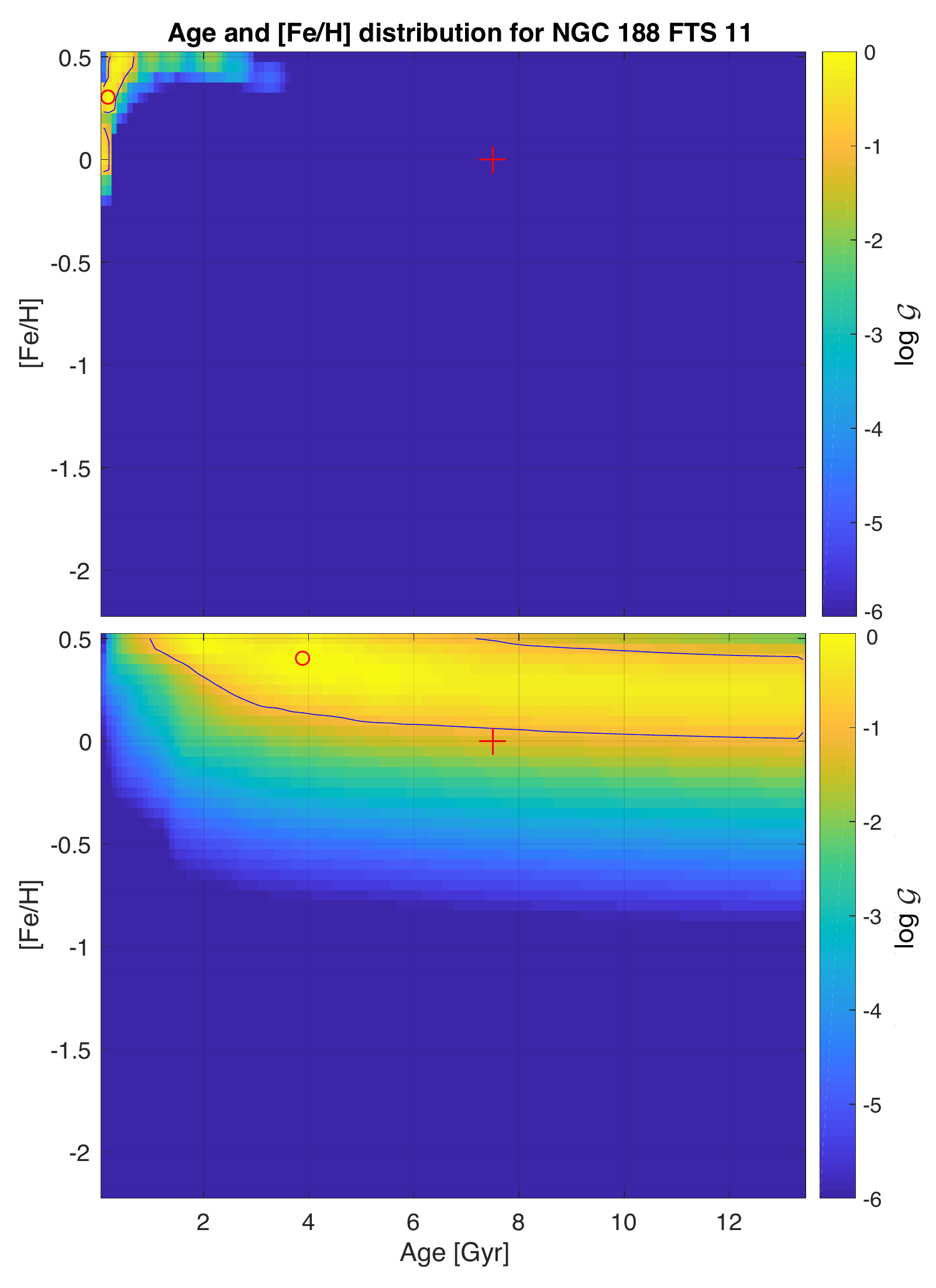}
  \caption{The $\mathcal{G}$ functions calculated for an example RGB star in NGC 188, using the $ugriz$ passbands. {\it Top:} the $\mathcal{G}$ function is calculated using the observed photometric {\referee uncertainties}. {\it Bottom:} the $\mathcal{G}$ function is calculated assuming both the photometric {\referee uncertainty} and an uncertainty of 0.1\,dex in the isochrones.}
  \label{fig:egRGB_NGC188}
\end{figure}

Figure~\ref{fig:NGC188_CMD} shows all of the observed data from \citet{2007AJ....133.1409F} compared to the isochrones. As expected, the $g$, $(g-r)$ CMD generally fits very well with the data. The other passbands fit less well, particularly in the case of the $u$ band, where the isochrone is consistently bluer than the data. For the bulk of the main sequence in the $(g-i)$ and $(g-z)$ CMDs, the photometry is broad and crosses the isochrone, suggesting that the observational uncertainties in these passbands are larger than thought -- although some of the broadening is due to binary systems appearing brighter than expected from their colours. The observed RGB is redder than the isochrone in all four passbands -- least in $(g-r)$, greatest in $(u-g)$.

\begin{figure}
  \centering
  \includegraphics[width=0.75\columnwidth]{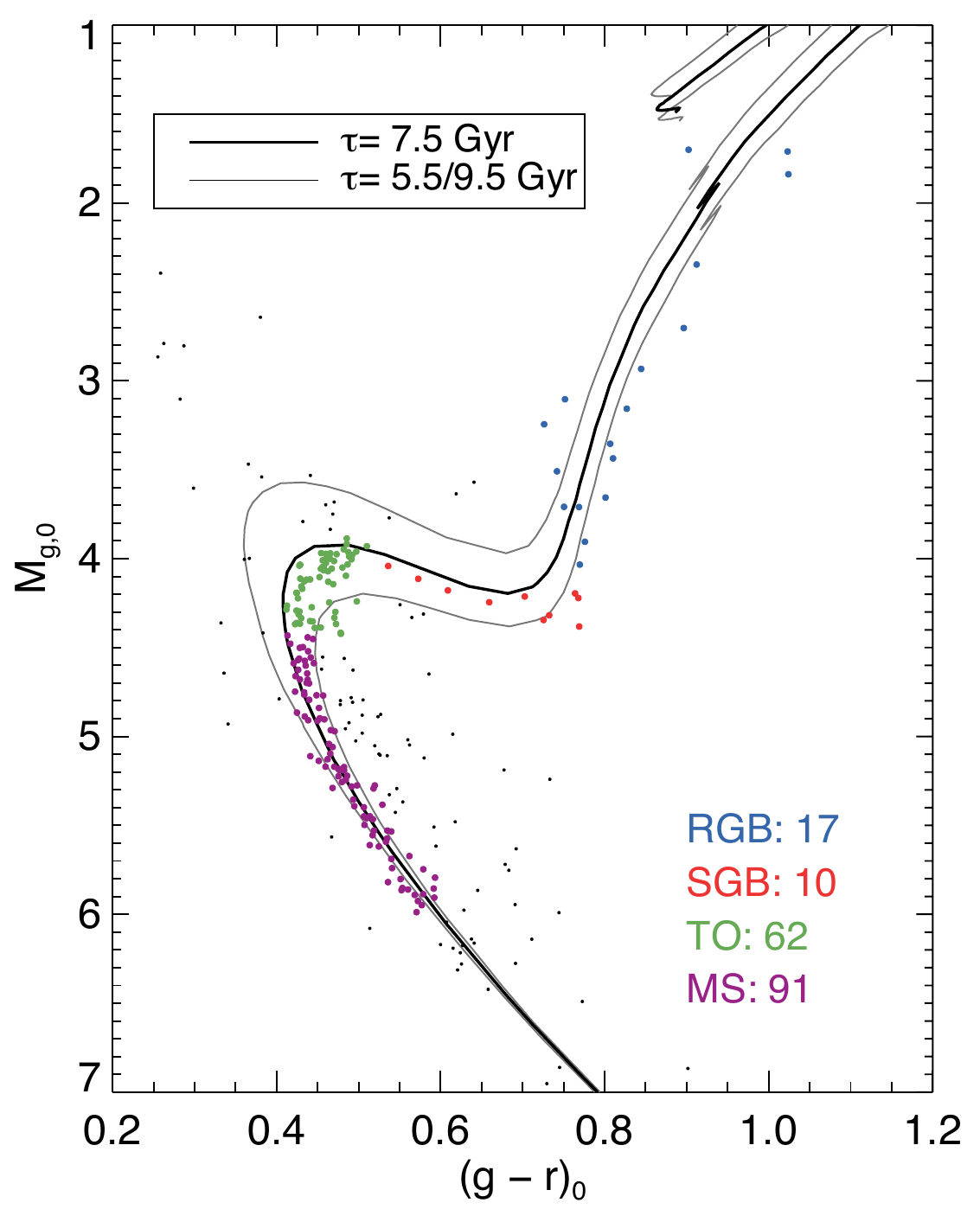}
  \caption{The $g$, $(g-r)$ CMD of NGC188, with stars coloured according to which of the four evolutionary stages they are in for the purposes of the box and whisker plots of Figs. \ref{fig:all_ages_NGC188} \& \ref{fig:all_fehs_NGC188}, and the discussion in Sect. \ref{sec:ngc188}. At the bottom right of the figure are the numbers of stars in each evolutionary stage. The small black dots are those stars which we excluded from further analysis. The isochrones plotted underneath are the same as those in Fig.~\ref{fig:NGC188_CMD}.}
  \label{fig:NGC188_groups}
\end{figure}

\begin{figure*}
\begin{minipage}{180mm}
  \centering
  \includegraphics[width=0.99\columnwidth]{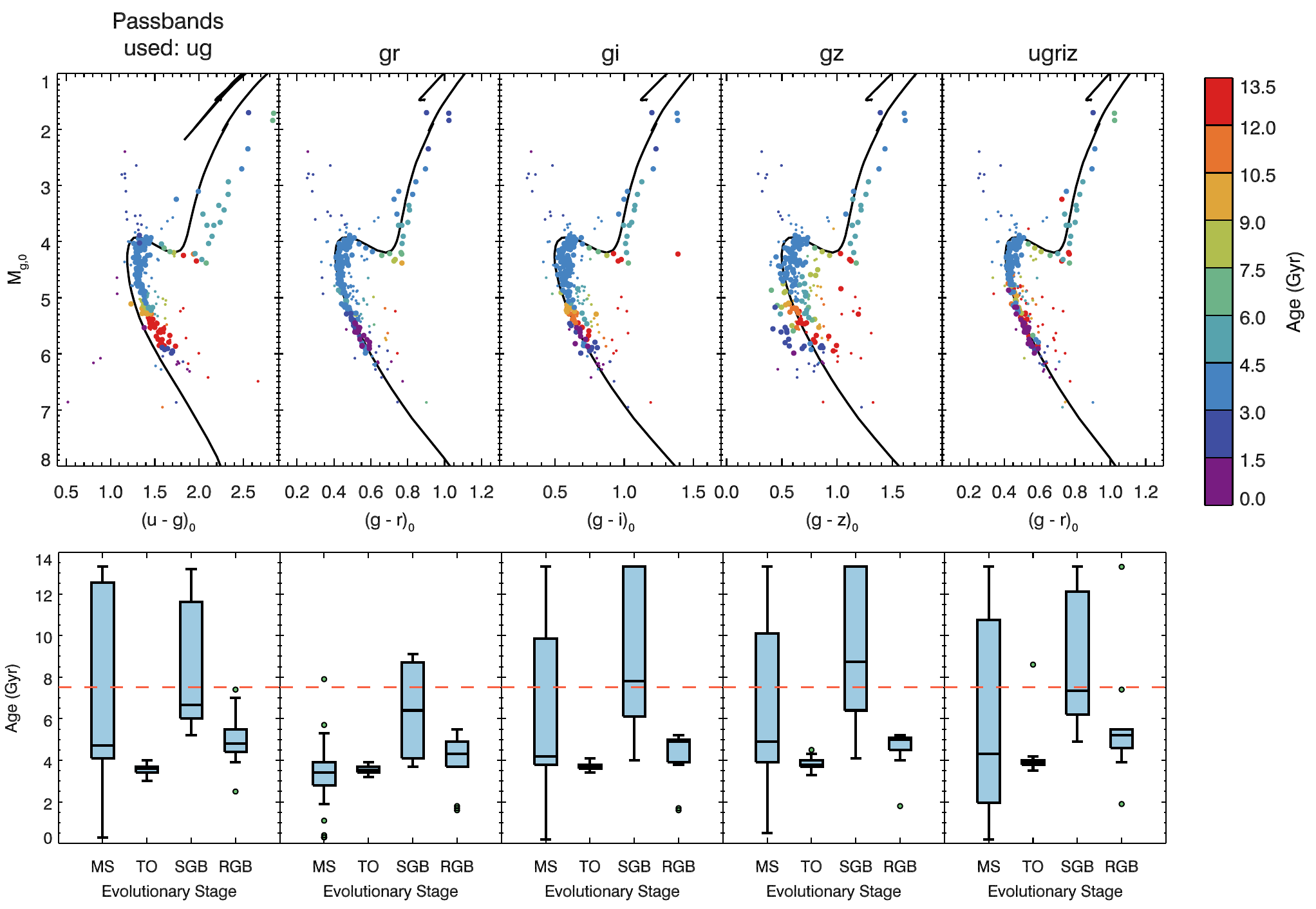}
  \caption{The most probable age calculated for each star, represented first individually, and grouped then by evolutionary type. {\it Top:} the CMDs of NGC 188 (as in Fig. \ref{fig:NGC188_CMD}), where each star is coloured by its most probable age. In each of the five plots, the age is calculated using the colour formed from the passbands printed above the plot. The far right plot uses all four colours ($(u-g)$, $(g-r)$, $(g-i)$, $(g-z)$), and is shown on the $g$, $(g-r)$ CMD. Printed for reference is the PARSEC isochrone with the given age (7.5\,Gyr) and metallicity (Solar) of the cluster. The larger circles are those selected as belonging to one of the four evolutionary stages and used in the bottom figures, whereas the smaller circles are the remaining probable members not selected for the box and whisker plots. {\it Bottom:} box and whisker plots of the most probable ages calculated for the stars in NGC 188 and grouped by evolutionary stage. The stars used are shown in different colours in Fig. \ref{fig:NGC188_groups}, and each $\mathcal{G}$ function is calculated using the same colours as in the plots above. Outliers are shown as green circles, and the red dashed line shows the `true' age.}
  \label{fig:all_ages_NGC188}
\end{minipage}
\end{figure*}

\begin{figure*}
\begin{minipage}{180mm}
  \centering
  \includegraphics[width=0.99\columnwidth]{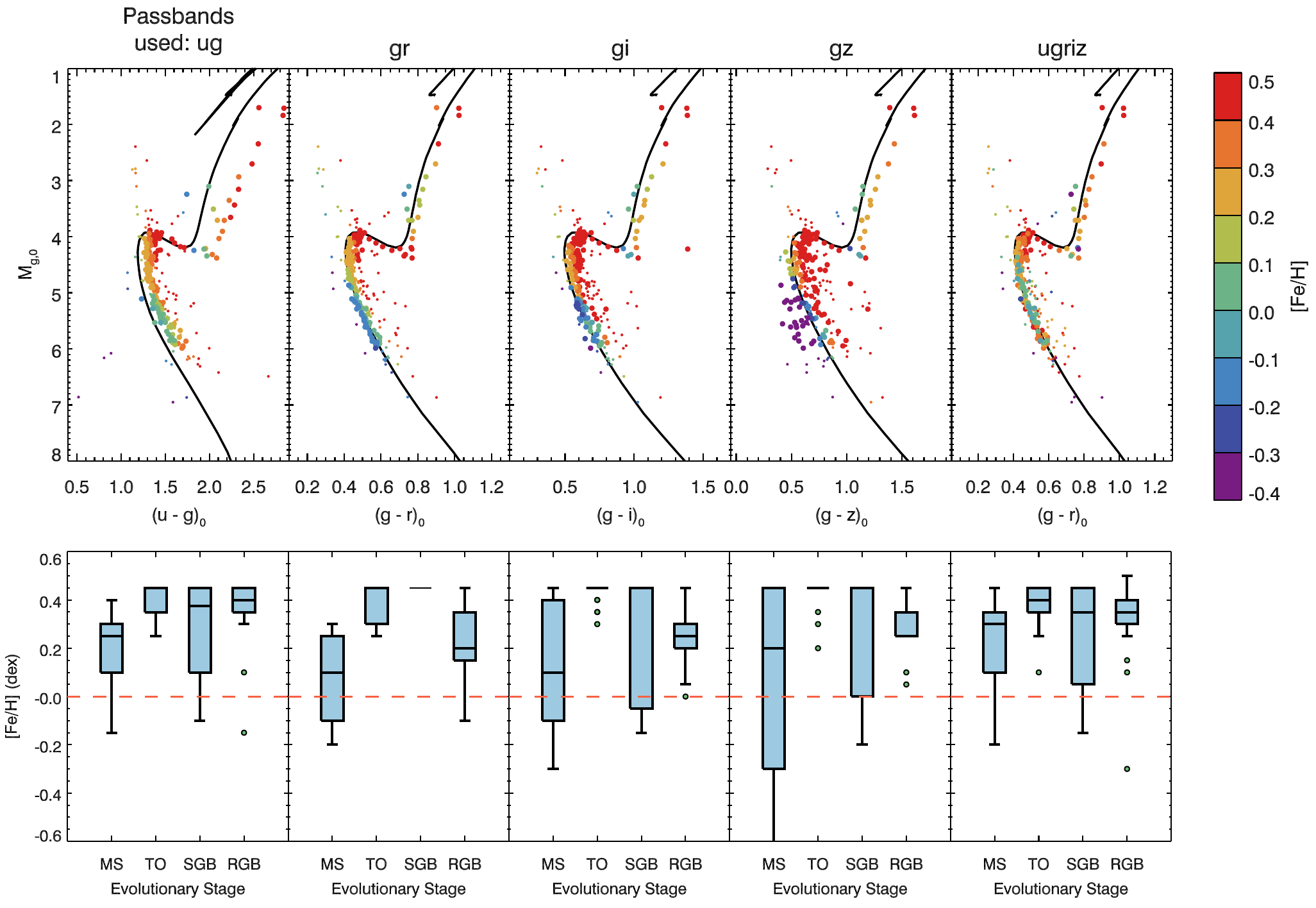}
  \caption{As in Fig. \ref{fig:all_ages_NGC188} but for [Fe/H] instead of age. {\it Top:} the CMDs of NGC 188, where each star is coloured by the most probable [Fe/H] value. Again only the stars shown as large circles are used to create the box and whisker plots below. {\it Bottom:} box and whisker plots of the most probable [Fe/H] values grouped by evolutionary stage. The whisker of the main sequence stars in the $(g-z)$ plot extends beyond the bottom of the plot to [Fe/H]$=-1.0$.}
  \label{fig:all_fehs_NGC188}
\end{minipage}
\end{figure*}

Given the uncertainties published by the photometric studies -- in the case of this \citet{2007AJ....133.1409F} data, all but the stars at the bottom of the main sequence have observational uncertainties of less than 0.02\,dex in all five passbands -- it is clear that the mismatch between the isochrones and photometry is much larger than the uncertainties in some passbands and for some evolutionary stages. This significantly skews the ages and metallicities estimated for individual stars using our method. In certain cases, this offset can lead to surprising, counter-intuitive results. An example of this is star NGC 188 FTS 11, a RGB star shown as the green diamond in Fig.~\ref{fig:NGC188_CMD}. In all four CMDs, the star appears to be redder than the isochrone of the cluster age and metallicity. As the star is a giant, we would expect the $\mathcal{G}$ function to poorly constrain the age, but provide a reasonable estimate of the metallicity (as in the RGB examples of Fig. \ref{fig:UBVPhotTest}). Due to the offset observed in the CMDs, we would also expect that this metallicity estimate would be more metal-rich than the chosen metallicity of the cluster.

The top panel of Fig.~\ref{fig:egRGB_NGC188} shows the $\mathcal{G}$ function calculated using all four colours, $(u-g)$, $(g-r)$, $(g-i)$, and $(g-z)$. The probable region is confined to a very small region at the top of the grid, with the most probable age and metallicity being 0.2\,Gyr and $+0.3$\,dex, respectively. Such a young age is unexpected, but can be understood by considering how the probabilities are calculated. The square of the observational uncertainty is used in the likelihood sum (for the full calculation, see Eq. \ref{eq:lik}), in the case of the colour terms with small photometric uncertainties (on the order of $\sim0.01$), the result is a very small probability for a given isochrone point when the offset between passbands is an order of magnitude greater than the uncertainty. The parallax is converted into a distance modulus and the likelihood is calculated, but with a parallax uncertainty of $\sigma_{\varpi}=0.1$, the uncertainty in distance modulus is $\sim0.4$\,mag. This uncertainty is 40 times greater than the colour uncertainty -- so when four colours are used in the calculation, these dominate the probability. The most likely solutions are then reasonable matches to the colours, but two or three magnitudes away in distance modulus from the observation. In the case of star NGC 188 FTS 11, the ``best fitting solution'' is a subgiant on a very young isochrone, one magnitude brighter than observed.

{\referee This is a typical example of where blindly calculating the $\mathcal{G}$ function (or similar isochrone fitting method) and taking the mode of the probability distribution function without looking at the probabilities more carefully can lead to the wrong answer.} In particular, to avoid the colour terms dominating the probability in such a manner, and to account for the offsets seen between the different passbands and the isochrones, we also recommend including a term in the calculation for uncertainties in the isochrones. If, for example, an ``isochrone uncertainty'' of $0.1$\,mag is added in quadrature to the observed uncertainty, the resulting $\mathcal{G}$ function is shown in the bottom panel of Fig. \ref{fig:egRGB_NGC188}. This result is a close match to the shape predicted by our theoretical tests, albeit offset from the true solution, as was expected from the CMDs.


We then calculated the $\mathcal{G}$ functions and recorded the most probable age and metallicity for each star, to determine how well the observed photometry matches across all evolutionary stages. The CMDs in Figures \ref{fig:all_ages_NGC188} \& \ref{fig:all_fehs_NGC188} show each star coloured by its most probable age and metallicity, respectively. Caution should be taken in Fig. \ref{fig:all_ages_NGC188} when considering the most likely ages of the main sequence and giant branch stars, remembering from our earlier tests that the age is poorly constrained in these regions (Sect.~\ref{sec:opt-nUV}). 

The first thing to note is that our determinations of age and metallicity do not match well to those made by \citet{2007AJ....133.1409F}, even in the turn-off region of the $g$, $(g-r)$ plot, where the isochrone was originaly fit. There are a few reasons this could be; firstly we are using a more recent version of the PARSEC isochrones. Secondly, \citet{2007AJ....133.1409F} fixed the [Fe/H] value from the literature, rather than fitting it to their data. Our most probable ages are free to vary with metallicity, so are likely to vary from theirs. Figure \ref{fig:egRGB_NGC188} implies that, for a fixed [Fe/H]$=+0.0$, the resulting $\mathcal{G}$ function age would be closer to their value. Another reason for the difference is the method of fitting -- in fitting an isochrone to the cluster by eye, one attempts to put this on the left side of the main sequence to avoid the binaries that naturally broaden the observations. In our case, we fit an isochrone to each individual star, so most of the stars in the main sequence and turn-off region will end up with slightly smaller ages and larger [Fe/H] values. This includes those stars that may well be binaries on the right of the main sequence, and we can see that these appear to be older and more metal-rich than the rest of the cluster -- an important reminder that unresolved binaries will have incorrect parameters, regardless of how well the isochrones fit the single star data. We will not consider the issue of binary stars further, but will return to this in future work.

Ignoring these offsets in $g$, $(g-r)$ as inevitable between different techniques, there are some other conclusions we can draw. Fig. \ref{fig:all_fehs_NGC188} has a noticeable gradient in derived metallicity along the main-sequence, with more metal-poor stars at the bottom. The fiducial line of the cluster's main sequence using this photometry is not well described by the shape of the isochrone, causing this variation. In both ages and metallicities, we get different values for the giant stars compared to the turn-off (they appear generally older and more metal-poor). Crucially the stars' parameters vary between each of the five plots; we get different answers by using different colours. This is most noticeable in the metallicities, where the lower part of the main sequence becomes progressively more metal-poor with redder colours. The four colours are combined in the far right plot of both figures, which shows a mixed bag of ages and metallicities and has a less prominent gradient.

We have chosen a subset of the NGC 188 stars as representative of their evolutionary stages, in the bottom panels of Figs. \ref{fig:all_ages_NGC188} \& \ref{fig:all_fehs_NGC188}, shown as the larger circles in the top panels and also shown in the CMD in Fig.~\ref{fig:NGC188_groups}, coloured by evolutionary stage. The stars were selected conservatively to avoid including outliers with unusual colours compared to the rest of the cluster stars, blue stragglers, and likely binaries on the main sequence. Each group of stars is shown as a box and whisker plot. These demonstrate the large discrepancy between the different stages, particularly in Fig. \ref{fig:all_ages_NGC188}, where broadly speaking the estimated ages are similar across the different colours used, but vary considerably with evolutionary stage. In particular, the turn-off stars and the subgiants show very different distributions, with the turn-off ages all clustering around one young age, whereas the subgiants are more spread out and are almost all older than the turn-offs. From our tests in Sect.~\ref{sec:tests} we concluded that only turn-off and subgiant stars should yield useful age estimates; however, we see here that they may produce systematically very different age estimates.

Variations among the evolutionary stages are also evident in [Fe/H] (Fig. \ref{fig:all_fehs_NGC188}), although here the more noticeable difference is between the different colours. {\referee This leads to the question, if each colour used produces a different preferred metallicity, which is the correct one to use?}

Analysing all the stars individually within the cluster highlights the issues with using isochrones for parameter determination. The metallicities of stars in open clusters are known to be homogeneous down to scales of $<0.05$\,dex (e.g., \citealt{2016MNRAS.463..696L}), far beyond the differences seen in our determinations, and stars born in clusters are known to have age spreads of at most tens of millions of years \citep{2003ARA&A..41...57L}. Furthermore, the parameters should match when derived using different photometric passbands, which is not the case for many of the individual ages and [Fe/H] estimates obtained in this example.

\section{{\referee Comparing photometry and isochrones}}
\label{sec:five}
In Sect.~\ref{sec:ngc188} we used stars from NGC 188 to test the success of predicting age and metallicity from photometry and faux--{\it Gaia} parallaxes. Our results show a mismatch between the observed photometry and predicted values from the isochrones, at various different evolutionary states and across the different colours.

{\referee The key result from Sect.~\ref{sec:ngc188} is that in order to obtain reliable results with these methods, the systematic differences between isochrones and photometric observations need to be taken into account. Our tests reveal that currently, this mismatch is $\sim 0.1$\,dex, which we compensate for by adding a corresponding uncertainty. However, even with this uncertainty, the results still vary by evolutionary type and photometric passband. For the method to work correctly, the offsets between isochrones and observations need to be an order of magnitude smaller -- on the same scale as the currently reported photometric uncertainties.

In the remainder of this section we examine the underlying causes for these mismatches, and suggest how future work can improve the situation.}

\subsection{Convection in stellar models}

Deriving all four possible cluster parameters (age, metallicity, distance, and extinction) from isochrone-fitting has long been known to be problematic due to the difference between the shape of the fiducial line of the cluster (the line fitted through the stellar observations of a single cluster on a CMD) and that of the isochrone. In particular, when an isochrone is fitted to the main sequence and turn-off, the observed red giant branch is often redder than that of the isochrone. {\referee The likely cause of this discrepancy is the treatment of convective motions in the underlying stellar models. Creating isochrones requires a model of the stellar atmosphere, which in low-mass stars is convective. Convection is an inherently 3D process, however three dimensional models of stellar atmospheres are too computationally expensive to use in stellar evolution models. For decades, stellar modellers have instead imitated the results of convection using the one dimensional mixing length theory \citep{1958ZA.....46..108B}. Typical implementations have one parameter, $\alpha_{MLT}$, which is calibrated to the Sun. In the case of the PARSEC models used in this paper, the Solar calibration gives $\alpha_{MLT}=1.74$ \citep{2012MNRAS.427..127B}.}

This calibration, however, has been found unable to describe all stars; in particular one value of $\alpha_{MLT}$ chosen for the Sun is unsuitable for stars on the giant branch. For some years now, more advanced models have been used to show this problem -- for example, the two dimensional radiative hydrodynamical models of \citet{1999A&A...346..111L} showed a variation in $\alpha_{MLT}$ of more than 0.4 between dwarfs and subgiants; increases in computational power since then have enabled three dimensional models \citep{2013A&A...557A..26M}. These models revealed that values of between 1.7 and 2.4 were needed to fully describe FGK-type stars with a range of metallicities \citep{2015A&A...573A..89M}.

The study of $\alpha_{MLT}$ has been setback many years by our inability to infer this parameter from observations. $\alpha_{MLT}$ is calibrated using a star's radius and mass, both of which are much more uncertain in stars other than the Sun. Small-scale studies of stellar radii and masses showed offsets from the Solar value (e.g., \citealt{1986ApJ...300..773D}), but with the arrival of large-scale asteroseismology surveys in the last decade, however, this is beginning to change. \citet{2012ApJ...755L..12B} showed using data from the {\it Kepler} mission that the Solar $\alpha_{MLT}$ when applied to giant stars, led to incorrect helium abundances. More recently, \citet{2017ApJ...840...17T} demonstrated a metallicity-dependence in the best fitting parameter, when compared to 3000 red giants in the APOKASC \citep{2014ApJS..215...19P} sample of stars with asteroseismology and spectroscopy.

{\referee New asteroseismic missions such as TESS and PLATO will provide even more data on the interiors of stars, so progress will be made in the near future towards better understanding the interiors of stars unlike the Sun. As the computational demands of three dimensional modelling become more manageable, better models of stellar convection will hopefully enable the production of more accurate isochrones in the near
future.}

\subsection{Stellar Opacities}

Further problems have been identified in other parts of the CMD. \citet{2008ApJS..179..326A} looked at SDSS $ugriz$ photometry of several globular and open clusters, and found not only mismatches between the best fitting isochrones required for RGB and main sequence stars, but also problems lower down the main sequence. For example, in their study of M67, they found that the isochrone colours were up to 0.5\,dex too blue at the bottom end. The same problem was discussed in detail in \citet{2003MNRAS.345.1015G}, who used $BVIK$ photometry of six open clusters to compare several sets of isochrones, specifically focusing on the main sequence. They found that none of the isochrone sets tested were fully able to match the entire length of the main sequence, in general, the models diverged lower down from the observations, with the isochrones being too blue. By comparing the different models in both the temperature-luminosity plane as well as using regular CMDs, they demonstrated that the problem is caused by both the colour transformations used to estimate the values in each photometric passband, and the underlying physics used in the models. In particular, they identify that missing sources of opacity in the stellar atmospheres of the models (a bigger problem in cooler stars) could be a significant contributor. 

Progress has been made in the area of stellar opacities since 2003, and the version of the Padova isochrones that we use here has more up-to-date opacity data with a specific low-temperature opacity code \citep{2009A&A...508.1539M}, however our comparisons in Fig.~\ref{fig:NGC188_CMD} show that there are still mismatches on the lower part of the main sequence. 

{\referee The work involved in improving our understanding of molecular opacities and their effects in stellar astronomy is ongoing (e.g., \citealt{2007ApJS..168..140S, 2009A&A...494..403L, 2014MNRAS.438.1741F}), and there is good reason to think that with time the uncertainty caused by missing sources of opacity will reduce to levels that allow accurate isochrone fitting in this region of the CMD.}

\subsection{Atomic diffusion}
Another well-known complication in stellar-modelling is that of atomic diffusion. Due to the diffusion of heavy elements from the stellar surface into the star whilst on the main-sequence, it has been found that a star's initial metallicity will vary from the measured metallicity. This has been shown in, for example, globular clusters (e.g., \citealt{2007ApJ...671..402K, 2014A&A...567A..72G}), where the [Fe/H] abundance at the turn-off can be 0.3\,dex lower than in giants (where due to mixing the initial abundance has been restored). Recently \citet{2017ApJ...840...99D} showed that by assuming a star's current abundance is the same as its initial abundance, calculations of the stellar age from isochrones could be overestimated by up to 30\%. In our approach [Fe/H] is not derived from spectroscopy but inferred from photometry, which is affected to a much smaller degree by changes in the surface abundance of the star, so we do not suffer from this problem. In particular, the changes in effective temperature and $\log{g}$ resulting from atomic diffusion are expected to be small and hence the photometric properties of the star should be largely unaffected.  The issues that we discuss elsewhere in this section will have a larger effect on the ages that we derive.

\subsection{The use of different photometric passbands}

As discussed, the different results produced using different photometric colours make it difficult to use the code on large datasets without some initial vetting. The answers produced by using different photometric passbands are investigated in detail in \citet{2015AJ....149...94H}, who also examine the open cluster NGC 188. \citet{2015AJ....149...94H} focus on fitting isochrones to whole clusters, rather than individual stars as we are doing. Using a Bayesian fitting code, the authors use different combinations of the $UBVRIJHK\text{s}$ photometric data (plus some extra information on each star, such as each star's membership probability) to derive a range of cluster parameters from three different isochrone sets. They ran a number of tests using different combinations of the selected passbands, from those with only two, to all eight passbands. The variation in the cluster parameters found in these tests highlights the large inconsistencies between different combinations. The problem extends across all three isochrone sets used, not just the PARSEC isochrones used here, suggesting the problem is universal. Their solution is to use the fits from the maximum number of available passbands, to minimise the effect of any one passband with a significant mismatch. On an individual star basis, however, this option may not be feasible, if the offset between passbands is too large (e.g., Sect.~\ref{sec:GBS}). We also found in our tests that although using more passbands gave smaller uncertainties, the results were not always closer to the truth.

The cause of offsets between passbands is not fully understood. Fig. \ref{fig:NGC188_CMD} in particular showed the offset in the $(u-g)$ colour, and we briefly mentioned in Sect. \ref{sec:four} the difficulties in making measurements in the near-UV -- these may impact how well the observations fit the isochrones. \citet{2015AJ....149...94H} also mention the difficulty of obtaining good stellar atmosphere models in the near-UV. The problems we faced in Sect. \ref{sec:GBS} in finding photometry with passbands defined similarly to the isochrones may well cause some of the problems seen in \citet{2015AJ....149...94H}. The problem extends to other work in the near-UV, however, as shown by \citet{2018PASP..130c4204B} -- where photometry of globular clusters from the Hubble Space Telescope are considered, and again the isochrones are found to be most discrepant in these shorter wavelength passbands. Opacity problems, like those discussed above, have been noted in the past to significantly affect the near-UV region of the spectrum \citep{2002A&A...391..195G}, more so than other regions. {\referee It is clear that careful calibrations of photometry, accurate filter curves, improved near-UV atmosphere models, and further work on the $T_{\rm{eff}}$-colour transformations are needed to make use of the wealth of information stored in the near-UV.}

\section{Summary}
 Photometric measurements of stars contain information about their intrinsic parameters: observables such as $T_{\rm{eff}}$ and $\log{g}$, but also the fundamental parameters age and metallicity, which are vital for studies of the stellar content of the Milky Way. Without additional information on the distance to the star, disentangling both of these parameters is impossible for the vast majority of field stars. {\it Gaia}, however, provides us with parallaxes for more than a billion stars, and in this paper we investigated what can be determined by combining astrometry and broadband photometry. To do this, we developed a 2D probability map (the $\mathcal{G}$ function) of an individual star’s age and metallicity, calculated using Bayesian estimation based on theoretical isochrones.

We have calculated these maps for a range of synthetic stars, covering different ages, metallicities, and evolutionary stages. We used a wide range of broadband photometric passbands in the near-UV, optical, and infrared, to determine those which provide the most information. From these investigations, we found that:
\begin{enumerate}
	\item Photometric data from the {\it Gaia} passbands ($G$, $G_\text{BP}$, and $G_\text{RP}$) and the NIR (e.g., 2MASS $JHK\text{s}$) results in a significant age--metallicity degeneracy that prevents a simultaneous determination of either.
	\item Including a passband in the near-UV, such as the Johnson $U$ or SDSS $u$ in the calculations allowed us to break the degeneracy and find unique solutions in many cases. 
	\item These unique solutions are possible for turn-off and subgiant stars, where uncertainties of $<1$\,Gyr and $<0.2$\,dex are feasible. 
	\item Main sequence or RGB stars do result in metallicity solutions with small uncertainties, but we do not find unique ages. This is because these evolutionary stages are not particularly age sensitive.
	\item For young stars ($\sim2$\,Gyr) we were able to reproduce ages for nearly all stellar types, due to the wider spread between the isochrones.
	\item In order to determine ages with uncertainties of less than $2$\,Gyr with $ugriz$ (or similar) photometry, we estimate that uncertainties on parallax of $\lesssim20\%$ are needed.
\end{enumerate}

A passband in the near UV such as the $U$ passband is so useful in determining age and metallicity because it covers the spectral region (mostly) below the Balmer jump. This region changes significantly with stellar metallicity, as we showed in Fig. \ref{fig:deltaspectra}. The Johnson $U$ is quite wide, covering below but also containing the Balmer jump and so is also sensitive to surface gravity in FGK-type stars. Instead, for a better metallicity discriminant, a narrower $U$ passband would be better, however then one suffers from loss of throughput. There are also difficulties with transformations to outside the Earth’s atmosphere, making $U$ bands in general trickier to work with. Hence there is not a wealth of historical data. SDSS, Luau-CFHT, and SkyMapper are examples of large-scale photometric surveys that are helping to change this, and we encourage future survey designers to consider the value of a passband in this region.

We also applied the $\mathcal{G}$ functions, along with the information gained on useful combinations of passbands, to groups of real stars. These two tests using real data are summarised below.

\begin{enumerate}
  	\item We considered the {\it Gaia} Benchmark stars \citep{2015A&A...582A..49H}, where we found 11 suitable candidates with \textsc{Hipparcos} parallaxes and archive photometry, with promising results. The ages we achieved were as expected from the theoretical tests for all stellar types. However, nearly all the metallicities we derived were offset from the literature (mostly our [Fe/H] values were higher). The reason for this is that the isochrone colours are offset from the observed colours. Furthermore, we also found differences in the ages and metallicities found with different combinations of passbands; in some cases the different solutions were incompatible with each other.
    \item We then studied in detail the open cluster NGC 188 using $u’g’r’i’z’$ photometric data from \citet{2007AJ....133.1409F} transformed to $ugriz$. We found clear issues when comparing these data with the isochrones that were found to be the best fit to the cluster data in \citet{2007AJ....133.1409F}. For example, the isochrones were chosen to match the position of the cluster turn-off, but are too blue for the observed giant branch stars. Similarly to the benchmark stars, we also found that the best fitting isochrone in one colour is offset from the data when examined in the other colours. The offset between passbands is so large that calculating ages/metallicities with incompatible colours leads to incorrect results on the edge of the isochrone grid due to the photometric uncertainties being much smaller than the offsets between colours and the parallax uncertainties. So to prevent this, we recommend including an “isochrone uncertainty” in the probability calculations of order $\sim0.1$\,mag, applicable to all colours or apparent magnitudes used. Finally we found that our most probable ages and metallicities for all cluster stars showed wide variation across the CMD, with a strong gradient in [Fe/H] apparent along the main sequence. Turn-off stars have most probable ages that are much younger than subgiant stars. In general all the giant stars in the cluster gave parameters that were too young and too metal-rich. 
\end{enumerate}

{\referee We have determined a number of potential areas for improvement.} In terms of the offsets seen at different evolutionary stages, there are several ways in which the isochrones are lacking. Stellar models behind the isochrones all use a mixing-length model for convection, which is calibrated to the Sun. This is known to be a poor match to giants, leading to incorrect surface parameters. For cooler stars, lacking opacity sources in the models can lead to isochrones that are too blue. The effect that atomic diffusion has on the surface metallicity of the star is not included in the models used here, although the result of this is not strongly visible in our data. If the star is an unresolved binary, the isochrones will not provide correct parameters.

It is less well understood why there are differences between results derived using different photometric passbands. As noted with the {\it Gaia} benchmark stars, the filter curves used to create the isochrones and observations can be slightly different, especially when using archive data from a range of original sources (all with small formal uncertainties). This should be less of an issue with more modern data from large surveys, where the filter curves have been well documented, and all the data comes from the same telescope. Other potential problems exist with the $T_{\rm{eff}}$-colour transformations used, and good stellar atmosphere models are needed for each photometric passband for isochrones to be correct. These are particularly lacking in the near-UV region. We have not provided an exhaustive list here, however, and acknowledge there may be other issues at work.

It is clear from this study that the crucial parameters of age and metallicity could be obtained for huge numbers of stars in the Galaxy, and beyond, with the right combination of passband sets. We suggest the use of near-UV passbands (with accurate calibration) in future large surveys, to provide the crucial information to break the age-metallicity degeneracy. We note, however, that these sort of studies are likely {\referee limited in range} until further efforts are made to improve stellar models and their corresponding isochrones. We encourage the Galactic archaeology community to support the stellar modelling teams already working to solve these difficult issues. {\referee In the short term, those using photometry as an input to Bayesian isochrone fitting based codes (to determine not only ages and metallicities, but also distances, extinction, $T_{\rm{eff}}$, and others) should consider the uncertainty on the isochrones as a crucial input.}

\begin{acknowledgements}
The authors wish to thank Christian Sahlholdt for supplying results from his age determination code. L.M.H., S.F., R.C. and T.B. were supported by the project grant `The New Milky Way’ from the Knut and Alice Wallenberg foundation. L.L. was supported by the Swedish National Space Board. SF was supported by the grant 2016-03412 from the Swedish Research Council. R.C. was supported by the project grant `IMPACT' from the Knut and Alice Wallenberg foundation and by funds from the eSSENCE Strategic Research Environment.
This work has made use of data from the European Space Agency (ESA) mission {\it Gaia} (\url{https://www.cosmos.esa.int/gaia}), processed by
the {\it Gaia} Data Processing and Analysis Consortium (DPAC, \url{https://www.cosmos.esa.int/web/gaia/dpac/consortium}). Funding for the DPAC has been provided by national institutions, in particular the institutions participating in the {\it Gaia} Multilateral Agreement. 
This publication makes use of data products from the Two Micron All Sky Survey, which is a joint project of the University of Massachusetts and the Infrared Processing and Analysis Center/California Institute of Technology, funded by the National Aeronautics and Space Administration and the National Science Foundation.
\end{acknowledgements}

%
\bibliographystyle{aa} 
\bibliography{references17.bib} 
%

\begin{appendix}

\section{Theory}
\label{section:maths}

\subsection{Parameters and observables}

\begin{figure}[t] 
\center\resizebox{0.9\hsize}{!}{\includegraphics{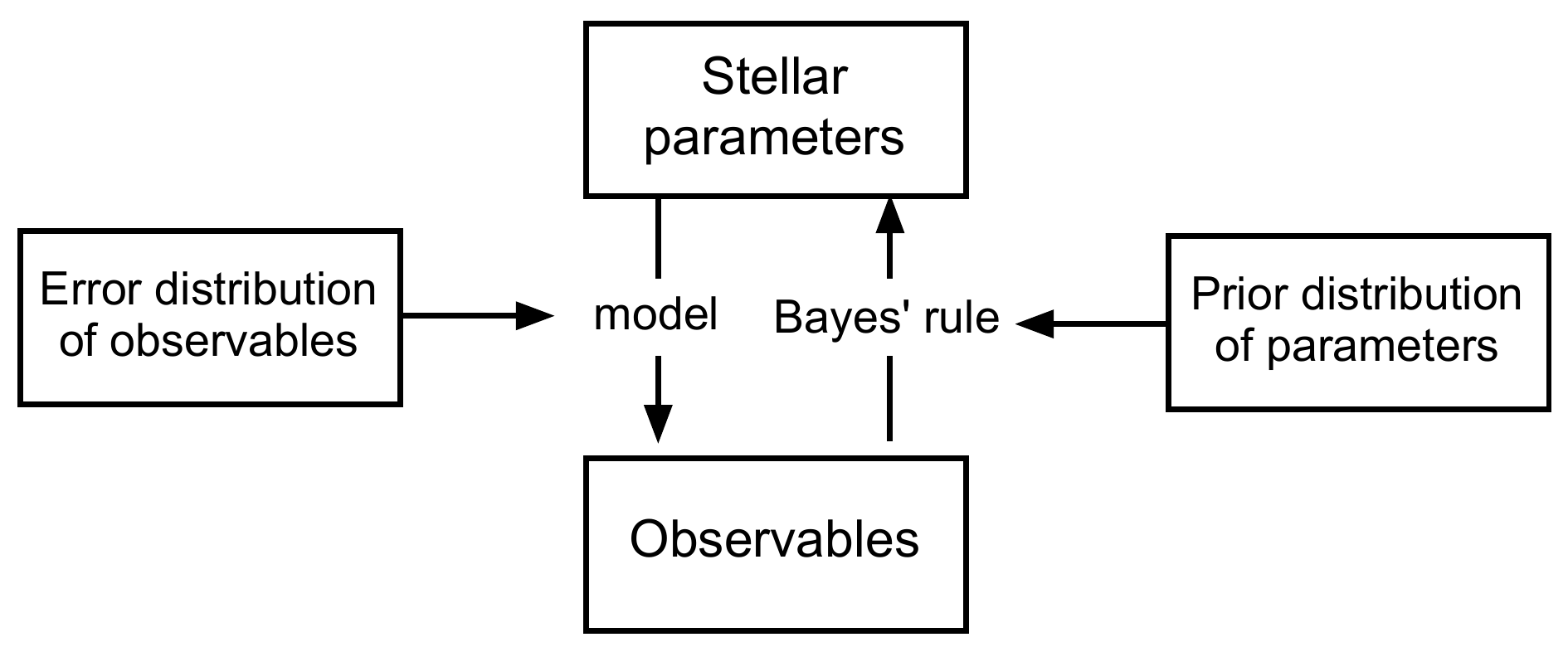}} 
\caption{Relations among the parameters and observables in Table~\ref{tab:params}.
}
\label{fig:bayes}
\end{figure}

Our goal is to derive information on certain intrinsic parameters of a star, primarily its age 
and metallicity, from observed photometric and astrometric properties such as its apparent magnitude, colour, and 
trigonometric parallax. Models of stellar evolution, stellar atmospheres, and extinction allow 
us to compute the expected values of the observables for a given set of model parameters. 
Inferences about the model parameters for given values of the observables can then be
obtained using Bayes' rule (Fig.~\ref{fig:bayes}).
  
Table~\ref{tab:params} summarises the parameters and observables considered in
this paper, and their notations. Although all the model parameters in the table are
needed, and Bayes' rule provides their six-dimensional joint probability density,
the results are simplified by marginalising over the less interesting ``nuisance'' 
parameters (Sect.~\ref{sec:gfunc}).
   
There are two mandatory observables, which are an apparent magnitude, here taken to be the 
{\it Gaia} integrated ($G$ band) magnitude, and a parallax. The other observables are optional, 
and a selection of them can be used. Each observable has an associated {\referee uncertainty} ($\sigma_G$, 
etc.). To simplify the mathematical formulation we allow only one apparent magnitude among 
the observables, i.e.\ $G$. In practice at least one optional parameter indicating effective 
temperature (e.g., a colour index) is required for the method to work. The observables may be 
represented by the vector $\vec{X}$, in which the first two elements are the mandatory observables 
$X_1=G$ and $X_2=\varpi$. The order of the remaining observables is arbitrary. 

It is important that the observables are chosen in such a way that one can reasonably assume 
Gaussian errors for all of them. The choice of parallax ($\varpi$) as one of the observables is a 
particularly important example, because any non-linear transformation of the parallax, such as 
distance $1/\varpi$, or distance modulus $\mu=-5\log(\varpi/(100~\text{mas}))$, would result 
in non-Gaussian error statistics. As far as practicable, the observables should also be statistically
independent of each other. For example, $\log g$ should not be used if it was computed using
the parallax value already included among the observables. The assumption of independent 
Gaussian errors is essential for the likelihood function, Eq.~(\ref{eq:lik}). 

The model, concisely written as $\vec{X}(m,\tau,\zeta,\alpha,\mu,A)$, actually consists of
three parts. The first part is the stellar model (isochrones and model atmosphere), which describes 
the potentially observable intrinsic properties of the star ($M_V$, $\log T_\text{eff}$, $\log g$, 
[Fe/H], [$\alpha$/Fe], $(B{-}V)_0$, $(V{-}I)_0$, etc.) as functions of the first four parameters (mass,
age, metallicity, and alpha-enhancement). The second part describes the expected values of the 
observed quantities as functions of the intrinsic properties and the remaining parameters
(distance modulus and extinction). The third part is the model of the observational errors,
which are assumed to be independent Gaussian noise with zero mean and standard deviation
$\sigma_i$ for observable $X_i$. In the following $\vec{X}$ denotes the expected (i.e., noise-free)
values predicted by the model, and $\vec{x}$ the actually measured values.

For given model parameters, the joint probability density function of the observables is then
a multi-variate normal distribution. Written as a function of the model parameters, for
given observables, this becomes the likelihood function
\begin{multline}\label{eq:lik}
L(m,\tau,\zeta,\alpha,\mu,A\,|\,\vec{x}) \\ 
=\exp\left[-\frac{1}{2}\sum_i \left(\frac{x_i-X_i(m,\tau,\zeta,\alpha,\mu,A)}{\sigma_i}\right)^2\right]\, .
\end{multline}
A multiplicative factor depending on the uncertainties $\sigma_i$ has been omitted in
this expression, which is acceptable since $\sigma_i$ are part of the ``given'' data.

\begin{table}
	\caption{Parameters and observables used in the calculation of the marginal likelihood function.}
	\centering\small
	\label{tab:params}
	\begin{tabular}{lllll}
		\hline
	 	\noalign{\smallskip}
		Model parameters & & \quad\quad & Observables & \\
		\noalign{\smallskip}
		\hline
		\noalign{\smallskip}
		initial stellar mass & $m$ && apparent magnitude & $G$ \\
		stellar age & $\tau$ && parallax & $\varpi$ \\
		metallicity & $\zeta$ && effective temperature & $\log T_\text{eff}$ \\
		alpha enhancement & $\alpha$ && gravity & $\log g$\\
		distance modulus & $\mu$ && metallicity & [Fe/H] \\
		extinction & $A$ && alpha enhancement & [$\alpha$/Fe] \\
		 & && reddening & $E_{B{-}V}$ \\
		 & && colour indices, e.g. & $(B-V)$ \\
		\noalign{\smallskip}
            	\hline
	\end{tabular}
	\tablefoot{
	$\zeta=\log_{10}(Z/Z_\odot)$ is the theoretical metallicity parameter in contrast to
	the observed metallicity denoted [Fe/H]; similarly, $\alpha$ is the theoretical alpha
	enhancement parameter in contrast to the observed quantity [$\alpha$/Fe].
	The distance modulus is $\mu=(m-M)_0$, i.e.\ excluding extinction.
	}
\end{table}

\subsection{The generalised $\mathcal{G}$ function}
\label{sec:gfunc}

The function $\mathcal{G}(\tau\,|\,\vec{x})$ introduced by \citet{2005A&A...436..127J} (hereafter JL05) 
describes the relative likelihood of different stellar ages $\tau$ based on a given set of observations 
$\vec{x}$ and theoretical isochrones. In the context of Bayesian estimation, $\mathcal{G}(\tau\,|\,\vec{x})$ is neither a posterior density, 
nor the likelihood function, but something in between: the marginal likelihood of $\tau$. Multiplied 
by the prior density $\psi(\tau)$ it gives, on normalisation, the posterior density of the age:
\begin{equation}
f(\tau\,|\,\vec{x}) = \frac{\mathcal{G}(\tau\,|\,\vec{x})\psi(\tau)}{\int_0^\infty \mathcal{G}(\tau'\,|\,\vec{x})
\psi(\tau')\,\text{d}\tau'} \, .
\end{equation}
$\mathcal{G}(\tau\,|\,\vec{x})$ is therefore proportional to the posterior density of $\tau$ for a 
flat prior. The stellar parameter of interest in JL05 was age, 
while all other parameters (mass, metallicity, ...) were treated as nuisance parameters. The 
$\mathcal{G}$ function was obtained by marginalising (averaging) over the nuisance parameters,
i.e., integrating the likelihood function multiplied by the prior density of the nuisance parameters.
$\mathcal{G}(\tau\,|\,\vec{x})$ may therefore be called the marginal likelihood of $\tau$.

In this paper we want to explore the possibility to derive both age and metallicity from {\it Gaia} data
supplemented with photometric information. We then need the joint relative likelihood of
age and metallicity ($\zeta$), represented by the generalised function 
$\mathcal{G}(\tau,\zeta,\alpha\,|\,\vec{x})$. For completeness we added the alpha enhancement 
parameter $\alpha$ as a third argument, although a fixed value such as\ $\alpha=0$ will usually be 
assumed.

The generalised $\mathcal{G}$ is the marginal likelihood of the parameters of interest
($\tau$, $\zeta$, $\alpha$), as is obtained by marginalising over the nuisance parameters,
for which their priors are needed. 
The nuisance parameters in this case are the initial mass ($m$), distance modulus ($\mu$), and
extinction ($A$). The prior density of $m$ is the initial mass function $\xi(m)$, assumed to be
known. The prior density of the remaining two parameters is written $\phi(\mu,A)$ and will
be discussed later. It is not unreasonable to assume that $\phi$ is independent of $\xi$, so 
that the joint prior density of $m$, $\mu$, and $A$ is $\xi(m)\phi(\mu,A)$. We have then
\begin{multline}\label{eq:G}
\mathcal{G}(\tau,\zeta,\alpha\,|\,\vec{x}) \\
\propto \int_A \int_\mu \phi(\mu, A)\,\text{d}\mu\,\text{d}A \int_m \xi(m)
L(m,\tau,\zeta,\alpha,\mu,A\,|\,\vec{x})\,\text{d}m  \, .
\end{multline} 
The normalisation of the $\mathcal{G}$ function is arbitrary, and we follow the convention
in JL05 to normalise it by the maximum value. 
The posterior joint density of $\tau$, $\zeta$, and $\alpha$ is proportional to 
$\mathcal{G}(\tau,\zeta,\alpha\,|\,\vec{x})$ times the joint prior of $\tau$, $\zeta$, 
and $\alpha$.

\subsection{Simplifications}

To simplify the problem somewhat, we assume that extinction can be ignored in the sense that
the observed apparent magnitudes and colours are already corrected for extinction. Thus, $A$ can
be removed from the set of model parameters, and we have
\begin{equation}\label{eq:Gs}
\mathcal{G}(\tau,\zeta,\alpha\,|\,\vec{x}) \propto \int_\mu \phi(\mu)\,\text{d}\mu \int_m \xi(m)
L(m,\tau,\zeta,\alpha,\mu\,|\,\vec{x})\,\text{d}m  \, .
\end{equation} 
Furthermore, we assume that the prior of the distance modulus has Gaussian form,
\begin{equation}\label{eq:phi}
\phi(\mu) \propto \exp\left[-\frac{w_\mu}{2}(\mu-\mu_0)^2\right]  \, ,
\end{equation} 
where $\mu_0$ and $w_\mu$ are fixed parameters. Note that we use the prior weight 
$w_\mu=\sigma_{\mu 0}^{-2}$ rather than the prior uncertainty $\sigma_{\mu 0}$ to 
characterise the prior. An advantage of this choice is that the case of a flat
prior in $\mu$ is covered by setting $w_\mu=0$ (with arbitrary $\mu_0$).

\subsection{Marginalising over $\mu$}

Inspection of the double integral in Eq.~(\ref{eq:Gs}) shows that the distance modulus $\mu$
only enters in the likelihood function via the first two observables $G$ and $\varpi$, and in 
the prior for $\mu$. That is, for $i>2$ we have 
$X_i(m,\tau,\zeta,\alpha,\mu)=X_i(m,\tau,\zeta,\alpha)$.
We can therefore write
\begin{multline}\label{eq:Gsa}
\mathcal{G}(\tau,\zeta,\alpha\,|\,\vec{x}) \propto \int_m \xi(m)\, U(m,\tau,\zeta,\alpha\,|\,\vec{x}) \\ \times \prod_{i>2}
\exp\left[-\frac{1}{2}\left(\frac{x_i-X_i(m,\tau,\zeta,\alpha)}{\sigma_i}\right)^2\right]\,\text{d}m \, ,
\end{multline} 
where
\begin{multline}\label{eq:Gsb}
U(m,\tau,\zeta,\alpha\,|\,\vec{x}) =  \int_\mu \phi(\mu) 
\exp\left[-\frac{1}{2}\left(\frac{G^\text{obs}-M_G(m,\tau,\zeta,\alpha)-\mu}{\sigma_G} \right)^2\right.   \\ 
-\frac{1}{2}\left(\frac{\varpi^\text{obs}-(100~\text{mas})\times 10^{-0.2\mu}}{\sigma_\varpi}\right)^2\Biggr]
\,\text{d}\mu  \, . 
\end{multline} 
Introducing the Gaussian prior from Eq.~(\ref{eq:phi}) we find that the integral can be simplified to
\begin{multline}\label{eq:Gs1}
U(m,\tau,\zeta,\alpha\,|\,\vec{x}) = \exp\left[-\frac{w_\mu w_G}{2w}(\mu_G-\mu_0)^2\right] \\
\times \int_\mu \exp\left[-\frac{w}{2}\left(\mu-\bar{\mu}\right)^2 
-\frac{1}{2}\left(\frac{\varpi^\text{obs}-(100~\text{mas})\times 10^{-0.2\mu}}{\sigma_\varpi}\right)^2
\right]\text{d}\mu  \, ,
\end{multline} 
where
$w = w_\mu + w_G$, $w_G=\sigma_G^{-2}$, $\bar{\mu} = (w_\mu\mu_0 + w_G\mu_G)/w$, 
and $\mu_G = G^\text{obs}-M_G(m,\tau,\zeta,\alpha)$.
It can be noted that $U(m,\tau,\zeta,\alpha\,|\,\vec{x})$ only depends on the first two observables,
$x_1=G^\text{obs}$ and $x_2=\varpi^\text{obs}$. 
Numerical integration is used to evaluate the integrals in Eqs.~(\ref{eq:Gs1})--(\ref{eq:Gsa}).

\section{Parallax uncertainty tests}
\begin{figure*}
\begin{minipage}{180mm}
  \centering
  \includegraphics[width=0.85\columnwidth]{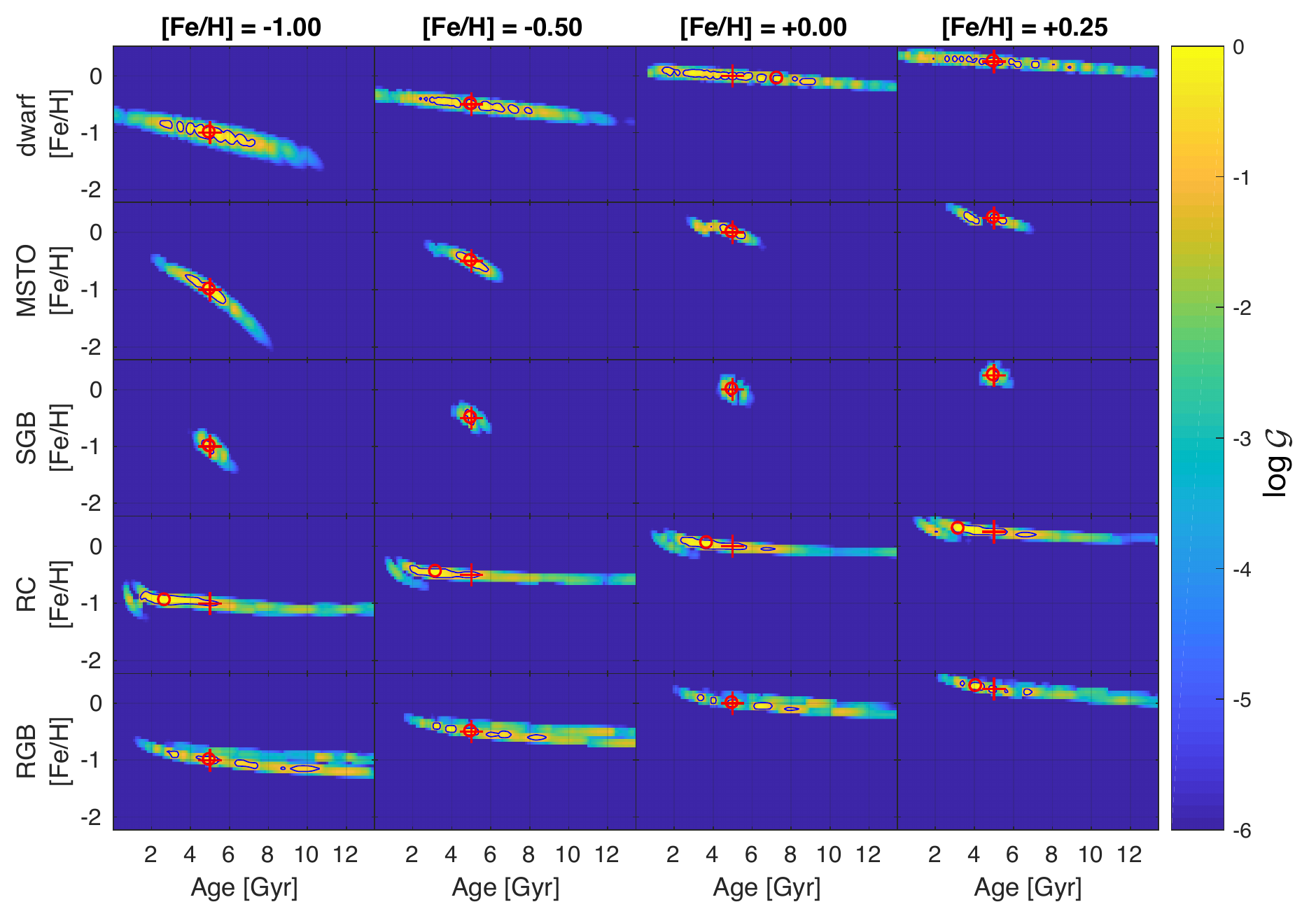}
  \caption{As in Fig. \ref{fig:2massPhotTest}, where the $\mathcal{G}$ functions have been calculated with both 2MASS $JHK\text{s}$ and SDSS $ugriz$ colours together, and the relative uncertainty of the parallax measurement is decreased to $1\%$.}
  \label{fig:highSNPhotTest}
\end{minipage}
\end{figure*}

\begin{figure*}
\begin{minipage}{180mm}
  \centering
  \includegraphics[width=0.85\columnwidth]{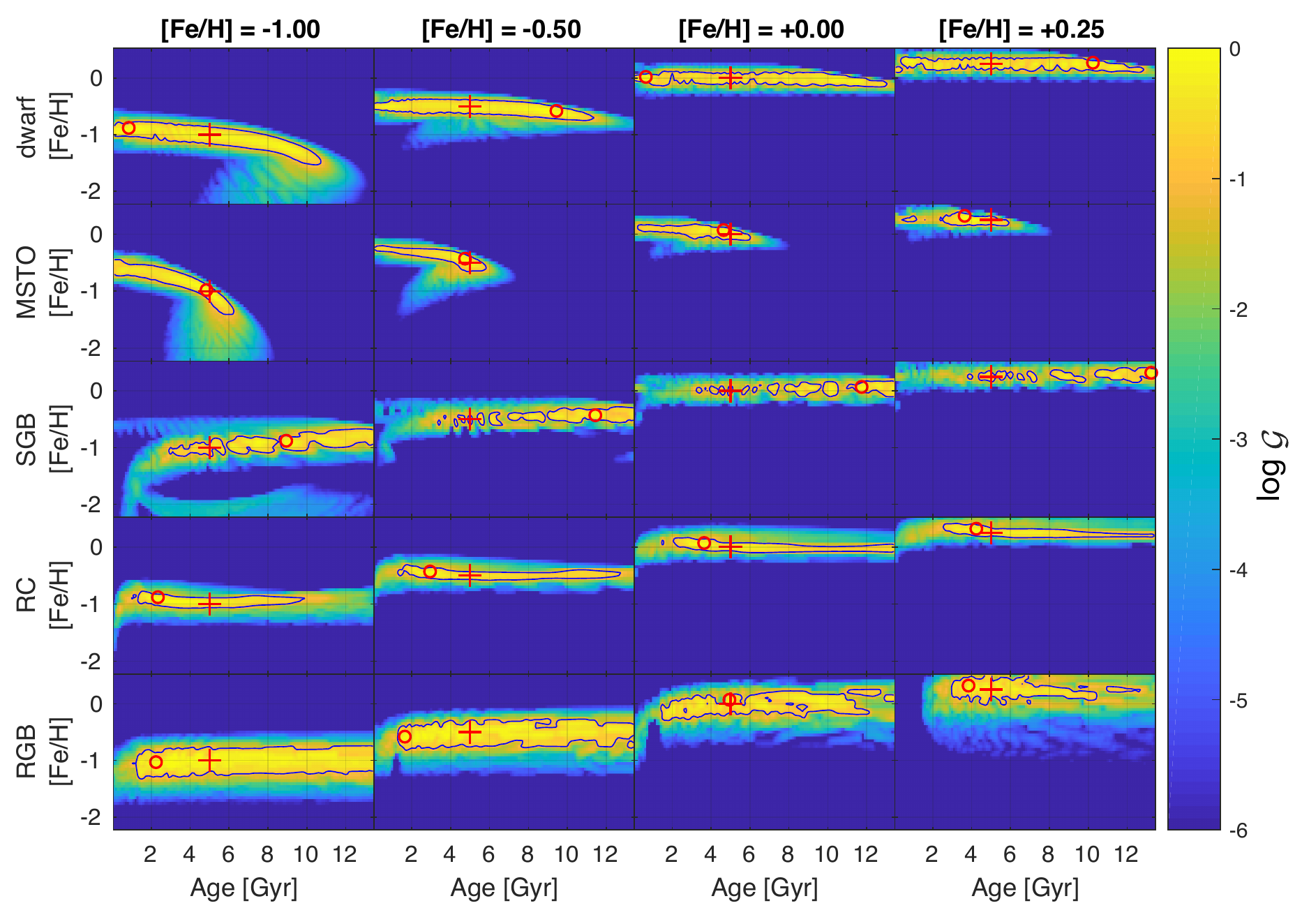}
  \caption{As in Fig. \ref{fig:2massPhotTest}, where the $\mathcal{G}$ functions have been calculated with both 2MASS $JHK\text{s}$ and SDSS $ugriz$ colours used together, and the relative uncertainty of the parallax measurement is increased to $50\%$.}
  \label{fig:lowSNPhotTest}
\end{minipage}
\end{figure*}

Here we include grids of $\mathcal{G}$ functions calculated using $ugrizJHK\text{s}$ colours as in Fig.~\ref{fig:ugrizPhotTest}, but where we have reduced and increased relative uncertainties on the parallaxes (Figs.~\ref{fig:highSNPhotTest} and \ref{fig:lowSNPhotTest} respectively).

\section{Tests with a stellar age of 2\,Gyr}
\begin{figure*}
\begin{minipage}{180mm}
  \centering
  \includegraphics[width=0.85\columnwidth]{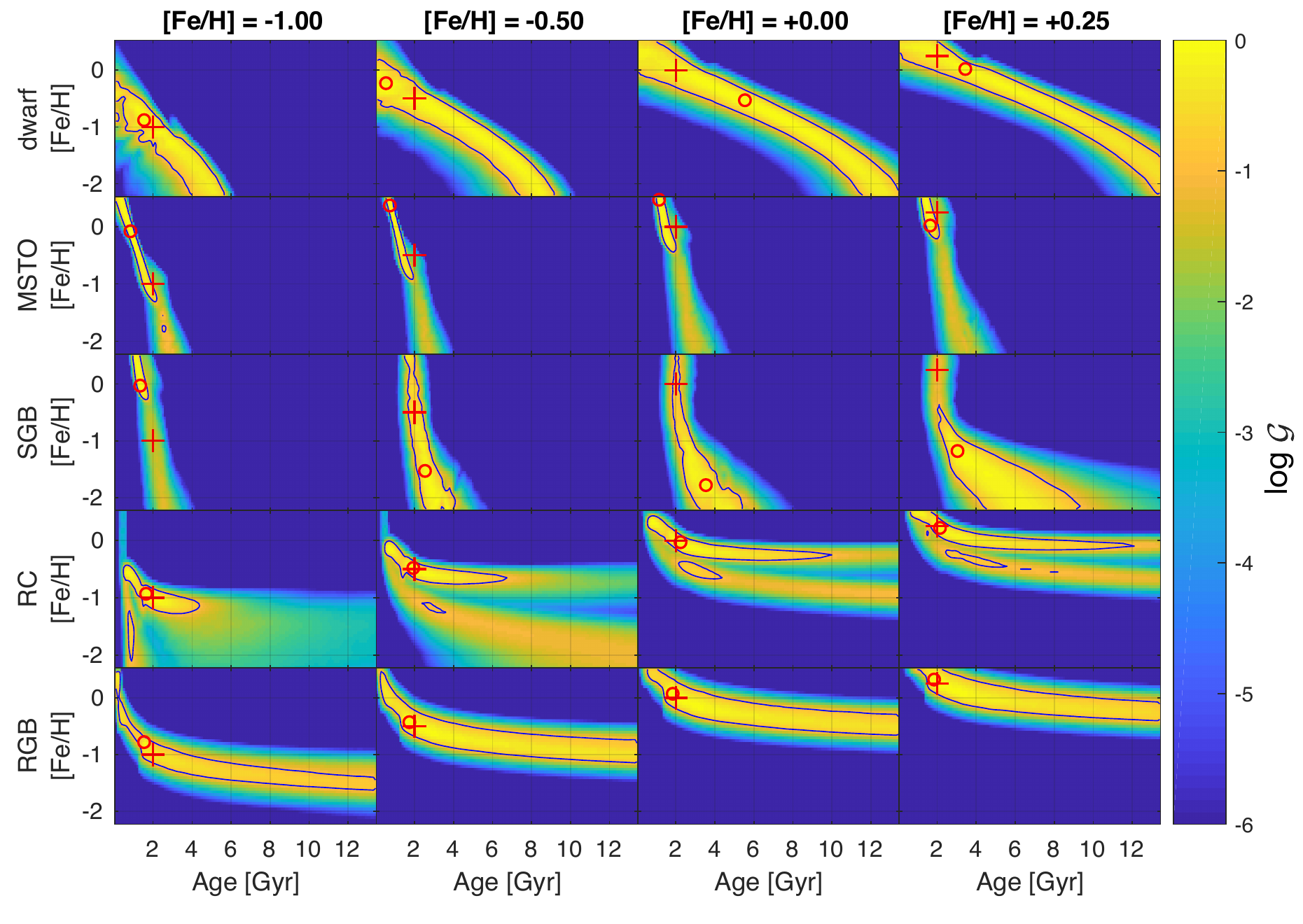}
  \caption{As in Fig. \ref{fig:2massPhotTest}, where the $\mathcal{G}$ functions have been calculated with all the 2MASS $JHK\text{s}$ colours together, and the true age of the stars tested is 2\,Gyr.}
  \label{fig:2mass2gyrPhotTest}
\end{minipage}
\end{figure*}

\begin{figure*}
\begin{minipage}{180mm}
  \centering
  \includegraphics[width=0.85\columnwidth]{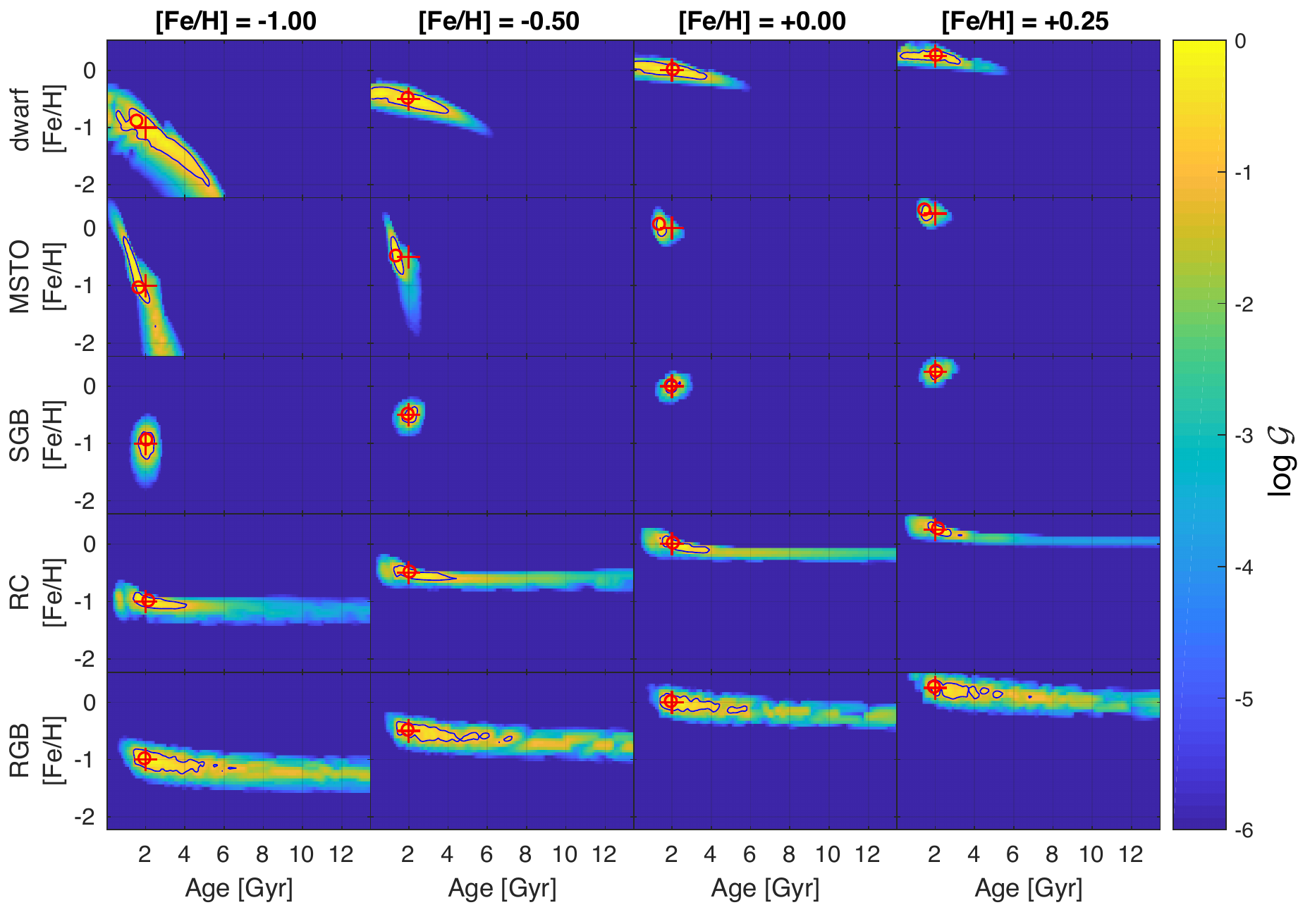}
  \caption{As in Fig. \ref{fig:2massPhotTest}, where the $\mathcal{G}$ functions have been calculated with both 2MASS $JHK\text{s}$ and $UBVRI$ colours together, and the true age of the stars tested is 2\,Gyr.}
  \label{fig:UBV2gyrPhotTest}
\end{minipage}
\end{figure*}

In this Appendix, we show a selection of $\mathcal{G}$ function grids for synthetic stars with an age of 2\,Gyr -- one with the 2MASS colours (Fig.~\ref{fig:2mass2gyrPhotTest}), and one with both 2MASS and $UBVRI$ (Fig.~\ref{fig:UBV2gyrPhotTest}).

\section{Other photometric passbands tested}
\begin{figure*}
\begin{minipage}{180mm}
  \centering
  \includegraphics[width=0.85\columnwidth]{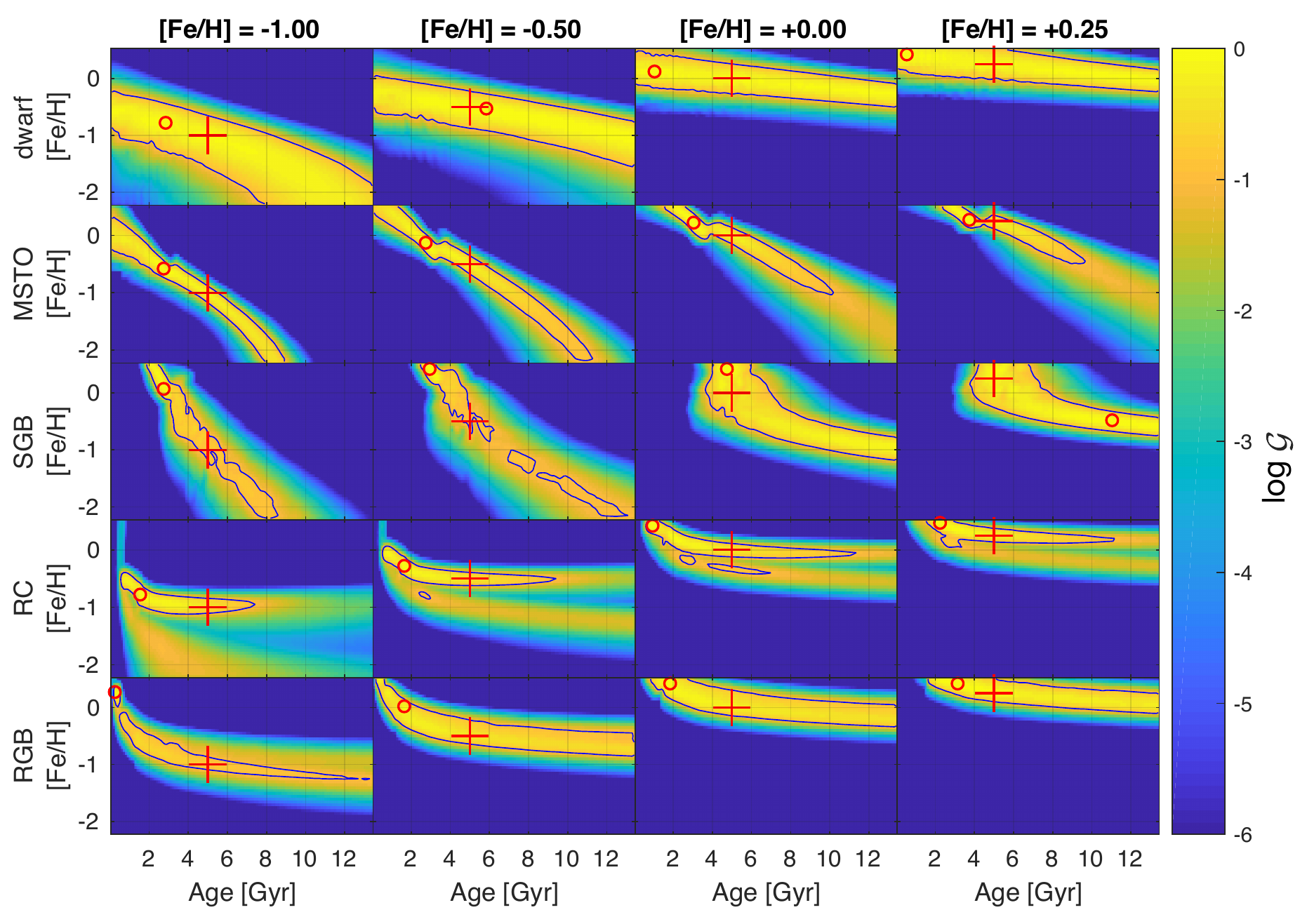}
  \caption{As in Fig. \ref{fig:2massPhotTest}, where the $\mathcal{G}$ functions have been calculated with the {\it Gaia} $G_\text{BP}$, $G_\text{RP}$ passbands.}
  \label{fig:GaiaNo2MASSPhotTest}
\end{minipage}
\end{figure*}

\begin{figure*}
\begin{minipage}{180mm}
  \centering
  \includegraphics[width=0.85\columnwidth]{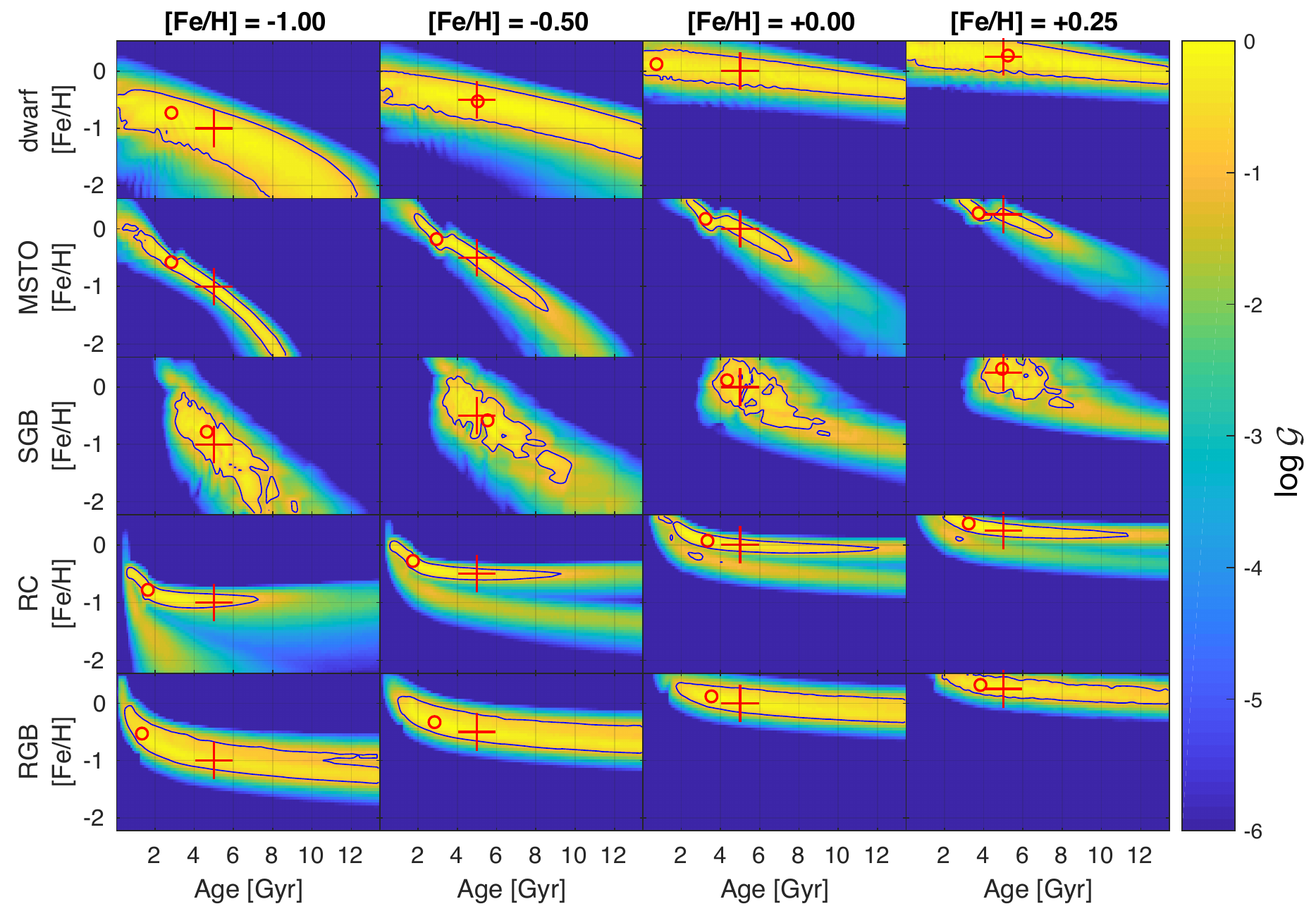}
  \caption{As in Fig. \ref{fig:2massPhotTest}, where the $\mathcal{G}$ functions have been calculated with both 2MASS $JHK\text{s}$ and {\it Gaia} $G_\text{BP}$, $G_\text{RP}$ colours, all together.}
  \label{fig:GaiaPhotTest}
\end{minipage}
\end{figure*}

\begin{figure*}
\begin{minipage}{180mm}
  \centering
  \includegraphics[width=0.85\columnwidth]{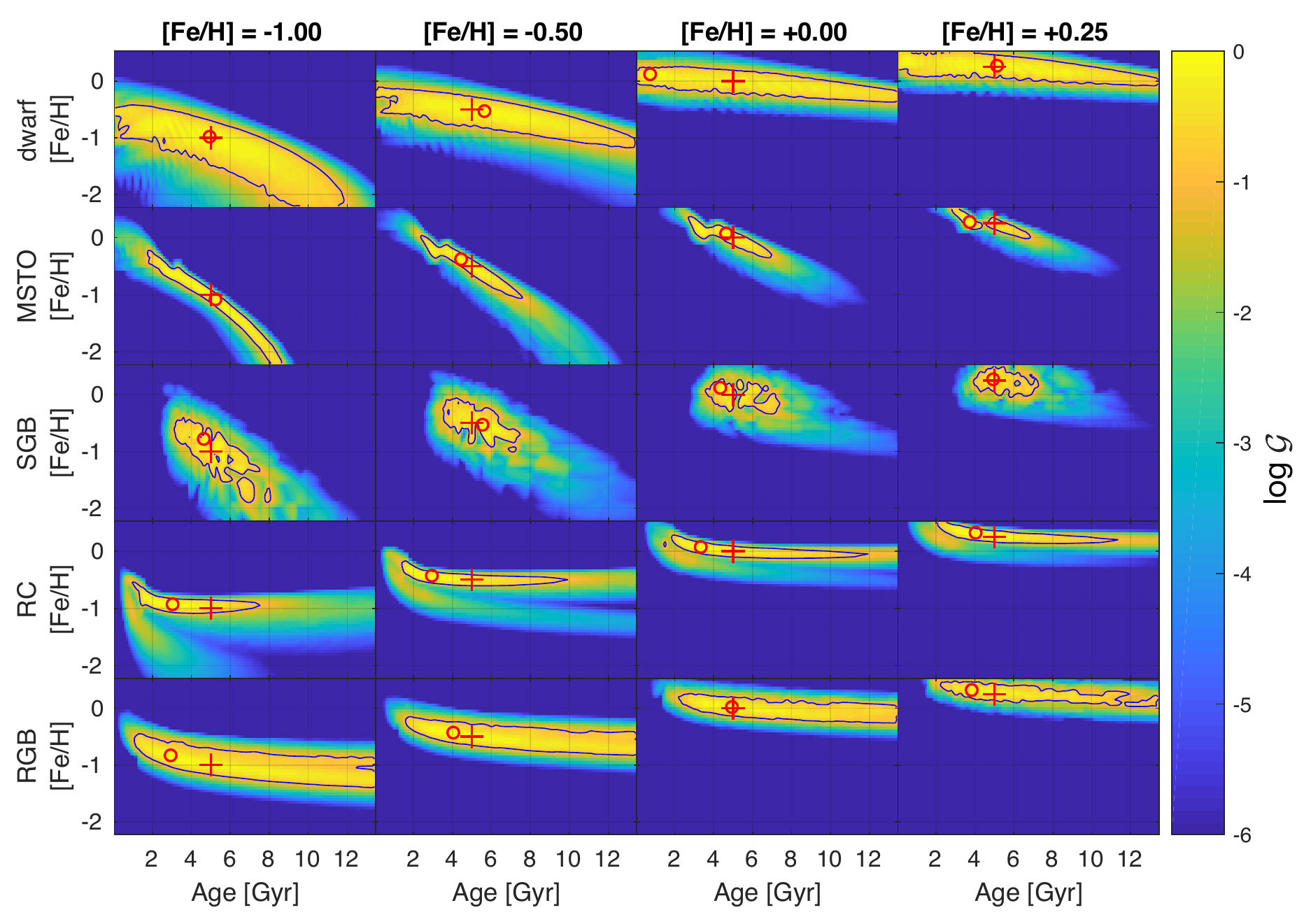}
  \caption{As in Fig. \ref{fig:2massPhotTest}, where the $\mathcal{G}$ functions have been calculated with both 2MASS $JHK\text{s}$ and $griz$ (PAN-STARRs) colours all together.}
  \label{fig:grizPhotTest}
\end{minipage}
\end{figure*}

\begin{figure*}
\begin{minipage}{180mm}
  \centering
  \includegraphics[width=0.85\columnwidth]{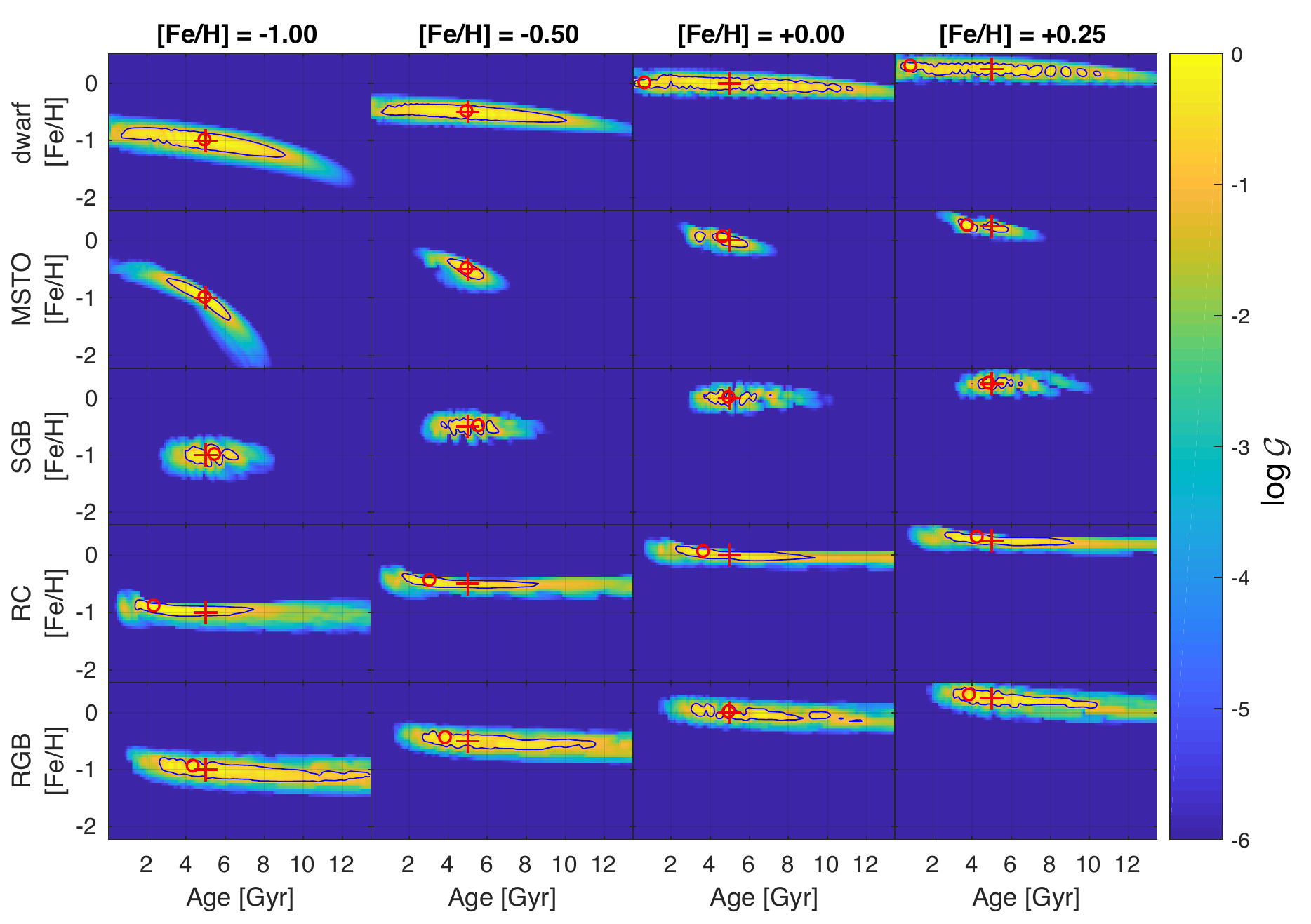}
  \caption{As in Fig. \ref{fig:2massPhotTest}, where the $\mathcal{G}$ functions have been calculated with all 2MASS and SDSS $ugriz$ colours together.}
  \label{fig:ugrizPhotTest}
\end{minipage}
\end{figure*}

Here we show additional grids of $\mathcal{G}$ functions calculated using different photometric passbands, specifically the {\it Gaia} $G_\text{BP}$ and $G_\text{RP}$ bands (Figs.~\ref{fig:GaiaNo2MASSPhotTest} and \ref{fig:GaiaPhotTest}), the PAN-STARRs $griz$ passbands (Fig.~\ref{fig:grizPhotTest}), {\referee and the SDSS $ugriz$ passbands (Fig.~\ref{fig:ugrizPhotTest})}. 

\section{Archive photometry used for the {\it Gaia} benchmark stars}
In Sect.~\ref{sec:GBS}, we calculated the $\mathcal{G}$ functions for a selection of the {\it Gaia} benchmark stars. The archive photometry used to do this is given in Table~\ref{tab:GBSphoto}.

\begin{table*}
\begin{minipage}{180mm}
	\caption{The photometry of the {\it Gaia} benchmark stars used in the tests in Sect.~\ref{sec:GBS}, with the respective references to the archives. All passbands are Johnson-Cousins, unless specified otherwise. Uncertainties for the archival data were not available, except in a couple of cases, where it is given in the table.}
	\centering
	\label{tab:GBSphoto}
	\begin{tabular}{lllrrrrl}
		\hline
	 	\noalign{\smallskip}
		Common & HD & Evol. & $U$ & $B$ & $V$ & $J$ & Photometry \\
		Name & Number & Stage & (mag) & (mag) & (mag) & (mag) & reference \\
		\noalign{\smallskip}
		\hline
		\noalign{\smallskip}
		$\beta$ Hyi & 2151 & MSTO & 3.52 & 3.41 & 2.79 & 1.79 & \citet{2002yCat.2237....0D} \\
        HD 22879 & 22879 & Dwarf & 7.15 & 7.22 & 6.67 & 5.59$\pm0.02$* & \citet{2010MNRAS.403.1949K} \& \citet{2003yCat.2246....0C} \\
        $\tau$ Cet & 10700 & Dwarf & 4.43 & 4.22 & 3.50 & 2.14 & \citet{2002yCat.2237....0D} \\
        $\beta$ Vir & 102870 & MSTO & 4.26 & 4.15 & 3.60 & 2.63 & \citet{2002yCat.2237....0D} \\
        Arcturus & 124897 & High-RGB & 2.46 & 1.18 & $-0.05$ & $-2.25$* &  \citet{2002yCat.2237....0D} \& \citet{2003yCat.2246....0C} \\
        $\mu$ Leo & 85503 & High-RGB & 6.50 & 5.10 & 3.88 & 1.93 & \citet{2002yCat.2237....0D} \\
        $\beta$ Gem & 62509 & RC & 3.00 & 2.14 & 1.14 & $-0.52$ & \citet{2002yCat.2237....0D} \\
        $\epsilon$ Vir & 113226 & RC & 4.45 & 3.71 & 2.79 & 1.31$\pm0.01$ & \citet{2002yCat.2237....0D} \& \citet{2012MNRAS.419.1637L} \\
        HD 107328 & 107328 & RC & 7.27 & 6.12 & 4.96 & 2.96 & \citet{2002yCat.2237....0D} \\
        Gmb 1830 & 103095 & Dwarf & 7.38 & 7.20 & 6.45 & 4.94* & \citet{2002yCat.2237....0D} \& \citet{2003yCat.2246....0C} \\
        61 Cyg B & 201092 & Dwarf & 8.63 & 7.40 & 6.03 & 3.55 & \citet{2002yCat.2237....0D} \& \citet{1966CoLPL...4...99J} \\
        \noalign{\smallskip}
        \hline
        \noalign{\smallskip}
	\end{tabular}
	\caption*{* In this case, the $J$ passband is the 2MASS passband. }
\end{minipage}
\end{table*}

\end{appendix}

\end{document}